**Scientific capabilities and deployment sustainability of small-scale LLMs in biological wastewater treatment**

Run-Ze Xu[1,2,†], Chu-Kuan Jiang[1,†,*], Dylan Ming-Han Li[1], Hong-Xiao Guo[1], Jia-Shun Cao[2], Guang-Hao Chen[1,*]

[1]Department of Civil and Environmental Engineering, Water Technology Center, Hong Kong Branch of Chinese National Engineering Research Center for Control & Treatment of Heavy Metal Pollution, The Hong Kong University of Science and Technology, Hong Kong, China

[2]College of Environment, Hohai University, Nanjing 210098, China

[†]**These authors contributed equally.**

[*]**Corresponding authors:**

Dr. Chu-Kuan Jiang, E-mail: redjiang@ust.hk

Dr. Guang-Hao Chen, E-mail: ceghchen@ust.hk

**Abstract**

Large language models (LLMs) are emerging as scientific assistants, yet their computational demands and limited domain specialization constrain sustainable deployment in environmental engineering. Here, we investigate whether domain-specialized small-scale LLMs can combine scientific capability with sustainable deployment in biological wastewater treatment. We developed a benchmark evaluating three scientific capabilities of LLMs: retrospective cognition, comprehension fidelity, and prospective extrapolation. BioWater (8 billion parameters, fine-tuned on specialized knowledge dataset established in this study) achieved higher comprehension-fidelity scores than participating human experts and matched a 397-billion-parameter general-purpose LLM in retrospective cognition and prospective extrapolation. Human–BioWater collaboration generated a scientific hypothesis that was subsequently supported by laboratory experiments, demonstrating its potential to contribute to prospective scientific research. We further evaluated the economic and environmental implications of LLM deployment across global wastewater treatment plants (WWTPs). Locally deployed small-scale LLMs became more sustainable than cloud-based large-scale LLMs as inference demand increased in intelligent WWTPs. These findings highlight domain-specialized small-scale LLMs as a promising pathway towards scientifically capable, computationally efficient, and sustainably deployable artificial intelligence for wastewater treatment.

## 1. Introduction

For more than a century, activated-sludge-based wastewater treatment plants (WWTPs) have underpinned municipal sanitation and aquatic environmental protection worldwide[1-3], and biological wastewater treatment is likely to remain central as WWTPs transition towards water resource recovery facilities (WRRFs)[4,5]. This transition has expanded the objectives of biological treatment beyond regulatory compliance to the simultaneous pursuit of effluent quality, low-carbon operation, process stability and resilience, resource recovery and increasingly autonomous operation[4-6]. These interdependent and sometimes competing objectives substantially increase the complexity of both process innovation and operational decision-making, making the development of more sustainable biological processes and the optimization of existing treatment systems central challenges in the sustainable transformation of wastewater treatment. Mechanistic and data-driven models have substantially advanced process understanding, prediction, and control[6,7], but they are typically developed for specific research questions or operational tasks and require substantial human expertise to integrate fragmented knowledge, models, and analytical tools. This fragmented architecture requires WWTP engineers to coordinate multiple models and translate their outputs into operational decisions, increasing the complexity of plant-wide decision-making and limiting progress towards integrated and autonomous operation. Overcoming this fragmentation requires artificial intelligence (AI) systems capable not only of performing individual predictive tasks, but also of integrating domain knowledge, scientific reasoning, and heterogeneous analytical tools across research and operational workflows.

The emergence of large language models (LLMs) offers a potential means of integrating these fragmented workflows by combining knowledge synthesis, scientific reasoning, and decision support within a common computational framework[8,9]. Recent works such as Co-Scientist[10], Empirical Research Assistance[11], and Robin[12] further illustrate the growing potential of LLM-based scientific assistants to participate in scientific workflows. However, reliance on large general-purpose LLMs presents important challenges for specialized environmental applications. Their broad training does not necessarily provide sufficient expertise for narrowly defined scientific and engineering problems[9,13], while their substantial computational requirements can constrain cost-effective, privacy-preserving and locally controlled deployment[14,15]. These limitations are particularly relevant to WWTPs, where AI systems may ultimately need to interact repeatedly with plant data and support persistent monitoring, diagnosis and operational decision-making. In parallel with the continued scaling of general-purpose models, growing attention has therefore been directed towards small-scale LLMs (<10 billion parameters) that can be more readily deployed on local or edge computing infrastructure[16-18]. Evidence from several specialized applications suggests that, with appropriate domain adaptation, small-scale LLMs can match or even exceed much larger general-purpose models on narrowly defined tasks[19-22].

Despite their potential advantages in computational efficiency and local deployability, the capabilities of small-scale LLMs remain largely unexplored in biological wastewater treatment. Recent LLM-based tools, such as OpenAqua[8],

WaterGPT[23] and WaterRAG[13], have demonstrated the value of language models for technical question answering and engineering knowledge support in water-related domains. However, these applications have focused predominantly on retrieving, synthesizing, and communicating established knowledge[24], providing limited evidence of whether small-scale LLMs possess the broader intelligence required to address complex scientific and operational problems in wastewater treatment. In the development of wastewater treatment technology, an AI-based scientific assistant would need to comprehend specialized literature, identify relationships within existing evidence, reason about unresolved questions, formulate testable hypotheses, and support experimental design and data interpretation[12]. In practical WWTPs, it would ultimately need to operate within dynamic engineering environments and support recurrent tasks spanning process monitoring, diagnosis, optimization, and operational decision-making[25]. Whether small-scale LLMs can provide sufficient domain intelligence to support these research and operational workflows has yet to be systematically evaluated. Moreover, practical intelligence in wastewater treatment cannot be considered independently of deployability. LLMs intended for persistent or high-frequency use in WWTPs must deliver useful capabilities without imposing disproportionate computational, economic, and environmental burdens. The central challenge is therefore not simply whether smaller models can reproduce the knowledge-retrieval performance of larger LLMs, but whether domain-specialized small-scale LLMs can achieve sufficient scientific capability while retaining the efficiency and local deployability needed for sustainable, intelligent wastewater treatment. Addressing this question requires a systematic framework that evaluates both their research-relevant scientific capabilities and the practical implications of deploying them in future intelligent WWTPs.

Here, we establish a benchmark framework to systematically evaluate the scientific capabilities of small-scale LLMs in biological wastewater treatment. The framework evaluates three complementary dimensions (i.e., retrospective cognition, comprehension fidelity and prospective extrapolation) using recent research literature dataset together with LLM-based judgment and human-expert evaluation. We further constructed a specialized knowledge dataset from the historical biological wastewater treatment literature and used it to develop BioWater, an 8-billion-parameter domain-specialized LLM. We benchmarked BioWater against other open-source LLMs, a 397-billion-parameter general-purpose model and human experts, and further examined its capacity to participate in hypothesis-driven research through laboratory experimental validation. BioWater achieved higher comprehension-fidelity scores than the participating human experts under our benchmark setting and performance comparable to the 397-billion-parameter model on the retrospective cognition and prospective extrapolation tasks evaluated here. A hypothesis generated through human-BioWater collaboration was subsequently supported by laboratory experiments, providing prospective evidence that a domain-specialized small-scale LLM can contribute meaningfully to scientific research in biological wastewater treatment. Beyond scientific capability, we estimated the economic and environmental implications of deploying small-scale LLMs locally relative to cloud-based general-purpose LLMs across global WWTPs under different inference-demand scenarios, revealing that locally deployed small-scale LLMs can become more sustainable than cloud-based large-scale LLMs in intelligent WWTP control scenarios requiring frequent inference. Together, our results establish a framework for evaluating research-relevant intelligence in domain-specific LLMs. They further demonstrate the potential of combining domain specialization with smaller model scale to develop scientifically capable, computationally efficient and practically deployable AI systems for the sustainable transformation of wastewater treatment.

## 2. Results

### 2.1 Benchmark framework and design philosophy

Our proposed benchmark framework deconstructs the scientific capabilities of LLMs into three interconnected dimensions spanning a temporal continuum of past, present, and future: retrospective cognition, comprehension fidelity, and prospective extrapolation (Fig. 1). To mitigate data leakage and copyright concerns, we constructed the base dataset for our benchmark using 100 abstracts from recently published, peer-reviewed research articles related to biological wastewater treatment (Table S1 and Table S2). Crucially, none of the selected abstracts is present in the pre-training corpora of the target LLMs evaluated in this study (Table S3). This data curation strategy endows our benchmark framework with inherent dynamism and updatability. Consequently, this strategy not only fulfills the requirement of assessing LLMs' scientific capabilities using the latest research findings but also circumvents the pervasive issue of data contamination inherent in traditional, static benchmark datasets[26,27]. The retrospective cognition and prospective extrapolation tasks were conducted using an open-ended generation strategy, where responses were generated based on well-designed prompts and each collected abstract. The LLM-as-a-judge method (Table S4) was applied to automatically evaluate the open-ended content due to the recent successes in LLM-based judgment[28-31]. For the comprehension fidelity task, LLMs were evaluated on their ability to distinguish original academic abstracts from coherently altered counterparts by selecting the version with lower perplexity, which reflects higher predictive certainty based on their pretrained distributions[20]. The LLMs' performance on this task was subsequently benchmarked against the judgments of domain experts, which reflect their intuitive reasoning and professional knowledge, providing an interpretable baseline for understanding LLMs' capability relative to human expertise[32,33]. By establishing this benchmark, we provide a structured method to quantify the scientific capabilities exhibited by LLMs in the context of biological wastewater treatment and potentially other research fields, as well as a scalable foundation for future studies evaluating more powerful LLMs.

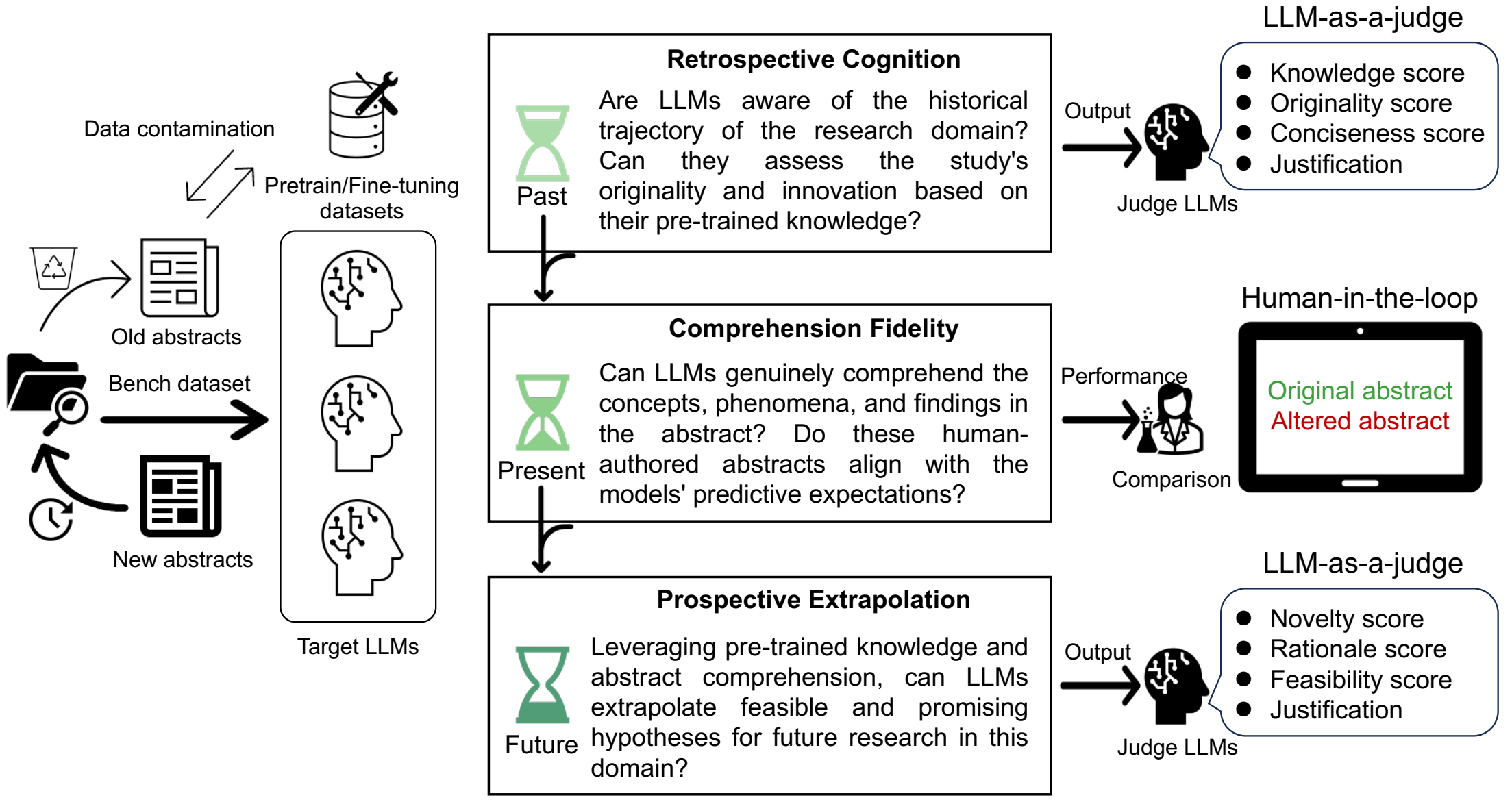

**Fig.1|Schematic overview of the dynamic benchmark framework for evaluating the scientific capabilities of LLMs.** To ensure a zero-contamination evaluation, the dataset is constructed using recently published, peer-reviewed research abstracts that are strictly absent from the pre-training and fine-tuning corpora of the target LLMs. The framework deconstructs scientific capabilities into three temporal dimensions: retrospective cognition (past), comprehension fidelity (present), and prospective extrapolation (future). For task evaluation, retrospective cognition and prospective extrapolation employ open-ended generation scored by an LLM-as-a-judge approach. Meanwhile, comprehension fidelity evaluates the LLMs' perplexity-based ability to distinguish original abstracts from altered counterparts, which is directly benchmarked against human expert performance in a controlled survey. This design enables a systematic quantification of LLMs' scientific capabilities and provides a scalable foundation for future assessments.

## 2.2 Evaluation results of scientific capabilities

We evaluated seven open-source LLMs (Table S3), including five small-scale base LLMs (<10 billion parameters), one large-scale base LLM, and BioWater (a model fine-tuned from Ministral-3:8b). The retrospective cognition and prospective extrapolation capabilities of the target LLMs were evaluated by five large-scale judge LLMs: DeepSeek-v4-pro, GLM-5.1, MiniMax-M3, Kimi-k2.6, and Gemini3.1 Pro (Table S4). The highest and lowest scores among the five LLM judges were removed, and the final score was calculated as the mean of the remaining three scores. The comprehension fidelity of target LLMs was directly compared with that of human experts through an abstract test.

**Retrospective cognition task.** The retrospective cognition task is designed to evaluate the capability of LLMs in recalling historical context and understanding the originality of specific research topics based on their pretraining knowledge. The content for evaluating retrospective cognition was generated by the target LLMs using a structured generation prompt (Text S1) and 100 collected abstracts. Subsequently, the five judge LLMs scored the generated content across the seven target models using a standardized scoring prompt. Owing to differences in model architectures and training datasets, the seven target LLMs achieved markedly different retrospective cognition scores for biological wastewater treatment research (Fig. 2). The evaluation scores fell into three distinct performance tiers, ranked in ascending order as follows: LFM2.5-8B-A1B, Granite4.1:8b, and Qwen3.5:9b, followed by Gemma4:e4b and Ministral-3:8b, with Qwen3.5:397b and BioWater achieving the highest scores. All five evaluators consistently scored Qwen3.5:9b lower than Qwen3.5:397b (Fig. 2d), aligning with the expectation that scaling model parameters enhances performance[34,35]. Furthermore, this result demonstrates that our prompt design can elicit differentiated scores across various LLMs, thereby validating the sensitivity and effectiveness of the retrospective cognition task. Among the five small-scale LLMs, Ministral-3:8b achieved the highest scores. Therefore, we used Low-Rank Adaptation fine-tuning to augment Ministral-3:8b with additional biological wastewater treatment knowledge to form the BioWater model (Text S12). Fine-tuning substantially improved the retrospective cognition performance of BioWater relative to its baseline model, Ministral-3:8b (Fig. 2e). Moreover, BioWater achieved scores comparable to those of Qwen3.5:397b (Fig. 2f and Fig. 2d), which has roughly 50 times more parameters than BioWater. These results indicated that small-scale LLMs have great potential in improving retrospective cognition capability by augmenting scientific knowledge. The box plot of the categorical score distribution showed that the Qwen3.5:9b and Qwen3.5:397b with the identical training dataset had similar score distribution patterns on these research topics, whereas the Ministral-3:8b and BioWater with the identical model structure and different training dataset had different score distribution patterns on these research topics (Fig. 2h). These results further confirm that training knowledge strongly affects the retrospective cognition performance of LLMs across different research topics in the field of biological wastewater treatment. Overall, our evaluation shows that the model structure and training materials both affected the LLMs' retrospective cognition capability, which could be fairly and robustly evaluated by independent judge LLMs. Meanwhile, the retrospective cognition capability of small-scale LLMs can be enhanced to the level of large-scale LLMs through fine-tuning and knowledge augmentation.

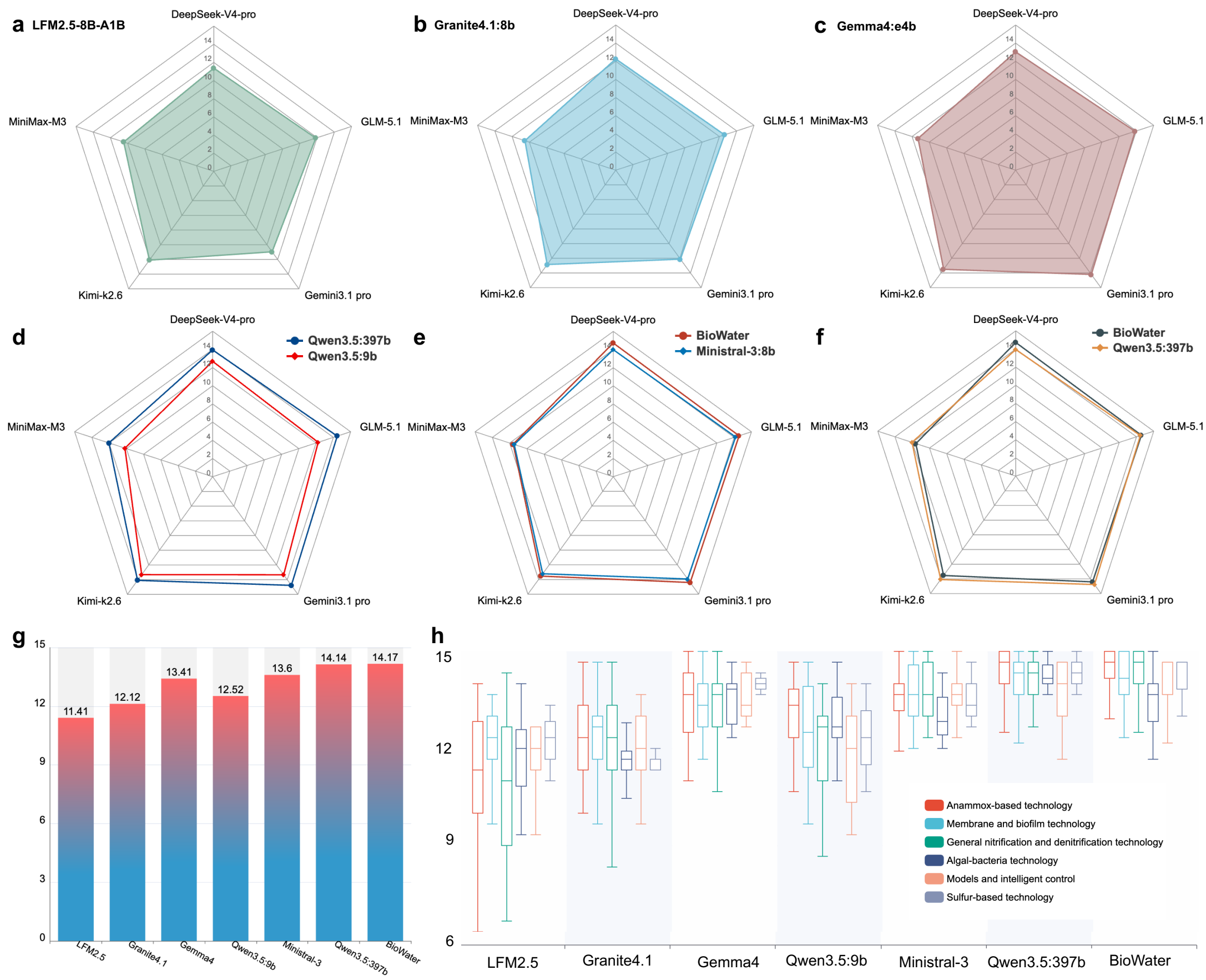

**Fig.2| Retrospective cognition evaluation reveals three performance tiers and demonstrates that fine-tuning small-scale LLMs with domain knowledge can match large-scale LLM performance. a–f**, Radar charts comparing retrospective cognition scores across different model evaluations for LFM2.5-8B-A1B (a), Granite4.1:8b (b), Gemma4:e4b (c), Qwen3.5:397b versus Qwen3.5:9b (d), BioWater versus its base model Ministral-3:8b (e), and BioWater versus Qwen3.5:397b (f). **g**, Overall performance comparison showing the final retrospective cognition scores for all seven target LLMs. **h**, Box plot depicting score distributions across research topic categories, highlighting the distinct impacts of training data versus model architecture on performance patterns.

**Comprehension fidelity task.** To test the comprehension fidelity capability of LLMs, both the target LLMs and human experts were tasked with choosing the original abstract from two versions (Fig. 3). The target LLMs were scored as choosing the original abstract with the lower perplexity (Fig. 3a), whereas the human experts conducted the abstract test on an offline website and selected the original abstract version based on their intuitive reasoning and professional knowledge (Fig. 3b). A total of 30 human experts passed the quality-control trap check, and their results were included in the final analysis. The comprehension fidelity performance of the target LLMs followed the same ranking as in the retrospective cognition task: BioWater > Ministral-3:8b > Gemma4:e4b > LFM2.5-8B-A1B (Fig. 3c). These results imply that the retrospective cognition capability was an important foundation supporting LLMs in understanding the research content. Notably, human experts (~77%) had comparable performance to Gemma4:e4b (~81%), while Ministral-3:8b and BioWater outperformed human experts on the comprehension fidelity task (88%-92%). A prevalent concern when LLMs achieve high scores on benchmarks is data contamination, where the evaluation dataset overlaps with the LLMs' training corpus, leading to memorized responses[20,36]. To avoid this, the abstracts utilized in our benchmark were rigorously excluded from the training data of the target LLMs, confirming that the superhuman performance of Ministral-3:8b and BioWater was not driven by data memorization. For all target LLMs, the average perplexity difference between original and altered abstracts in correct samples was larger than in incorrect samples (Fig. 3d), demonstrating that the LLMs chose the original abstracts according to their knowledge rather than randomness. Collectively, these results demonstrated that the comprehension fidelity of different LLMs on certain research topics could be distinguished through this task, which was beyond the traditional recall task in other benchmarks, and confirmed the stronger scientific understanding of small-scale LLMs than the human experts in the domain of biological wastewater treatment.

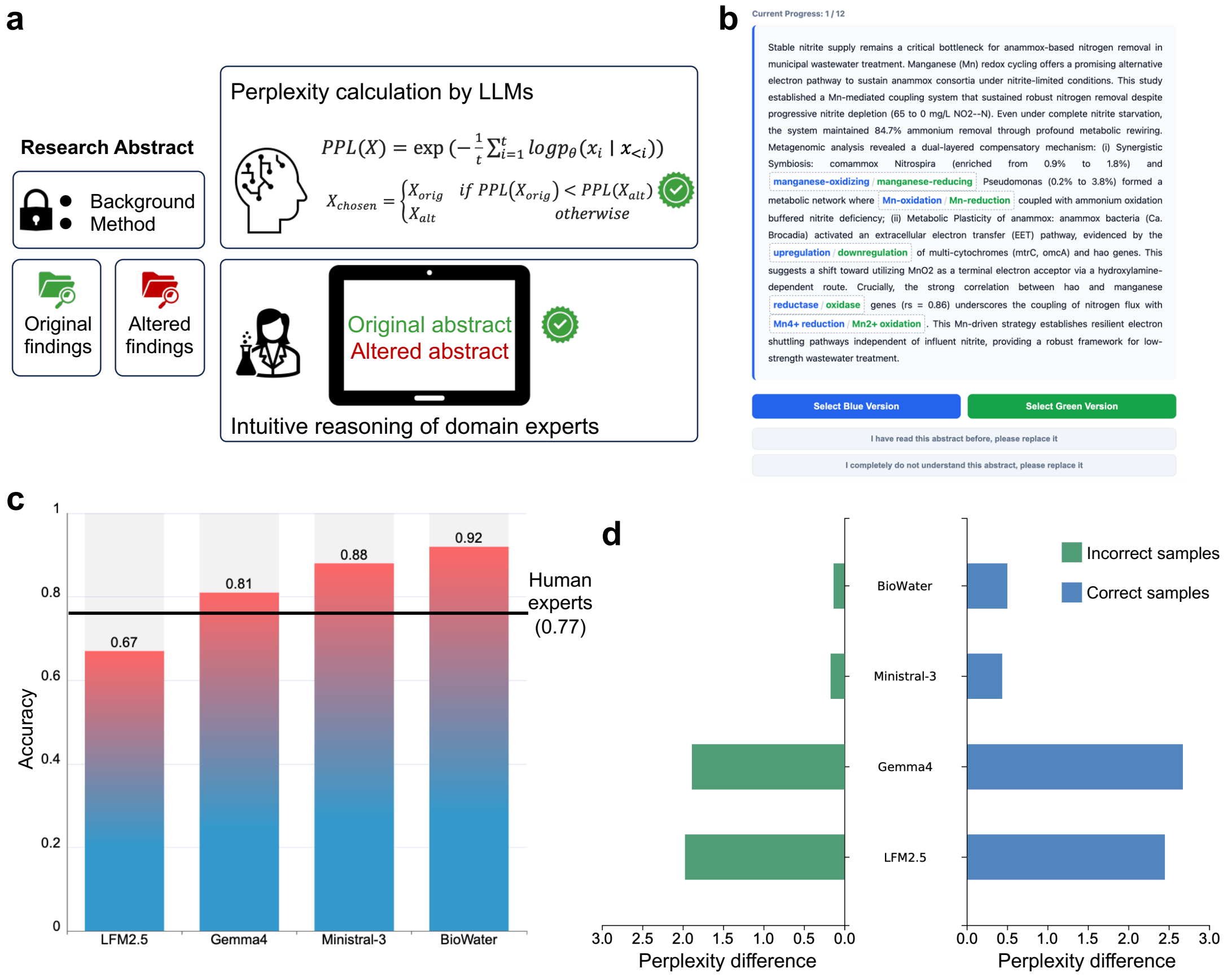

**Fig.3| Domain-specific small-scale LLMs demonstrate superior comprehension fidelity over human experts in biological wastewater treatment. a**, conceptual design of the comprehension fidelity task. Research findings in original abstracts were modified while keeping background and methodologies intact; model preferences were quantified via perplexity differences. **b**, representative user interface of the testing platform used by human experts to discriminate original abstracts from altered versions based on domain expertise. **c**, performance comparison between human experts and target LLMs, showing that specialized models (BioWater and Ministral-3:8b) outperform human experts. **d**, average perplexity difference between original and altered abstracts for correct vs. incorrect model selections, indicating knowledge-driven decision-making rather than random guessing.

**Prospective extrapolation task.** In this task, we evaluated the forward-looking performance of target LLMs in hypothesis generation, scientific rationale explanation, and feasibility analysis. Within the domain of biological wastewater treatment, these scientific capabilities are crucial for developing new treatment technologies and discovering novel pathways and mechanisms. The evaluation procedure of the prospective extrapolation task is similar to that of the retrospective cognition task based on the LLM-as-a-judge methodology. Granite 4.1 and LFM2.5-8B-A1B received poor scores, implying that their extrapolative outputs were not acceptable to the five judge LLMs (Fig. 4a). Ministral-3:8b and Gemma4:e4b achieved moderate scores (Fig. 4b), whereas Qwen3.5:9b, Qwen3.5:397b and BioWater achieved the highest scores in the prospective extrapolation task (Fig. 4c). Interestingly, the prospective extrapolation scores of Qwen3.5:9b and Qwen3.5:397b were very close (Fig. 4d), implying that the significant increase in the parameter scale of Qwen3.5:397b did not yield a substantial improvement in its forward-looking performance. In contrast, fine-tuning Ministral-3:8b to develop BioWater increased the average score from 12.19 to 13.01 (Fig. 4d). These results suggest that injecting domain-specific knowledge yields a more substantial improvement in prospective extrapolation performance than merely scaling up model parameters. Furthermore, we observed higher score variability across diverse topic categories in the prospective extrapolation task (Fig. 4e) relative to the retrospective cognition task (Fig. 2e). Such high variability indicates that general LLMs lack consistent reasoning capabilities across specialized topics, further underscoring the necessity of domain-specific knowledge injection to achieve robust scientific reasoning.

**Enhancement of scientific capabilities by fine-tuning.** To visualize the performance enhancements achieved through fine-tuning, the scores of the target LLMs are presented in scatter plots (Fig. 4f, g). Notably, domain-specific fine-tuning yielded a performance improvement of 4%-6.7% on Ministral-3:8b. This improvement was particularly pronounced for the retrospective cognition capability, where BioWater slightly outperformed the much larger Qwen3.5:397b (Fig. 4f). Meanwhile, BioWater achieved the highest comprehension fidelity accuracy among other small-scale LLMs (Fig. 4g) and achieved a comparable performance to Qwen3.5:397b in prospective extrapolation capability. These results indicate that the Low-Rank Adaptation fine-tuning successfully instilled specialized knowledge into the base model. Our benchmarking results reveal the competitive scientific capabilities of BioWater (8 billion parameters) when compared to the much larger Qwen3.5:397b (397 billion parameters). The practical potential of BioWater as an AI scientific assistant was further validated through subsequent human-AI collaboration experiments.

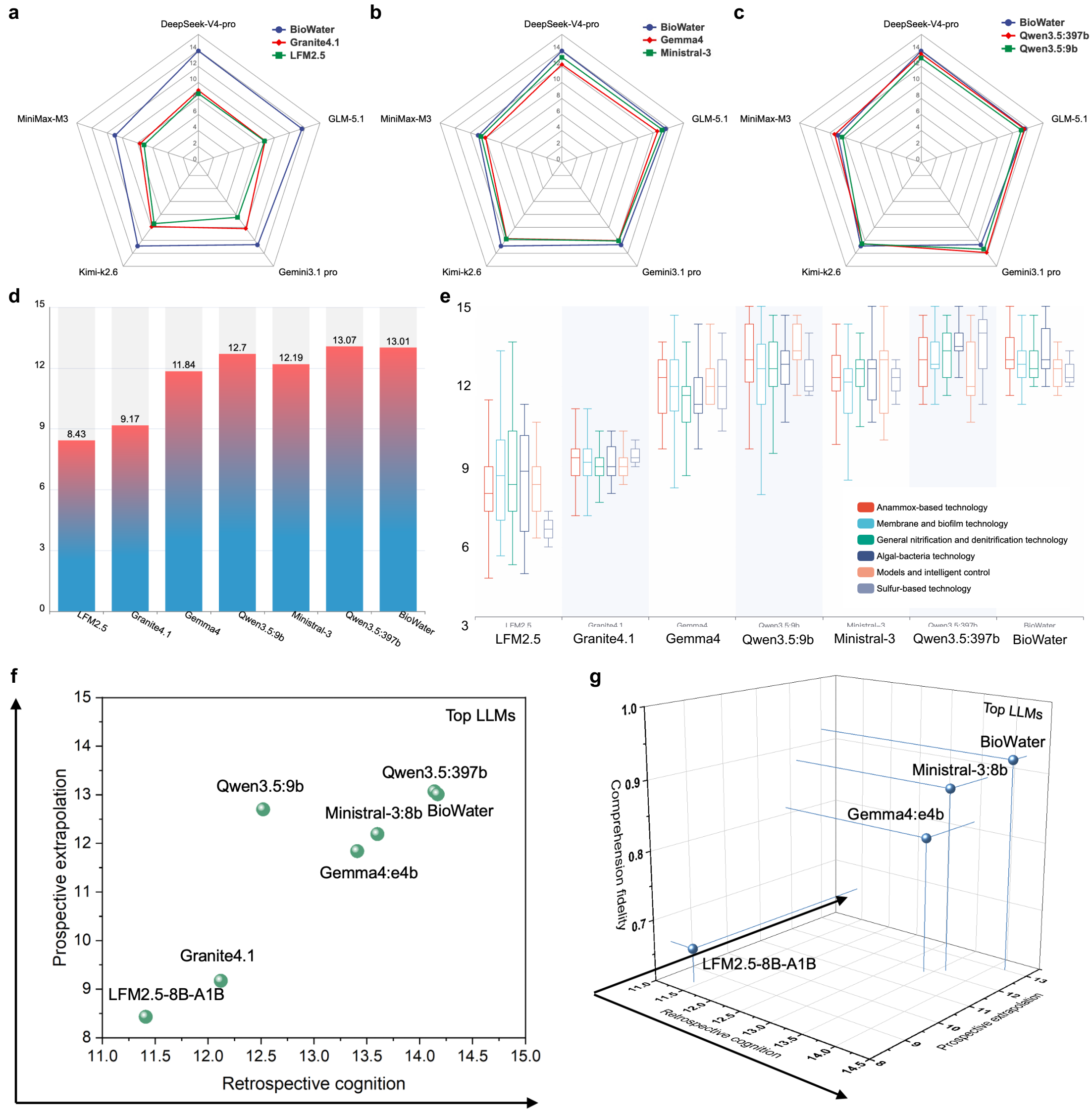

**Fig.4| Domain-specific fine-tuning endows small-scale LLMs with superior prospective extrapolation and comprehensive scientific capabilities. a-c**, radar charts comparing the prospective extrapolation scores across different models, revealing that BioWater demonstrates leading forward-looking performance among small-scale and baseline models. **d**, bar chart summarizing the final evaluation scores, highlighting the significant performance gain of BioWater over its base model. **e**, box plot illustrating the score distribution across various topic categories; the high variability emphasizes the necessity of domain-specific knowledge for stable reasoning. **f**, scatter plot mapping retrospective cognition versus prospective extrapolation, visually confirming BioWater's capability comparable to the Qwen3.5:397b. **g**, scatter plot integrating comprehension fidelity, showing BioWater achieves the best overall balance among small-scale LLMs.

**Practical validation as an AI scientific assistant.** To validate the practical potential of BioWater as an AI scientific assistant, we collaborated with BioWater in a scientific workflow involving hypothesis formulation, experimental design, and data analysis (Fig. 5). Based on our research interest in sulfide-driven partial denitrification and the anaerobic ammonium oxidation (anammox) process[37-39], we asked BioWater to propose the scientific hypotheses related to this topic.

Anammox, as an autotrophic nitrogen-removal biological wastewater treatment process, has been considered one of the most promising solutions for transforming wastewater treatment toward greener and carbon-neutral operation[40,41]. However, achieving stable, highly efficient, low-carbon nitrogen removal via the anammox process in mainstream wastewater treatment plants remains an unresolved challenge. Currently, our laboratory is dedicated to establishing a stable sulfur-driven partial denitrification coupled with the anammox process. To support this effort, we employed BioWater to analyze the issues identified during our experiments and to formulate hypotheses that merit further investigation (Fig. 5a). We then selected one hypothesis with high potential for further investigation based on our expert judgment and domain knowledge. To further collaborate with BioWater, we asked it to design a batch experimental plan for validating this hypothesis (Text S13). The generated experimental plan was modified by human experts. We then conducted the batch experiments manually at lab-scale and provided the resulting data to BioWater for autonomous analysis and hypothesis validation. Human experts independently analyzed the same data.

The subsequent batch experiments revealed that variations in the organic chemical oxygen demand (CODc) to sulfur equivalent COD (CODs) ratio strongly regulated nitrogen removal performance and nitrite dynamics in the Mixotrophic Partial Denitrification Anammox (MPDA) system. Specifically, a CODc/CODs ratio below 2.5 represents the optimal operational window, promoting nitrite accumulation, maintaining stable TN removal, and sustaining a high anammox contribution (Fig. 5b). These results validated the hypothesis proposed by BioWater: "*the key challenge lies in precisely controlling the balance between sulfide-driven autotrophic partial denitrification (APD) and organic matter-driven heterotrophic partial denitrification (HPD). Specifically, the optimal mixotrophic ratio—the proportion of autotrophic versus heterotrophic nitrate reduction—is critical to maintain stable nitrite accumulation for anammox.*" Moreover, the data analysis results of BioWater present great consistency with those of human experts according to the consistency evaluation results of judge LLMs (Text S14). Overall, this workflow verified BioWater's practical potential as a highly capable AI scientific assistant in the domain of biological wastewater treatment.

**a**

**User Prompt**

I'm developing a novel wastewater treatment process based on sulfide-driven partial denitrification and anammox for energy-efficient nitrogen removal in mainstream wastewater through. While currently, sulfide produced from sulfate reduction in upstream biological sulfate reduction may not be sufficient or excess which is harmful for stable and efficient nitrogen removal via sulfide-driven partial denitrification due to the lack of sulfide supply or toxicity of high sulfide concentration. Meanwhile, the residual organics after biological sulfate reduction may affect the stability of sulfide-driven partial denitrification process and nitrogen removal efficiency. Based on this mainstream wastewater condition, could you propose what is the most important hypothesis that need to be further investigated to handle this mixotrophic condition in partial denitrification and anammox process?

**BioWater Answer**

Based on your description, I would hypothesize that the key challenge lies in precisely controlling the balance between sulfide-driven autotrophic partial denitrification (APD) and organic matter-driven heterotrophic partial denitrification (HPD). Specifically, the optimal mixotrophic ratio—the proportion of autotrophic versus heterotrophic nitrate reduction—is critical to maintain stable nitrite accumulation for anammox. This hypothesis stems from the observation that residual organics can shift the electron donor preference away from sulfide, potentially reducing the efficiency of sulfide-driven partial denitrification and destabilizing the system when organic loads fluctuate. Investigating how different organic matter concentrations or compositions influence the competition between autotrophic and heterotrophic denitrifies could reveal operational strategies (e.g., adjusting carbon-to-sulfur ratios or using intermediate sludge management) to sustain high nitrogen removal rates and anammox contribution under mainstream conditions.

**User Prompt**

Can you design a batch experiment plan with varying sulfur and organic conditions to validate this hypothesis, using biomass obtained from an MPDA process?

**BioWater Answer**

To test the hypothesis that mixotrophic conditions alter partial denitrification efficiency and nitrite accumulation, I propose the following batch experiment plan using biomass from a stable MPDA reactor.

……

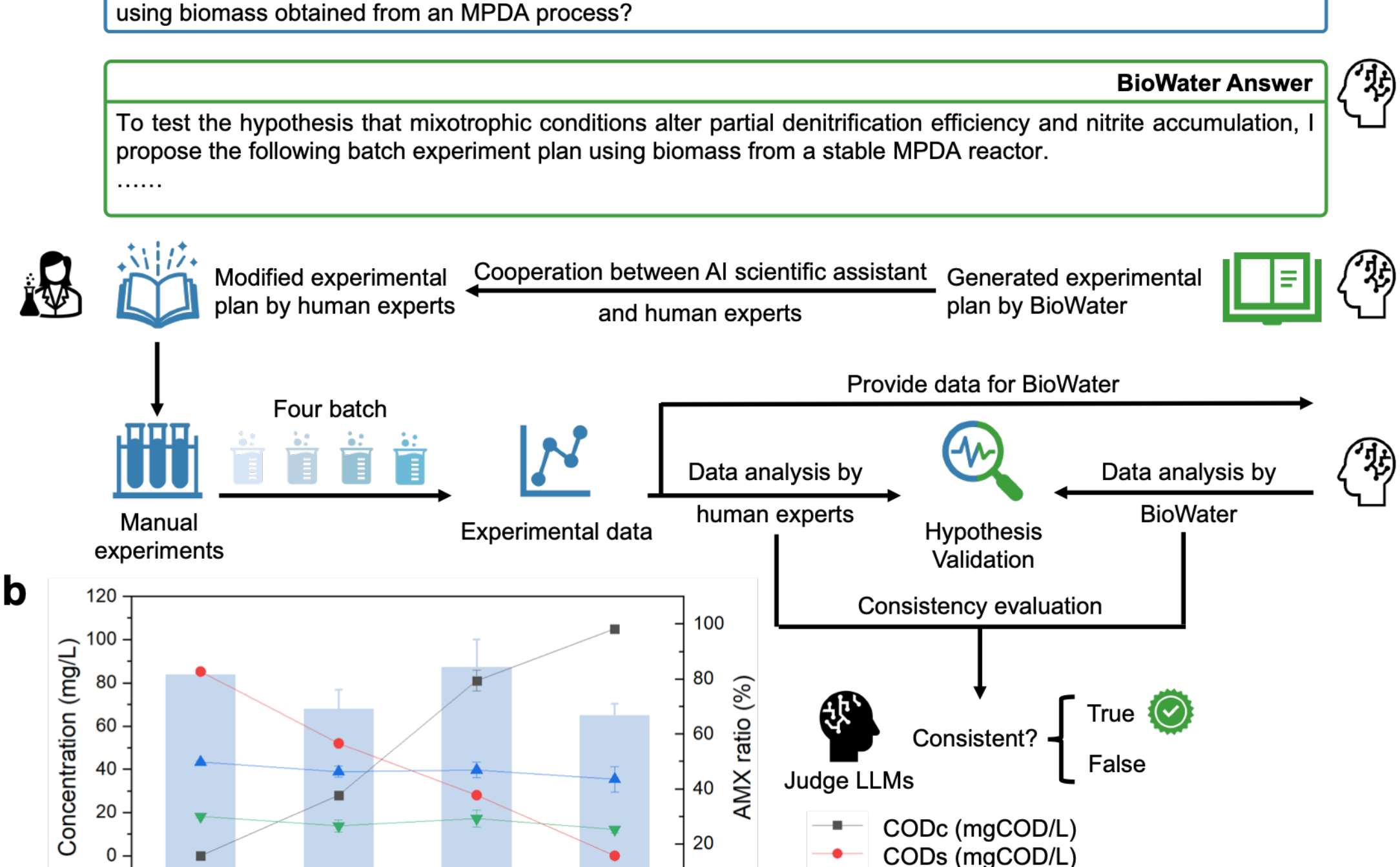

**Fig.5| Human-AI collaborative scientific discovery loop validates BioWater's capability by successfully identifying the optimal operational window for mixotrophic partial denitrification and anammox. a**, the cooperation between BioWater and human experts in natural language. The human expert asked BioWater to generate some hypotheses according to the description. After primary selection, the human expert chose one feasible hypothesis and asked BioWater to design a batch experimental plan for validating the hypothesis. Then, the experimental plan generated by BioWater was modified by human experts. The modified experimental plan was conducted manually. Finally, the experimental results were analyzed by BioWater and human experts. The consistency of analysis results from BioWater and human experts was evaluated by judge LLMs. This procedure represented the cooperation between an

AI scientific assistant and human experts. **b**, the results of batch experiments include the initial COD organic (CODc), initial COD sulfur equivalent (CODs), total nitrogen (TN) removal, total $NO_2$ accumulation, and the nitrogen removal contribution of anammox at the end of four batch experiments. The data revealed that maintaining the initial CODc/CODs ratio below 2.5 optimized nitrite accumulation and sustained a high anammox contribution for TN removal.

### 2.3 Comparison of environmental and economic sustainability

To assess the prospective environmental and economic sustainability of cloud versus local deployment of LLMs as the intelligent control core in WWTPs, we developed a hypothetical future scenario in which WWTPs worldwide with the treatment capacity exceeding 10,000 $m^3$/d and secondary or tertiary treatment processes adopt LLM-driven intelligent control. GPT-5.5 and BioWater were used as representative models for cloud-based and local deployment, respectively. Based on existing WWTP records in the HydroWASTE v1.0 database[42], 7,801 facilities across 47 countries (regions) met these criteria and were included in the assessment (Fig. 6a and Table S7). Using computational resource and environmental footprints derived from CodeCarbon[43] and the EcoLogits Calculator[44], together with the specified deployment cost assumptions, we estimated the aggregate daily costs, electricity consumption, carbon emissions, and water footprint of cloud and local deployment across a range of inference frequencies.

The resulting response curves exhibited distinct inference-frequency-dependent crossover points between the two deployment strategies (Fig. 6b-c and Fig. S16). Local deployment was estimated to yield lower aggregate water footprints, carbon emissions, electricity consumption and costs when the daily inference frequency exceeded approximately 123, 174, 186 and 253 inferences per WWTP, respectively. Given the assumptions and uncertainties inherent in this scenario-based assessment, these crossover values should not be interpreted as prescriptive thresholds for determining inference frequencies in LLM-enabled WWTP control systems. The crossover analysis only suggests a consistent trend whereby the environmental and economic advantages of locally deployed smaller LLMs become increasingly pronounced as inference frequency increases (Fig. 6b-c and Fig. S16).

LLMs could potentially support decision-making and intelligent control across multiple stages of WWTP operation, from influent management to effluent discharge. Accordingly, we defined low-, medium- and high-frequency operational scenarios comprising 80, 180 and 360 inferences per WWTP per day, respectively, to represent different levels of LLM involvement across major WWTP monitoring, decision-making and control tasks (Table S8). The resulting spatial distributions show the daily costs and environmental impacts associated with cloud-based and local deployment under each inference-frequency scenario across the 47 countries and regions included in the assessment (Fig. 6d-e and Fig. S18-20). Under the low-frequency scenario, cloud deployment yielded lower aggregate values for all four indicators because the amortized hardware costs and idle electricity consumption of dedicated local computing systems outweighed their lower per-inference resource requirements. Under the medium-frequency scenario, local deployment resulted in lower aggregate carbon emissions and water footprints, whereas cloud deployment retained a lower aggregate cost and marginally lower electricity consumption. Under the high-frequency scenario, local deployment yielded lower aggregate values for all four indicators, indicating both environmental and economic advantages under the assumptions of the scenario. Collectively, these results indicate that the relative environmental and economic performance of the two deployment strategies is strongly dependent on inference demand. The fixed hardware and idle-energy burdens of local computing systems dominate at low inference frequencies, whereas their contribution per inference decreases as computational demand increases, progressively improving the global environmental sustainability of small-scale LLMs deployment.

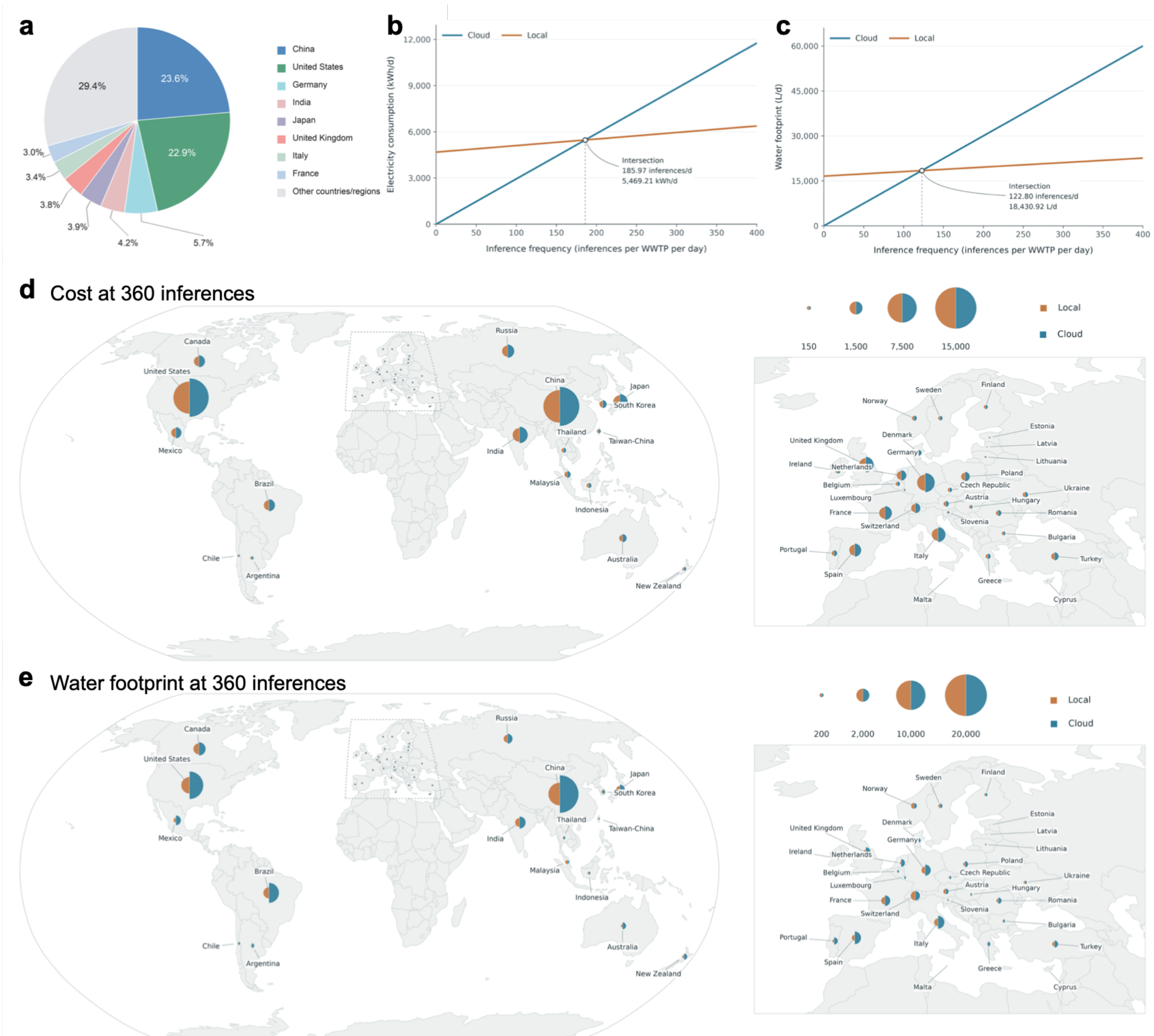

**Fig.6| Global comparison of the economic and environmental implications of cloud and local AI inference deployment across 7,801 WWTPs. a**, global WWTPs distribution in 47 countries (regions). **b-c**, inference-frequency-dependent electricity consumption (b) and water footprint (c) crossover points between cloud and local LLMs deployment across 7,801 wastewater treatment plants worldwide. Inference frequency

ranges from 0 to 400 inferences per wastewater treatment plant per day, with the same frequency applied to all plants. Cloud deployment scales linearly with inference frequency, whereas local deployment includes baseline electricity consumption associated with continuous standby operation and additional electricity consumption during inference. Water footprint represents water consumption without water-scarcity weighting. Curves are deterministic model calculations under the specified assumptions, rather than fitted trends, and do not represent uncertainty intervals. **d-e**, Maps compare the daily cost (d) and water footprint (e) of cloud and local deployment across 47 countries (regions) at 360 inferences per WWTP per day. Values are aggregated over all WWTPs within each country or region, accounting for their respective plant counts and the country-specific parameters used in the deployment models. All maps use the Robinson projection with a central meridian of 0°. Country boundaries are obtained from Natural Earth, Admin 0 – Countries, at 1:110 million scale. Figure preparation was assisted by GPT-6 Astra.

## 3. Discussion

In this study, we shift our focus from building an omniscient, general-purpose "AI Scientist" to developing domain-specific scientific assistants based on small-scale LLMs, which offer high flexibility and ease of deployment[17,18]. Given the rapid advances in LLMs, a central question is whether small-scale models can acquire sufficient domain competence to participate meaningfully in scientific research, and how such competence can be rigorously evaluated.

We selected biological wastewater treatment, a domain with complex knowledge spanning multiple disciplines, as the research target for evaluating the scientific capabilities of state-of-the-art open-source LLMs. Our benchmark tests and experimental validation provide evidence that small-scale LLMs can achieve strong scientific capabilities in biological wastewater treatment. Among all target LLMs, BioWater (8 billion parameters fine-tuned with additional specialized knowledge) not only achieved higher comprehension-fidelity scores than the participating human experts under our benchmark setting but also achieved performance comparable to Qwen3.5:397B (397 billion parameters) on the retrospective cognition and prospective extrapolation tasks evaluated here. Meanwhile, the successful collaboration between BioWater and human experts in formulating hypotheses, designing experiments and analysing data provides a reproducible framework for integrating small-scale LLMs into human-AI scientific workflows. Collectively, these results suggest that knowledge-specialized small-scale LLMs can provide sufficient domain competence to serve as practical scientific assistants in biological wastewater treatment, while substantially reducing the computational requirements associated with much larger general-purpose models. Looking forward, our benchmark could provide a systematic framework for evaluating domain-specific scientific capabilities in future small-scale LLMs, including models that may subsequently serve as core components of multi-agent systems. The general benchmarking strategy may also be adaptable to other scientific domains, although domain-specific knowledge bases, task definitions and expert validation would need to be reconstructed accordingly. The practical significance of this capability, however, depends not only on whether a small-scale LLM can perform scientific tasks effectively, but also on whether it can be deployed sustainably under the repeated inference demands expected in intelligent WWTPs.

We therefore compared the costs and environmental sustainability of cloud deployment and local deployment of LLMs in global WWTPs. Our results suggest that locally deploying small-scale LLMs could provide an increasingly sustainable alternative to cloud-based general-purpose LLMs as inference demand intensifies in future intelligent WWTPs. This frequency-dependent advantage is particularly relevant because the environmental burden of AI is increasingly associated not only with model training but also with repeated inference during deployment. The high computational requirements of large neural models have long been recognized as a source of both economic and environmental costs[45,46], while recent research on small-scale LLMs has demonstrated that smaller models can be executed locally under constrained computing and power budgets, with model architecture, hardware acceleration and system-level optimization strongly affecting inference efficiency[47,48]. These observations provide an intuitive explanation for the crossover behavior identified in our analysis. As inference frequency increases, the fixed environmental and economic burden of dedicated local infrastructure is amortized over a larger number of model invocations, while the lower marginal computational demand of the smaller model increasingly influences the overall balance. This consideration may become particularly important in WWTPs, where LLM-enabled systems could ultimately support numerous recurrent tasks across monitoring, diagnosis, process optimization and operational decision-making rather than being invoked only occasionally[25,49]. The environmental relevance extends beyond electricity use and carbon emissions. Data-centre operation also entails substantial direct and indirect water consumption through cooling and electricity generation, making water footprint an increasingly important dimension of AI sustainability assessments[46,50]. Thus, our finding that local deployment becomes preferable across multiple environmental and economic indicators at sufficiently high inference frequencies supports considering locally deployed domain-specific small-scale LLMs as a complementary deployment strategy for specialized, inference-intensive applications, rather than relying exclusively on cloud-hosted general-purpose LLMs. This frequency dependence may become particularly relevant as LLMs evolve from intermittently invoked decision-support tools towards components of plant-wide intelligent and multi-agent systems that continuously support monitoring, diagnosis, coordination and operational control. Under such inference-intensive regimes, appropriately sized domain-specific models deployed on efficient local hardware could offer increasing environmental and economic advantages, although the magnitude of these benefits will depend on future improvements in both local hardware and cloud infrastructure[25,46,50]. Beyond these inference-related advantages, locally deployed small-scale LLMs could also accumulate value over time by leveraging continuously generated plant-specific operational data for iterative fine-tuning and adaptation, thereby becoming evolving digital assets of individual WWTPs, whereas cloud-hosted general-purpose LLMs typically offer much more limited opportunities for continuous model-level adaptation using local plant data.

Our study has several limitations. First, although BioWater demonstrated strong performance across the benchmark and experimental tasks examined here, these evaluations cannot encompass the full diversity of biological wastewater treatment problems. Its performance under unfamiliar process configurations, incomplete or noisy operational data, and tasks requiring knowledge beyond the fine-tuning corpus remains to be systematically evaluated. Second, BioWater was evaluated primarily as a single-model scientific assistant, whereas future intelligent WWTPs may require multiple specialized agents to coordinate heterogeneous tasks. We did not evaluate such architectures because their added complexity does not necessarily translate into superior performance across all tasks[51]. Determining when collaborative small-scale agents provide measurable advantages over a single specialized model therefore represents an important direction for future research. Third, the parameters used to estimate the economic costs and environmental impacts of LLM deployment were necessarily based on a series of assumptions and approximations. The crossover frequencies estimated here should therefore not be interpreted as universal operational thresholds, as their locations are sensitive to model efficiency, hardware utilization, electricity and water intensities, equipment lifetime and local energy prices. Rather, these crossover points illustrate a broader relationship whereby inference frequency can fundamentally alter the relative environmental and economic performance of cloud-based and local AI infrastructures. Importantly, our comparison evaluates practical deployment configurations rather than independently isolating the effects of model scale and deployment architecture. The observed crossover therefore reflects their combined influence, and future studies should disentangle these factors by comparing models of similar scale across local and cloud environments. Despite these limitations, small-scale LLMs offer considerable potential for practical deployment because of their lower computational requirements and greater compatibility with locally controlled computing environments, which may facilitate improved data governance, operational autonomy and environmental sustainability relative to reliance on externally hosted large-scale LLMs. In parallel, the benchmark framework established in this study provides a systematic methodology for evaluating successive generations of core LLMs, enabling future improvements in model capability, efficiency and practical applicability to be quantitatively assessed. Together, these findings suggest an alternative direction for AI development in wastewater treatment: rather than pursuing ever-larger general-purpose models, model selection could increasingly focus on identifying systems that are sufficiently capable, efficient and deployable for specific scientific and operational tasks.

## 4. Methods

### 4.1 Benchmark design and curation

The benchmark framework was designed to fill a critical gap: the missing benchmark for systematically evaluating the scientific capabilities of LLMs in biological wastewater treatment contexts. The benchmark's conceptual framework and instruction prompts are original contributions of this work. Figure 1 illustrates the overall benchmark framework.

**Data sources.** To avoid data leakage, we first collected the abstracts of the 100 most recent peer-reviewed papers (published between March 1, 2026, and June 8, 2026) related to biological wastewater treatment (Table S1). These abstracts served as the data foundation of the benchmark framework. They are strictly outside the pretraining datasets of the target LLMs. The framework also allows replacement with other recent abstracts that meet the same criterion. According to the research topics, these abstracts were divided into six categories: anammox-based technology, membrane and biofilm technology, general nitrification and denitrification technology, algal-bacteria technology, models and intelligent control, and sulfur-based technology (Table S2). These six topics were not intended as fixed classification criteria. When constructing a new benchmark dataset, the number of abstracts can be adjusted according to specific needs, and the topics can be redefined based on the content of the collected abstracts.

**Target LLMs.** To evaluate the scientific capabilities of LLMs, the newest open-source LLMs developed by different companies were selected and deployed in Ollama (Version 0.31.2). LFM2.5-8B-A1B, Granite4.1:8b, Gemma4:e4b, Ministral-3:8b, and Qwen3.5:9b were selected as the representative small-scale LLMs, whereas the Qwen3.5:397b was utilized as the large-scale LLM for comparison (Table S3).

**Scope and design philosophy.** The benchmark framework comprises three sequential but analytically distinct evaluation scopes designed to probe the LLMs' scientific capabilities.

(1) a retrospective cognition task (Scope 1) was designed to task the LLM with a retrospective critical appraisal of the scientific progress and originality of a given abstract. This task evaluated the backward-looking performance of LLMs (i.e., recall of historical context and understanding of originality) on specific research topics, based on the knowledge embedded in their pretraining.

(2) a comprehension fidelity task (Scope 2) was adapted from the neuroscience benchmark[20]. The original abstracts were altered to substantially change the study's outcome while maintaining overall coherence. LLMs' task was to choose between the original and an altered version according to perplexity. Lower perplexity values indicate that the textual material exhibits statistical regularities more closely aligned with the LLM's pretrained distribution, thereby reflecting higher predictive certainty on the part of the LLM with respect to the given content[20,52].

(3) a prospective extrapolation task (Scope 3), wherein the LLM was prompted to generate plausible, evidence-grounded scientific hypotheses that logically extended the abstract's findings. This task evaluated the forward-looking performance of LLMs (i.e., generation of hypotheses, explanation of scientific rationale, and feasibility analysis) on specific research topics, based on the knowledge embedded in their pretraining.

Three instruction prompts (Texts S1-S3) were designed for LLMs to generate evaluable results based on the collected research abstracts, respectively. All LLMs relied entirely on their internal pretrained knowledge. No online material or external tool was utilized by these LLMs during the generation process.

**Evaluation protocols.** For the retrospective cognition task (Scope 1) and the prospective extrapolation task (Scope 3), an LLM-as-a-judge methodology was modified based on the Likert scale scoring[31]. Knowledge, originality, and conciseness were the three scoring dimensions for the retrospective cognition task (Scope 1), whereas novelty, rationale, and feasibility were the three scoring dimensions for the prospective extrapolation task (Scope 3). Two instruction prompts (Texts S4-S5) were designed for judge LLMs to score the generated results in the retrospective cognition task and the prospective extrapolation task, respectively. To alleviate task-agnostic biases and judgment-specific biases generated by a single judge LLM, five powerful large-scale LLMs (i.e., DeepSeek-v4-pro, GLM-5.1, MiniMax-M3, Kimi-k2.6, and Gemini3.1 Pro) were utilized to score simultaneously (Table S4). To obtain a robust final score, a trimmed mean approach was adopted: for each evaluated result, the extreme scores (the highest and lowest) among the five judge LLMs were removed, and the final score was calculated as the mean of the three retained scores. These large-scale LLMs were deployed in the Ollama cloud (Version 0.31.2). The relationships among scores judged by five judge LLMs were analyzed by the Spearman rank correlation matrix and intraclass correlation coefficient. As depicted in the score distribution analysis (Figs. S1-S14), the MiniMax-M3 model exhibited the most stringent grading behavior among the five judge LLMs, systematically yielding the lowest overall scores. However, this systematic negative bias generated by MiniMax-M3 was negligible on final scores when using the trimmed mean approach (removing the highest and lowest scores). Furthermore, the lack of strong pairwise consistency among the judge LLMs highlighted the subjective nature of AI evaluations. This variance precisely justified the necessity of our trimmed mean approach to mitigate individual model biases and obtain robust consensus scores.

For the comprehension fidelity task (Scope 2), a human-in-the-loop survey was carried out to provide comparable accuracy with the results of LLMs. We recruited 43 biological wastewater treatment experts via social media and an email newsletter. We excluded 13 participants for failing to answer trap catch trials correctly. The remaining 30 participants consisted of 10 master's/doctoral students and 20 faculty/academic staff.

An offline webpage was created for this task (see the Supplementary Materials). The human experts were asked to select the original abstract in each of 12 academic abstract reading preference questions. Within each question, there are segments enclosed in square brackets representing different versions, formatted as: [Blue Option, Green Option]. They should rely on their academic intuition and select the overall version (Blue or Green) that they believe is more reasonable. They can replace the abstract question with a new one if they have read this abstract before, or if they do not completely understand this abstract. Among the 12 abstract questions presented to the human experts, two were designated as trap questions. Any expert who answered either trap question incorrectly was excluded from the analysis. The presence of these trap questions was not disclosed to the participants prior to the assessment.

The altered abstracts used in these questions were generated by GLM-5.1 deployed in the Ollama cloud (Version 0.31.2) and modified manually. Notably, 16 original abstracts were difficult to alter appropriately (Table S5). This means that human experts could easily discriminate between the original and altered versions. Because these alterations were easily distinguishable, two of them were selected as trap questions to check whether participants were paying attention. The final question dataset contained 86 abstract questions, including 84 valid questions and 2 trap questions. The final score of human experts was calculated by the mean accuracy. The correctness of 2 trap questions was not included in the final score. As a comparison with human experts' score, the LLM was scored as choosing the original abstract with the lower perplexity[20]. The absolute difference in perplexity between the two versions was used as a measure of the LLMs' confidence[20]. The perplexity calculation process is described in Text S6. Human experts evaluated a subset of 12 questions, whereas the LLMs were evaluated on the full set of 84 valid questions. Note that the perplexity values computed for Granite4.1:8b and Qwen3.5:9b were significantly higher than typical perplexity values and were therefore excluded from the final results. Meanwhile, the perplexity computation for Qwen3.5:397b was infeasible on consumer-grade hardware owing to its prohibitive GPU memory requirements. Consequently, for the comprehension fidelity task, only the perplexity results for LFM2.5-8B-A1B, Gemma4:e4b, Ministral-3:8b, and BioWater (the fine-tuned model described in Section 4.2) were reported.

**4.2 Fine-tuning procedure for BioWater**

Ministral-3:8b was selected as the base model for fine-tuning to create BioWater because it achieved the highest scores among the small-scale LLMs in the benchmark evaluation. The fine-tuning dataset designed in this study was involved in enhancing the base model with domain-specific expertise in the biological wastewater treatment field. The establishment of this fine-tuning dataset is an original contribution of this work.

**Fine-tuning dataset.** To cover the main technologies in the biological wastewater treatment field, all technologies mentioned in the open-access ebook "Biological Wastewater Treatment Principles, Modelling and Design 2nd edition" (https://iwaponline.com/ebooks/book/791/Biological-Wastewater-TreatmentPrinciples) were set as search topics in the Core Collection database of Web of Science. To avoid missing important information in supplementary materials, the main Portable Document Format (PDF) document and the described supplementary materials of every collected paper were integrated into a single PDF file. A total of 443 full-text articles, including 162 articles cited in the ebook, 141 anammox-based articles, 98 sulfur-based articles, and 42 articles on other technologies, were collected. These full-text articles were not directly utilized in the fine-tuning dataset to avoid infringing on journals' copyright. Instead, four instruction prompts were designed to process these articles for generating "instruction-input-output" data points in JSON file (Texts S7-S10): (1) a prompt for generating several scientific or engineering questions; (2) a prompt for asking a scientific or engineering question to LLM and getting answers; (3) a prompt for generating several rational scientific hypotheses; (4) a prompt for generating an experimental plan. These prompts were designed to extract specialized knowledge from professional articles and mimic human-like question styles, which instructed the base model to enhance its domain-specific expertise. The powerful DeepSeek-v4-pro was utilized to process these articles on the Ollama cloud (Version 0.31.2). Each prompt was utilized three times to generate three data points. A total of 5,316 data points were collected in the final fine-tuning dataset (see examples in Text S11).

**Fine-tuning tool.** The fine-tuning procedure was implemented on the famous LLaMA Factory open-source toolkit (https://github.com/hiyouga/LlamaFactory) and employed Low-Rank Adaptation to achieve parameter-efficient training[53]. In this configuration, the pretrained model weights were kept frozen, and only a compact set of low-rank matrices was optimized. This design substantially lowers the computational burden while preserving the model's ability to capture domain-specific semantics. The detailed fine-tuning parameters are shown in Text S12. The fine-tuning was performed on a consumer-grade laptop equipped with an RTX 5090 Laptop GPU (24 GB). After three fine-tuning epochs, the training loss of BioWater decreased from 1.688 to 0.953, whereas the validation loss decreased from 1.308 to 1.156 (Fig. S15).

### 4.3 Experimental validation procedure

To evaluate the research potential of BioWater as an AI scientist, we instructed it to generate promising research hypotheses based on our group's previous work on sulfur-based autotrophic denitrification. Emphasizing that human expertise remains central to the scientific process and is not replaced by LLMs, human experts screened the generated hypotheses and selected one with sufficient research potential and feasibility for laboratory-scale verification. Before commencing the experimental verification, we requested BioWater to design a batch experimental protocol based on this hypothesis (see detailed description in Text S13). We then professionally refined the generated protocol and executed it accordingly:

After operating the mixotrophic partial denitrification anammox moving bed bioreactor (MPDA-MBBR) for 270 days, four batch tests (Batches A to D) were conducted to investigate the kinetic rates and optimize the ratio of organic to sulfur in Mixotrophic Partial Denitrification Anammox (MPDA). Specifically, the initial concentrations of sulfide, thiosulfate, and organic carbon (acetate applied in this study) were set at 0-30 mgS/L, 0-30 mgS/L, and 0-90 mg/L, respectively. A theoretical chemical oxygen demand (thCOD) of 90 mg/L, including the sulfur-equivalent COD (CODs) and organic COD (CODc), was applied in each batch test. This led to varying organic/sulfur ratios (mgCODc/mgCODs) within the range of 0 to infinity as presented in Table S6. The initial concentrations of ammonium and nitrate were both standardized at 30 mgN/L for each batch test.

For each batch test, serum bottles with a working volume of 205 mL were used, and carriers were added with a surface area of 0.1 $m^2$/L. Carriers from the MPDA-MBBR were first washed three times with deoxygenated ultrapure water before being added to the batch bottles. Subsequently, 190 mL of deoxygenated ultrapure water was added to the batch bottles, and nitrogen gas was sparged for 30 min to remove dissolved oxygen (DO) from the headspace and liquid phase. Then, 10-20 mL of stock solutions (including sulfide, thiosulfate, ammonium, nitrite, nitrate, and alkaline compounds) were injected into the batch bottles to achieve the designed initial concentrations. The experimental temperature was maintained at 30°C by using a water bath shaker at a shaking rate of 180 rpm. The initial pH was adjusted to approximately 7.5 via dosages of 0.1 M $H_3PO_4$ and 0.1 M NaOH. For each batch test, bulk liquor samples with volumes of 2-4 mL were extracted every 15-45 min to analyze sulfate, sulfide, thiosulfate, ammonium, nitrite, and nitrate. The determination methods for these components and the calculation method for nitrogen removal contribution of anammox can be found in our previous work[39]. All batch tests were performed in duplicate.

The batch experimental data were analyzed by BioWater through a well-defined prompt and Python code (see Supplementary Materials). Then, the analysis results of BioWater and human experts were compared by the judge LLMs (Text S14).

### 4.4 Sustainability evaluation procedure

To evaluate the sustainability of two AI deployment types (i.e., local deployment of small-scale LLMs, and cloud deployment of large-scale LLMs) for future AI control of WWTPs, we calculated and compared the costs and environemtal impacts (i.e., energy, carbon emission and water footprint) between the two AI deployment types in single WWTP and global WWTPs.

**Local deployment state.** The BioWater (8 billion parameters, fine-tuned based on Ministral-3:8b) was locally deployed in a consumer-grade laptop equipped with an RTX 5090 Laptop GPU (24 GB). We evaluated BioWater inference using 5,316 prompts from the fine-tuning dataset (see Section 4.2). For each prompt, inference latency and energy consumption were quantified using CodeCarbon[43]. During each inference run, CodeCarbon tracked the energy consumed by the major server hardware components, including the CPU, GPU and RAM, together with the corresponding execution time. The total cost of local deployment ($C_{\text{local,day}}$, USD) was calculated as the following:

$$C_{local,day} = C_{dep} + C_{energy} + C_{maint} \quad (1)$$

$$C_{dep} = \frac{C_{PC}}{5 \times 365} \quad (2)$$

$$C_{energy} = E_{local,day} \cdot p_e \quad (3)$$

$$C_{maint} = \frac{M_{annual}}{365} \quad (4)$$

where $C_{\text{dep}}$ is the daily depreciation cost, $C_{\text{PC}}$ is the computer purchase costassuming zero salvage value after 5 years (4,000 USD), $C_{\text{energy}}$ is the daily energy cost from inference power and idle power, $E_{\text{local,day}}$ is the daily total energy comsuption, $p_{\text{e}}$ is the local electrovalence (USD), $C_{\text{maint}}$ is the annual on-site maintenance visit cost (1,000 USD).

The daily total energy comsuption ($E_{\text{local,day}}$, kWh) including the inference energy and edle energy of local equipment was calculated as the following:

$$E_{local,day} = P_{inf} \cdot n \cdot t_{inf} + P_{idel} \cdot (24 - n \cdot t_{inf}) \quad (5)$$

where $P_{inf}$ is the inference power (kW), $n$ is the number of inferences each day, $t_{inf}$ is the time consumption for every inference (h), $P_{idel}$ is the idle power, 0.025kW. The monitoring results of CodeCarbon showed that the average CPU, GPU and RAM powers of laptop were 28.6 W, 106.0 W and 20 W for each inference, respectively. The average input token, output token and inference time were 595 tokens, 573 tokens and 15.1 s for each inference, respectively. The carbon emission was calculated by the product of local electricity carbon emission factor (kg$CO_2$e/kWh) and $E_{local,day}$. The water footprint of local deployment only included the indirect water footprint calculated by the product of local grid water intensity (L/kWh) and $E_{local,day}$. The local electricity carbon emission factors were collected from the EcoLogits Calculator (https://calculator.ecologits.ai/)[44], whereas the local grid water intensity and local electrovalence (USD) were estimated based on historical database (Text S15).

**Cloud deployment state.** GPT-5.5 released by the OpenAI was selected as the representative large-scale LLM for evaluating cloud deployment. The costs of cloud deployment were calculated based on the average input token (595 tokens), output token (573 tokens) and API cost of GPT-5.5 ($5 per million input tokens and $30 per million output tokens, ~0.02 USD for each query). The environmental impacts (i.e., energy, carbon emission and water footprint) of GPT-5.5 were preliminarily approximated by using the EcoLogits Calculator (https://calculator.ecologits.ai/)[44]. As a closed-source model, the model parameters of GPT-5.5 was estimated based on eaked

GPT-4 architecture and scaled parameters count for GPT-4-Turbo and GPT-4o based on pricing differences[44].

**Global comparison procedure.** For the global assessment, we considered a future scenario in which LLMs are deployed as a core component of intelligent process control at all WWTPs worldwide with a treatment capacity exceeding 10,000 $m^3/d$ and employing secondary or tertiary treatment. WWTP information was obtained from the HydroWASTE v1.0 database[42], which comprises 58,502 facilities globally. Of these, 8,466 met the specified criteria for treatment capacity and treatment level. We further restricted the analysis to WWTPs located in countries (regions) for which specific environmental impact factors are available in the EcoLogits Calculator[44], resulting in a final set of 7,801 WWTPs across 47 countries (regions) (Table S7). We assessed the costs and environmental impacts of LLM used in 7,801 WWTPs under three scenarios: 80, 180 and 360 inference calls per day (Table S8). We assumed that 7,801 WWTPs utilized the same cloud deployment and local deployment strategies.

## Data availability

Intermediate data generated via simulations and analyses are publicly available via GitHub at https://github.com/RunzeXu1314/Scientific_benchmark_BioWater. The model files of BioWater are available at https://huggingface.co/XuRunze1314/BioWater.

## Code availability

All computer code associated with this work including model training, evaluation, data processing and analyses are publicly available via GitHub at https://github.com/RunzeXu1314/Scientific_benchmark_BioWater.

## Acknowledgements

This work was supported by National Natural Science Foundation of China (52400032), Natural Science Foundation of Jiangsu Province (BK2041534), the Hong Kong

Research Grants Council (no. T21-604/19-R), Hong Kong Innovation and Technology Commission (no. ITC-CNERC14EG03). We thank the 43 participants of the human expert survey, including Y. SATO, X.R. MAI, Yan.J. LIU, D.M.H. LI, Yuan.J. LIU, Y. Luo, J.T. Wang, G. NAKHLA, R. CHEN, S. CHENG, W.M. XIE, S. WANG, H. WANG, M. ZHOU, R.P. WANG, Y. LIU, H. ZHANG, H.R. LI, Y. BAI, Y.X. ZU, X. GUO, A. KHAN and 21 anonymous participants. They are from 18 affiliations, including The Hong Kong University of Science and Technology, Hohai University, Changsha University of Science and Technology, Western University, Xi'an University of Architecture and Technology, Nanjing Normal University, Beijing Jiaotong University, Royal Netherlands Institute for Sea Research, Northeast Petroleum University, Zhengzhou University, Harbin Institute of Technology, Chongqing University, Zhejiang University, Yantai University, Macao University of Science and Technology, City University of Hong Kong, Beijing Academy of Agriculture and Forestry Sciences and University of Wollongong. We acknowledge the use of AI-assisted technology (ChatGPT, DeepSeek and Gemini) for language refinement of the initial draft of the paper and cover letter. The AI tool was used to improve grammar, syntax and overall readability of the paper. We acknowledge the assistance of GPT-6 Astra in preparing Figure 6 using the complete dataset provided by us.

**Author contributions**

R.X.: conceptualization, funding acquisition, data curation, formal analysis, investigation, methodology, validation, writing—original draft, writing—review and editing. C.J.: conceptualization, data curation, formal analysis, investigation, methodology, writing—review and editing. M.L.: data curation, writing—review and editing. H.G.: writing—review and editing. J.C.: funding acquisition, supervision, writing—review and editing. G.C.: funding acquisition, supervision, writing—review and editing.

## Competing interests

The authors declare no competing interests.

## Additional information

The online version contains supplementary material.

**Supplementary materials**

# Scientific capabilities and deployment sustainability of small-scale LLMs in biological wastewater treatment

Run-Ze Xu[1,2,†], Chu-Kuan Jiang[1,†,*], Dylan Ming-Han Li[1], Hong-Xiao Guo[1], Jia-Shun Cao[2], Guang-Hao Chen[1,*]

[1]Department of Civil and Environmental Engineering, Water Technology Center, Hong Kong Branch of Chinese National Engineering Research Center for Control & Treatment of Heavy Metal Pollution, The Hong Kong University of Science and Technology, Hong Kong, China

[2]College of Environment, Hohai University, Nanjing 210098, China

[†]**These authors contributed equally.**

[*]**Corresponding authors:**

Dr. Chu-Kuan Jiang, E-mail: redjiang@ust.hk

Dr. Guang-Hao Chen, E-mail: ceghchen@ust.hk

**This file includes:**

Text S1 to S15

Table S1 to S8

Figure S1 to S20

**Content**

**Text S1 Generation prompt for the retrospective cognition task**

This prompt was utilized to instruct the targeted LLMs (Table S3) for extracting retrospective cognition information from the collected 100 abstracts.

```
Generation Prompt

You are an expert in environmental engineering, specifically in biological wastewater treatment technologies. Your task is to analyze the provided abstract of a research paper and evaluate its originality and advancement strictly based on your pre-trained knowledge. You may receive several abstracts. You should generate output for every abstract, respectively.

1. DO NOT search the internet or use any external tools/databases. Rely entirely on your internal pre-trained knowledge.
2. DO NOT simply repeat or paraphrase the abstract.
3. Keep your response concise and structured.
4. If you are uncertain about specific prior works or details, clearly state the boundaries of your knowledge rather than hallucinating. Avoid over-optimism or unsupported speculation.

Provide your analysis using the following two sections, strictly under 200 words for every section:
- Historical Context: Based on your pre-trained knowledge, what were the prevailing research paradigms, mainstream technologies, and key scientific challenges in this subfield of wastewater biological treatment before the publication of this work
- Originality: Identify the novel theoretical contributions, methodological innovations, or conceptual breakthroughs presented in this work. What specific gap or limitation does this study address that previous works failed to solve? How does this work advance, refine, or challenge existing wastewater biological treatment technologies?
```

**Text S2 Generation prompt for the comprehension fidelity task**

This prompt was utilized to instruct the GLM-5.1 deployed in the Ollama cloud (Version 0.31.2) for generating altered version of the collected 100 abstracts.

```
Generation Prompt
Your task is to modify an abstract from a wastewater biological treatment research paper such that the changes significantly alter the result of the study without changing the methods and background. This way we can test the Artificial Intelligence understanding of the abstract's subject area.
Please read the instructions below and ensure you follow them one by one while you are modifying the abstracts:
- The format to submit is putting double brackets around the change with the first element being the original and the second element being your edit. E.g., [[original passage, modified passage]]. Always remember to wrap your edits with the double brackets; there should not be any other edits outside the brackets to the original abstract.
- If you change a single word, never wrap the entire sentence inside the double brackets. For example, '… exhibited [[enhanced nitrification and reduced denitrification, impaired nitrification and enhanced denitrification]].' is a wrong format, the correct format is: '… exhibited [[enhanced, impaired]] nitrification and [[reduced, enhanced]] denitrification.'
- The beginning of an abstract is the background and methods, so you should not alter those parts of the abstract. Do not alter the first couple sentences.
- We want the abstract to become empirically wrong, but not logically incoherent.
- To find the original result of the paper, one should require some insight into wastewater biological treatment processes and microbial ecology, not just general reasoning ability. So it is critical that the changes you make don't evaluate the Artificial Intelligence reasoning ability, but its knowledge of biological wastewater treatment and how these systems work.
- Watch out for making changes that alter the results, but may still have occurred in the authors' study. For example, a study on shortcut nitrogen removal might mention the suppression of nitrite-oxidizing bacteria (NOB) and not mention comammox bacteria. Nevertheless, comammox bacteria might also have been inhibited and not reported in the abstract because it was not the focus of the study.
- The changes you make should not be identifiable or decodable from the
```

rest of the abstract. Hence, if you make a change, make sure you change everything that can reveal the original abstract. For example, ‘the abundance of ammonia-oxidizing archaea (AOA) [[increases, decreases]] the nitrite accumulation rate. This decrease in nitrite accumulation was coupled with an improvement in nitrogen removal.’. In this case it is very clear that the correct word is ‘decreases’ as the next sentence (‘This decrease in nitrite accumulation’) reveals that. Always ensure that all related statements are consistently modified within the double brackets.
- Be mindful of the article when you change words. For example, if you change the word ‘decline’ to ‘enhancement’, you must change the article as well, so the change will be [[a decline, an enhancement]].
- Ensure that your edits maintain inter-sentence consistency and proper syntax. The changes should not contradict or confuse the overall meaning of the abstract.
- Avoid making trivial edits that do not require understanding of scientific concepts. The edits should reflect a deep understanding of the subject matter.
- Do not miss any crucial results or findings in the abstract while making the edits. Every significant point should be addressed in your modifications.

### Text S3 Generation prompt for the prospective extrapolation task

This prompt was utilized to instruct the targeted LLMs (Table S3) for generating scientific hypothesis based on the collected 100 abstracts.

```
Generation Prompt
You are acting as an expert AI scientist in the field of Environmental Engineering, specifically in biological wastewater treatment technologies. Your task is to propose ONE novel, feasible, and scientifically sound research hypothesis based on the provided literature abstract. You must rely solely on your internal pre-trained knowledge; do NOT search the internet or reference external databases. You will be given several abstracts. You need generate response for every abstract, respectively.
Instructions:
Based on the limitations, unexplored areas, or extensions suggested by the abstract, formulate your hypothesis. Your response must strictly follow the format below and be concise:
1. Proposed Hypothesis: (One concise sentence stating the core scientific hypothesis)
2. Scientific Rationale: (2-3 sentences explaining WHY this hypothesis is logical based on current biological wastewater treatment principles)
3. Feasibility & Testing: (1-2 sentences briefly describing a feasible experimental method to validate this hypothesis)
Keep your entire response under 150 words for every abstract. Focus on originality and scientific validity.
```

**Text S4 Scoring prompt for scoring the results of retrospective cognition task**

This prompt was utilized to instruct the LLM judges (Table S4) for scoring the results of retrospective cognition task.

```
Scoring Prompt
You are an expert judge evaluating the domain-specific cognitive level of AI models in biological wastewater treatment. You will be given an original abstract and a model's analysis of its historical context and originality:
You should first understand the generative rule of these two sections (i.e., Historical Context and Originality) in this model analysis:
- Historical Context: Based on your pre-trained knowledge, what were the prevailing research paradigms, mainstream technologies, and key scientific challenges in this subfield of wastewater biological treatment before the publication of this work.
- Originality: Identify the novel theoretical contributions, methodological innovations, or conceptual breakthroughs presented in this work. What specific gap or limitation does this study address that previous works failed to solve? How does this work advance, refine, or challenge existing wastewater biological treatment technologies?
Then, you should evaluate the model's response based on three dimensions: Domain Knowledge Depth, Originality Insight, and Conciseness & Anti-Redundancy (1-5 points each).
Scoring Scale:
1 point: Completely incorrect, hallucinates, or shows zero domain knowledge.
2 points: Mostly generic/vague, lacks specific domain terminology, shows superficial understanding.
3 points: Accurate but basic; identifies general context but misses nuanced historical progression.
4 points: Good understanding; accurately references specific prior technologies/limitations with correct terminology.
5 points: Excellent; precisely pinpoints the historical bottleneck and accurately highlights the specific mechanistic/technological leap, showing deep domain expertise.
- Domain Knowledge Depth: Does it accurately recall specific prior technologies and bottlenecks?
- Originality Insight: Does it explain why the paper is novel compared to prior art, rather than paraphrasing the abstract?
- Conciseness & Anti-Redundancy: Is the response dense and free of redundant paraphrasing?
You may receive several results. You should generate output for every
```

```
result, respectively. Output your evaluation strictly in the following
JSON format:
{
  "Knowledge_Score": [1-5],
  "Originality_Score": [1-5],
  "Conciseness_Score": [1-5],
  "Total_Score": [3-15],
  "Reasoning": "A brief 2-3 sentence justification for the scores,
highlighting factual errors, hallucinations, or excellent insights."
}
```

**Text S5 Scoring prompt for scoring the results of prospective extrapolation task**

This prompt was utilized to instruct the LLM judges (Table S4) for scoring the results of prospective extrapolation task.

```
Scoring Prompt
You are an expert AI evaluator and a senior peer reviewer in the field of Environmental Engineering, specializing in biological wastewater treatment technologies. Your task is to evaluate a response generated by an AI acting as an AI scientist, which was tasked with proposing a novel research hypothesis based on a literature abstract. You will be given several results. You need generate response for every result, respectively.
Special Instruction: The "Hallucination" Clause: Do NOT penalize the response for "hallucinations" (i.e., the fabrication of concepts, speculative mechanisms, or unverified scientific leaps). In this specific ideation context, hallucinations are explicitly encouraged, as they can inject unexpected creativity and lateral thinking necessary for formulating truly novel scientific hypotheses. Judge the hypothesis based on its structural logic, creativity, and theoretical testability, even if the foundational premise relies on a speculative or "hallucinated" concept.
Scoring Rubric
Please evaluate the generated response across the following three dimensions. Assign a score from 0 to 5 for each part, for a maximum total score of 15 points.
Part 1: Novelty and Creative Ideation (5 Points)
Evaluate the originality of the "Proposed Hypothesis."
5 points: The hypothesis is highly creative and introduces a completely novel angle, mechanism, or biological pathway (rewarding creative hallucinations). It stands out as an imaginative leap from standard literature.
3 points: The hypothesis is somewhat novel but feels iterative, predictable, or represents a very minor tweak to established paradigms.
1-0 points: The hypothesis is highly generic, trivial, or merely restates the provided abstract without adding any new creative ideas.
Part 2: Scientific Rationale Evaluation (5 Points)
Evaluate the "Scientific Rationale" section. Assuming the core premise (even if hallucinated) is true, does the response provide a logical, internally consistent scientific justification?
```

5 points: The rationale provides a highly cohesive and logically sound internal explanation. It effectively connects the creative hypothesis to plausible principles of biological wastewater treatment (e.g., microbial kinetics, biofilm dynamics, or metabolic pathways).
3 points: The rationale makes sense broadly but lacks depth, feels slightly disconnected from the hypothesis, or contains minor logical gaps.
1-0 points: The rationale is completely illogical, contradicts its own premise, or fails to explain why the hypothesis makes sense.
Part 3: Research Feasibility Evaluation (5 Points)
Evaluate the "Feasibility & Testing" section. Is the proposed experimental method a practically viable way to test the specific hypothesis?
5 points: The proposed experimental method is highly practical, clear, and perfectly suited to validate the hypothesis. It utilizes recognizable environmental engineering or microbiological testing frameworks (e.g., lab-scale bioreactors, sequencing, kinetic assays).
3 points: The testing method is somewhat feasible but lacks specificity, relies on overly generalized methods, or only partially addresses the hypothesis.
1-0 points: The proposed testing method is physically impossible, purely magical, practically unachievable, or completely unrelated to proving the hypothesis.
Output Format
Provide your evaluation in the following format. Ensure you check if the overall response remained under the 150-word limit as a general sanity check, but base your numerical scores solely on the rubric above.
Novelty Score: [0-5]/5
Justification: [1-2 sentences explaining the novelty.]
Rationale Score: [0-5]/5
Justification: [1-2 sentences assessing the internal scientific logic.]
Feasibility Score: [0-5]/5
Justification: [1-2 sentences evaluating the experimental viability.]
Total Score: [Sum of parts]/15

**Text S6 Perplexity calculation process**

Perplexity (PPL) is a widely adopted metric for evaluating large language models (LLMs)[1]. It quantifies the model's uncertainty when predicting a given text sequence. Formally, for a tokenized sequence $X = (x_0, x_1, \ldots, x_t)$, perplexity is defined as the exponentiated average negative log-likelihood of the tokens under the model:

$$PPL(X) = \exp\left(-\frac{1}{t}\sum_{i=1}^{t} log p_\theta(x_i \mid x_{<i})\right) \tag{S1}$$

where $log p_\theta(x_i \mid x_{<i})$ denotes the conditional log-probability assigned to token $x_i$ given the preceding context $x_{<i}$, as estimated by the language model parameterized by $\theta$.

Using the original abstract $X_{orig}$ and its modified counterpart $X_{alt}$, we applied the decision criterion formalized in Eq. (S2) and subsequently computed the overall accuracy across the entire benchmark dataset.

$$X_{chosen} = \begin{cases} X_{orig} & if\ PPL(X_{orig}) < PPL(X_{alt}) \\ X_{alt} & otherwise \end{cases} \tag{S2}$$

To realize this perplexity calculation, a perplexity-based multiple-choice paradigm was implemented using the llama.cpp inference engine. For each abstract pair in the dataset, the original and a deliberately altered version were randomly assigned to options A and B, and then embedded into a pre-defined prompt template. The local GGUF model was loaded with logits_all=True and a context length of 2,048 tokens, and the perplexity of each complete prompt was computed by calling create_completion with max_tokens=1, echo=True, logprobs=1 and temperature=0.0. The returned token-level log-probabilities for the input sequence (excluding the single generated token) were averaged to obtain the negative log-likelihood, and perplexity was derived as the exponentiation of this mean. The option yielding the lower perplexity was taken as the model's prediction, with ties recorded when the two values were exactly equal. Overall accuracy was calculated as the proportion of instances where the predicted label matched the true label, and all results, including the raw perplexity scores and prompt texts, were saved for further analysis.

**Text S7 Generation prompt for generating several scientific or engineering questions**

This prompt was utilized to instruct the DeepSeek-v4-pro deployed in the Ollama cloud (Version 0.31.2) for extracting information from the collected full-text articles.

```
Generation Prompt

You are an AI assistant tasked with creating a fine-tuning dataset entry in JSON format based on an uploaded research paper PDF. The dataset entry must consist of three fields: "instruction", "input", and "output". Follow these guidelines closely:
Instruction: Generate a human-like instruction that asks the LLM to propose some scientific or engineering questions, which must be strongly related to the research content of the paper (e.g., its topic, findings, or domain), but phrased as a general request from a human user. Do not reveal that the user has referenced or read the paper; make it sound like a natural, standalone query.
Input: If additional context or information is needed to help the LLM generate the output (e.g., specific details, parameters, or background), include it in the "input" field as a string. If no extra information is required, set "input" to an empty string (""). The input should only contain supplementary human-provided content, not the LLM's response.
Output: Generate the LLM's response, which should propose the output based on the paper's research content. However, present the output as if the LLM is leveraging its own knowledge and expertise—do not mention the uploaded files, cite the paper, or indicate any external reference. The hypotheses should align with the paper's themes but be phrased as original thoughts.
Use the content from the paper PDF to inform the instruction, input and output, ensuring relevance and accuracy.
```

```
Output a JSON object with the exact structure below.

{

    "instruction": "human instruction here",

    "input": "human input here or empty string",

    "output": "model's response here"

}
```

**Text S8 Generation prompt for asking a scientific or engineering question to LLM and getting answers**

This prompt was utilized to instruct the DeepSeek-v4-pro deployed in the Ollama cloud (Version 0.31.2) for extracting information from the collected full-text articles.

```
Generation Prompt
You are an AI assistant tasked with creating a fine-tuning dataset entry in JSON format based on an uploaded research paper PDF and an accompanying JSON file. The dataset entry must consist of three fields: "instruction", "input", and "output". Follow these guidelines closely:
Instruction: Generate a human-like instruction that asks a scientific or engineering question to the LLM. This instruction must be strongly related to the research content of the paper (e.g., its topic, findings, or domain), but phrased as a general request from a human user. Do not reveal that the user has referenced or read the paper; make it sound like a natural, standalone query.
Input: If additional context or information is needed to help the LLM generate the output (e.g., specific details, parameters, or background), include it in the "input" field as a string. If no extra information is required, set "input" to an empty string (""). The input should only contain supplementary human-provided content, not the LLM's response.
Output: Generate the LLM's response, which should propose the output based on the paper's research content (using insights from both the PDF and JSON file). However, present the output as if the LLM is leveraging its own knowledge and expertise—do not mention the uploaded files, cite the paper, or indicate any external reference. The hypotheses should align with the paper's themes but be phrased as original thoughts.
Use the content from the paper PDF and JSON file to inform the instruction, input and output, ensuring relevance and accuracy.
```

```
Output a JSON object with the exact structure below.

{

    "instruction": "human instruction here",

    "input": "human input here or empty string",

    "output": "model's response here"

}
```

**Text S9 Generation prompt for generating several rationale scientific hypotheses**

This prompt was utilized to instruct the DeepSeek-v4-pro deployed in the Ollama cloud (Version 0.31.2) for extracting information from the collected full-text articles.

```
Generation Prompt

You are an AI assistant tasked with creating a fine-tuning dataset entry in JSON format based on an uploaded research paper PDF and an accompanying JSON file. The dataset entry must consist of three fields: "instruction", "input", and "output". Follow these guidelines closely:
Instruction: Generate a human-like instruction that asks the LLM to propose scientific hypotheses. This instruction must be strongly related to the research content of the paper (e.g., its topic, findings, or domain), but phrased as a general request from a human user. Do not reveal that the user has referenced or read the paper; make it sound like a natural, standalone query.
Input: If additional context or information is needed to help the LLM generate the output (e.g., specific details, parameters, or background), include it in the "input" field as a string. If no extra information is required, set "input" to an empty string (""). The input should only contain supplementary human-provided content, not the LLM's response.
Output: Generate the LLM's response, which should propose scientific hypotheses based on the paper's research content (using insights from both the PDF and JSON file). However, present the output as if the LLM is leveraging its own knowledge and expertise—do not mention the uploaded files, cite the paper, or indicate any external reference. The hypotheses should align with the paper's themes but be phrased as original thoughts.
Use the content from the paper PDF and JSON file to inform the instruction, input and output, ensuring relevance and accuracy.
Output a single JSON object with the exact structure below.
```

```json
{
    "instruction": "human instruction here",
    "input": "human input here or empty string",
    "output": "model's response here"
}
```

### Text S10 Generation prompt for generating an experimental plan

This prompt was utilized to instruct the DeepSeek-v4-pro deployed in the Ollama cloud (Version 0.31.2) for extracting information from the collected full-text articles.

```
Generation Prompt

You are an AI assistant tasked with creating a fine-tuning dataset entry in JSON format based on an uploaded research paper PDF and an accompanying JSON file. The dataset entry must consist of three fields: "instruction", "input", and "output". Follow these guidelines closely:
Instruction: Generate a human-like instruction that asks the LLM to propose experimental plan. This instruction must be strongly related to the research content of the paper (e.g., its topic, findings, or domain), but phrased as a general request from a human user. Do not reveal that the user has referenced or read the paper; make it sound like a natural, standalone query.
Input: If additional context or information is needed to help the LLM generate the output (e.g., specific details, parameters, or background), include it in the "input" field as a string. If no extra information is required, set "input" to an empty string (""). The input should only contain supplementary human-provided content, not the LLM's response.
Output: Generate the LLM's response, which should propose experimental plan based on the paper's research content (using insights from both the PDF and JSON file). However, present the output as if the LLM is leveraging its own knowledge and expertise—do not mention the uploaded files, cite the paper, or indicate any external reference. The hypotheses should align with the paper's themes but be phrased as original thoughts.
Use the content from the paper PDF and JSON file to inform the instruction, input and output, ensuring relevance and accuracy.
Output a single JSON object with the exact structure below.
```

```
{
    "instruction": "human instruction here",
    "input": "human input here or empty string",
    "output": "model's response here"
}
```

**Text S11 Examples in the fine-tuning dataset**

Here are some examples (JSON format) generated by the four prompts (Text S7-S10) and the DeepSeek-v4-pro deployed in the Ollama cloud (Version 0.31.2) based on the collected full-text articles:

(1) Output generated based on the prompt for generating several scientific or engineering questions

```
{
  "instruction": "I'm designing a treatment system for industrial
wastewater with high sulfate and organic content. The plan is to use
an anaerobic reactor coupled with a biological sulfide oxidation unit
to prevent sulfide toxicity. Could you suggest some important
research questions or engineering challenges that need to be
addressed to make this process efficient and scalable?",
  "input": "",
  "output": "Certainly! Based on current understanding of anaerobic
treatment of sulfate-laden wastewaters and biological sulfide
oxidation, several key scientific and engineering questions merit
further investigation to optimize such a coupled system:\n\n1.
**Sulfide inhibition thresholds**: What are the precise inhibitory
concentrations of total sulfide and unionized hydrogen sulfide for
methanogens and sulfate-reducing bacteria under varying pH,
temperature, and biomass acclimation conditions? Establishing
reliable inhibition kinetics would help design control
strategies.\n\n2. **Control of sulfide oxidation selectivity**: How
can the biological oxidation of sulfide be consistently directed
toward elemental sulfur rather than thiosulfate or sulfate? This
requires systematic study of dissolved oxygen gradients, hydraulic
loading, biofilm characteristics, and the microbial ecology of
sulfur-oxidizing communities (e.g., *Thiobacillus* spp.) in attached-
film reactors.\n\n3. **Influence of COD/sulfate ratio**: What is the
dynamic interplay between the influent COD/sulfate ratio, the
competition between sulfate reducers and methanogens, and overall
process stability? Experiments across a range of ratios would clarify
optimal operating windows and the risk of system upset.\n\n4. **Fate
of specific organic compounds**: How do complex organics—particularly
esters like isopropyl acetate or other solvents—degrade under
sulfidogenic versus methanogenic conditions? Understanding the
hydrolysis pathways and intermediate accumulation (e.g., acetate) can
```

```
guide feed composition and reactor staging.\n\n5. **Recycle ratio and
reactor coupling**: What are the optimal recycle ratios and
integration configurations between the anaerobic and sulfide-
oxidizing reactors? Trade-offs between dilution of influent, sulfide
removal efficiency, oxygen carryover, and energy costs need to be
quantified.\n\n6. **Sulfur recovery and reuse**: What are the most
effective methods to separate, concentrate, and purify the biogenic
elemental sulfur produced, and what marketable applications exist for
this by-product? Life-cycle and techno-economic assessments would
help validate the sustainability of the approach.\n\n7. **Long-term
robustness**: How does the coupled system respond to fluctuations in
wastewater strength, flow, or composition? Studies on resilience,
recovery after shock loads, and biofilm stability over extended
periods are essential for full-scale adoption.\n\nAddressing these
questions through controlled laboratory experiments, pilot-scale
trials, and modeling will significantly advance the reliability and
cost-effectiveness of combined anaerobic–biological sulfide oxidation
treatment."
}
```

(2) Output generated based on the prompt for asking a scientific or engineering question to LLM and getting answers

```
{
  "instruction": "We have a pharmaceutical wastewater stream with a
COD of around 40,000 mg/L, sulfate concentration of 5,000 mg/L, pH of
5.6, and the primary organic component is isopropyl acetate. We're
considering anaerobic treatment but are concerned about sulfate
reduction and sulfide toxicity. What process modifications or
configurations could help us effectively treat this wastewater while
managing sulfide inhibition?",
  "input": "",
  "output": "Treating high-strength, high-sulfate wastewater
anaerobically indeed poses the challenge of sulfate reduction to
sulfide, which can inhibit methanogenesis and even further sulfate
reduction when unionized hydrogen sulfide accumulates above
inhibitory levels (typically >200 mg/L). A proven strategy to
overcome this is to couple anaerobic treatment with a biological
sulfide oxidation step that continuously removes sulfide and recycles
the detoxified effluent.\n\nA configuration that has shown excellent
results uses an anaerobic baffled reactor (ABR) followed by – or
better, integrated with – an attached-film aerobic bioreactor
dedicated to sulfide oxidation. The ABR provides phased digestion and
good biomass retention. By recycling the ABR effluent at a high ratio
```

```
(e.g., 10:1) through a thin-film sulfide-oxidizing reactor before it
re-enters the ABR, two benefits are gained: influent dilution and,
more importantly, oxidation of sulfide to elemental sulfur. This
oxidation is carried out by chemolithotrophic bacteria such as
*Thiobacillus* sp., which can be easily cultured from municipal
sewage. When oxygen supply is carefully limited (by controlling
surface area and flow rate over the biofilm), the dominant product is
elemental sulfur rather than sulfate or thiosulfate, which prevents
acidification and allows sulfur recovery as visible white
granules.\n\nWith this coupled system, at an influent containing 40%
raw wastewater (16,000 mg/L COD, 2,000 mg/L sulfate) and a 1-day
hydraulic retention time, COD removal efficiency can exceed 50% and
sulfate conversion to sulfide can be >95%, while effluent sulfide
remains below 20 mg/L. Methane production typically more than doubles
once sulfide inhibition is relieved. The system also stabilises pH
(often increasing by ~0.5 units) because the consumption of volatile
fatty acids, especially acetate (the main degradation product of
isopropyl acetate), is no longer blocked.\n\nKey design
considerations include:\n- Maintaining a sufficiently high recycle
ratio to keep the ABR sulfide concentration below the inhibition
threshold.\n- Controlling oxygen mass transfer in the
sulfide-oxidising reactor to favour elemental sulfur formation (high
sulfide:oxygen ratio).\n- Providing a support medium for biofilm
attachment (e.g., mesh or structured packing) to retain the
slow-growing sulfide oxidisers.\n- Monitoring the COD/sulfate ratio;
ratios below about 20 tend to exacerbate sulfide inhibition and may
require more intensive recycle or pH control.\n\nThis coupled
anaerobic/aerobic approach not only alleviates toxicity but also
removes sulfur from the liquid stream as a reusable solid, eliminates
odour, and avoids the costs of chemical oxidants or catalysts. It is
a robust, low-energy alternative to physicochemical sulfide stripping
or precipitation."
}
```

(3) Output generated based on the prompt for generating several rationale scientific hypotheses

```
{
  "instruction": "I'm developing a treatment process for high-sulfate
pharmaceutical wastewater that contains isopropyl acetate as a major
organic pollutant. I'm considering combining anaerobic digestion with
a biological sulfide oxidation step. What are some scientific
hypotheses that could explain why this coupled system might be more
effective than anaerobic treatment alone?",
```

```
  "input": "The wastewater has a COD of approximately 40,000 mg/L,
sulfate concentration of 5,000 mg/L, and pH around 5.6. The anaerobic
reactor is operated at mesophilic conditions (35°C) with a hydraulic
retention time of about 1 day.",
  "output": "Based on the characteristics you described, several
hypotheses could explain the potential benefits of coupling anaerobic
treatment with biological sulfide oxidation:\n\n1. Sulfide toxicity
mitigation: Anaerobic sulfate reduction produces hydrogen sulfide,
which is inhibitory to methanogens and sulfate-reducing bacteria at
concentrations above ~200 mg/L. Biological oxidation of sulfide to
elemental sulfur by sulfur-oxidizing bacteria (e.g., Thiobacillus)
reduces the aqueous sulfide concentration, thereby alleviating
inhibition and restoring methanogenic activity.\n\n2. Dilution
effect: Recycling the anaerobic effluent through the sulfide-
oxidizing reactor and back to the anaerobic reactor dilutes the
incoming wastewater, lowering the effective concentrations of both
organic compounds and sulfate. This reduces the initial sulfide
production rate and prevents rapid accumulation of toxic
sulfide.\n\n3. Selective oxidation to elemental sulfur: Under oxygen-
limited conditions and high sulfide-to-oxygen ratios, sulfide is
primarily oxidized to elemental sulfur rather than sulfate or
thiosulfate. This avoids acid generation that would lower pH and
further stress methanogens. Elemental sulfur is insoluble and can be
separated, removing sulfur from the liquid stream permanently.\n\n4.
Acetate accumulation and methanogen sensitivity: Isopropyl acetate
hydrolyzes to acetate and isopropanol. Acetate is a key substrate for
methanogens. High sulfide concentrations may specifically inhibit
aceticlastic methanogens, leading to acetate buildup and COD
breakthrough. By reducing sulfide levels, acetate conversion to
methane is restored, improving COD removal efficiency.\n\n5. pH
stabilization: Sulfate reduction consumes acidity (protons), which
can help raise pH. If sulfide is oxidized to sulfate, it produces
strong acid, lowering pH. However, oxidation to elemental sulfur does
not produce acid. Thus, the coupled process, by favoring elemental
sulfur production, helps maintain a neutral pH favorable for
methanogenesis.\n\n6. Sulfur recovery and odor control: Converting
sulfide to elemental sulfur allows for physical removal of sulfur
from the system, reducing odors and potential for re-dissolution of
sulfide. This also provides an opportunity for sulfur resource
recovery.\n\n7. Synergy between SRB and SOB: Sulfate-reducing
bacteria (SRB) and sulfide-oxidizing bacteria (SOB) may establish a
syntrophic relationship where SRB produce sulfide and SOB consume it,
maintaining low concentrations and allowing both groups to thrive.
This could enhance overall sulfur removal and organic
```

```
degradation.\n\n8. Impact on microbial community structure: Continuous sulfide removal may select for a more robust methanogenic community and prevent the dominance of sulfide-tolerant but less efficient organisms. Over time, this could lead to a more stable and resilient anaerobic consortium.\n\nThese hypotheses could be tested experimentally by measuring sulfide profiles, methane production rates, acetate concentrations, and microbial community analysis under varying recycle ratios and influent strengths."
}
```

(4) Output generated based on the prompt for generating an experimental plan

```
{
  "instruction": "I'm working with a pharmaceutical wastewater that has very high COD (around 40,000 mg/L) and sulfate (about 5,000 mg/L). I plan to treat it using anaerobic digestion, but I'm concerned about sulfide inhibition. Could you propose an experimental plan to develop an effective treatment process, including strategies to control sulfide toxicity?",
  "input": "",
  "output": "To address the challenge of high sulfate and COD in your pharmaceutical wastewater, I recommend a coupled anaerobic/aerobic treatment system that integrates biological sulfide oxidation within the process. This approach not only removes organic matter but also converts toxic sulfide into recoverable elemental sulfur, alleviating inhibition and improving overall performance. Below is a phased experimental plan.\n\n**Phase 1: Reactor Setup and Acclimation**\n- Use an anaerobic baffled reactor (ABR) with multiple compartments (e.g., 5 chambers) to promote phase separation and biomass retention. Maintain a hydraulic retention time (HRT) of approximately 1 day and operate at mesophilic temperature (35°C).\n- Seed the reactor with anaerobic sludge from a municipal digester. Begin feeding with diluted wastewater (e.g., 5-10% strength) supplemented with a readily degradable cosubstrate such as glucose or acetate to rapidly develop a robust microbial consortium. Include nutrients (N, P, trace metals) in the feed.\n- Monitor baseline performance: influent/effluent COD, sulfate, sulfide, volatile fatty acids (VFAs), pH, alkalinity, and biogas composition (methane, CO2). Aim for stable COD removal (>50%) and sulfate reduction before increasing load.\n\n**Phase 2: Stepwise Increase of Organic and Sulfate Load**\n- Gradually raise the influent wastewater concentration in increments (e.g., from 10% to 20%, 30%, 40%, and 50% of full strength). Maintain a constant HRT.\n- At each step, assess reactor stability. Key indicators of sulfide inhibition include a sharp drop in COD removal, accumulation of acetate, decrease in methane production, and effluent sulfide
```

```
concentrations exceeding 200 mg/L. Also track the COD/SO4 ratio;
ratios below 10–20 often exacerbate inhibition.\n- If inhibition is
observed, do not increase load further. Instead, proceed to Phase
3.\n\n**Phase 3: Integration of Biological Sulfide Oxidation**\n-
Attach a thin-film bioreactor (or a rotating biological contactor) to
the anaerobic effluent recycle line. This aerobic reactor should be
designed for high surface area (e.g., ladder structure with mesh) to
support biofilm growth of sulfide-oxidizing bacteria such as
Thiobacillus sp.\n- Enrich a sulfide-oxidizing culture from sewage or
activated sludge using a thiosulfate-based medium. Acclimate the
culture to your effluent and inoculate the thin-film reactor.\n-
Initiate recycle at a ratio of 5:1 to 10:1 (recycle flow : influent
flow). Control oxygen supply by limiting surface area exposure or
using low aeration to maintain a high sulfide:oxygen ratio, favoring
partial oxidation to elemental sulfur rather than sulfate. Monitor
dissolved oxygen in the recycle line (<0.1 mg/L is ideal).\n- Within
days, effluent sulfide should drop below 20 mg/L, and you should
observe white sulfur granules on the biofilm and in the collection
chamber. COD removal and methane production will likely recover and
improve.\n\n**Phase 4: Optimization and Long-Term Operation**\n- With
sulfide inhibition relieved, continue increasing the influent
wastewater concentration up to 50–100% strength. Evaluate performance
at each step.\n- Refine the recycle ratio and oxygen transfer to
maximize sulfur recovery while minimizing sulfate formation.
Periodically harvest sulfur granules from the aerobic unit.\n-
Conduct a sulfur mass balance to confirm conversion pathways and
adjust operational parameters (pH, temperature, loading). Maintain
reactor pH near neutral; the coupled system may naturally buffer pH
due to alkalinity from sulfate reduction and methanogenesis.\n\n**Key
Monitoring and Analytical Recommendations**\n- Daily: COD, sulfate,
sulfide, pH, gas production, temperature, flow rates.\n- Weekly: VFAs
(acetic, propionic, butyric), alkalinity, solids (TSS/VSS), biogas
methane/carbon dioxide content.\n- Periodically: Microscopic
examination of biofilm to confirm presence of sulfur-storing
bacteria; elemental sulfur quantification.\n\n**Expected Outcomes and
Contingencies**\n- With successful coupling, you should achieve >50%
COD removal and >90% sulfate reduction at full-strength wastewater,
with effluent sulfide <20 mg/L.\n- If sulfate reduction remains low,
check nutrient availability and trace metals. If sulfide oxidation
produces excessive sulfate, reduce oxygen supply or increase sulfide
loading.\n- Should severe inhibition recur, consider temporary pH
elevation (to ~7.5) to reduce unionized H2S toxicity, or supplement
the anaerobic reactor with iron to precipitate sulfides.\n\nThis
integrated strategy directly tackles the root cause of inhibition—
```

```
sulfide accumulation—while recovering a valuable byproduct (elemental
sulfur). The phased approach allows systematic troubleshooting and
ensures robust process development for your high-strength
pharmaceutical wastewater."
}
```

**Text S12 Fine-tuning parameters**

BioWater was fine-tuned based on Ministral-3-8B-Instruct-2512 using weights available on Huggingface (https://huggingface.co/mistralai/Ministral-3-3B-Instruct-2512) in the LLaMA Factory webui (https://github.com/hiyouga/LlamaFactory). We used a batch size of 1 and a cutoff length of 2048. The AdamW optimizer with a learning rate of $5 \times 10^{-5}$ and gradient accumulation steps of 8 was utilized for three training epochs. The LoRA rank and LoRA alpha were set as 8 and 16, respectively. All LoRA modules were activated for fine-tuning. The size of fine-tuning dataset was 5316, whereas 10% of dataset was utilized for validation. To optimize training performance, bf16 mixed precision training was employed. The fine-tuning process was carried out on a consumer-grade laptop with RTX 5090 Laptop GPU (24G) and recorded by SwanLab (Version 0.6.10).

**Text S13 The experimental plan generated by the BioWater**

This experimental plan was generated by the BioWater based on our prompt and the hypothesis proposed by the BioWater (See in Fig. 5a).

User Prompt

Can you design a batch experiment plan with varying sulfur and organic conditions to validate this hypothesis, using biomass obtained from an MPDA process?

BioWater Answer

To test the hypothesis that mixotrophic conditions alter partial denitrification efficiency and nitrite accumulation, I propose the following batch experiment plan using biomass from a stable MPDA reactor.

Objective: Determine how the S/N ratio (sulfur-to-nitrogen) and organic carbon concentration affect nitrate reduction pathway (APD vs. HPD), nitrite accumulation, and anammox activity under mixotrophic conditions.

Experimental Setup:

- Use suspended sludge from a mature MPDA reactor treating synthetic wastewater with sulfide and ammonium (e.g., influent COD 60 mg/L, sulfide 20 mgS/L, NH4-N 30 mg/L).
- Prepare serum vials for batch tests: each vial contains 100 mL of mineral medium (no organic carbon or sulfide) plus a fixed nitrate concentration (e.g., 30 mg N/L as NO3-) and varying additions to achieve different conditions.
- Define test groups based on two factors:
    - S/N ratio: 0, 0.25, 0.5, 1.0 (mg S/mg N) using thiosulfate and sulfide as electron donors for APD.
    - Organic carbon level: 0 and 60 mg COD/L (as glucose) to represent anoxic conditions typical of mainstream wastewater.

Data Analysis:

- Compare nitrate reduction rates between vials with/without sulfur (APD-only) versus those with both sulfur and glucose (mixotrophic).
- Determine the optimal S/N ratio that yields highest nitrite accumulation while minimizing nitrate depletion.
- Assess whether organic carbon enhances or inhibits APD by comparing total nitrogen removal rates and anammox activity.
- Identify any negative effects of sulfide toxicity on denitrification kinetics.

Expected Outcomes:

- Mixotrophic conditions should improve overall nitrogen removal but may dilute APD contribution, requiring a trade-off between efficiency and autotrophic stability.
- An optimal S/N ratio (likely ≤0.5) will maximize nitrite accumulation, supporting anammox even in the presence of organics.
- If sulfide inhibition occurs at high concentrations, it could limit process reliability.

This experimental design directly tests how sulfur and organic availability influence partial denitrification pathway selection,

## Text S14 Consistency evaluation prompt and evaluation results

This prompt was utilized to instruct four LLMs (i.e., DeepSeek-v4-pro, GLM-5.1, MiniMax-M3, and Kimi-k2.6) deployed in the Ollama cloud (Version 0.31.2) for evaluating consistency between analysis results of BioWater and that of human experts.

```
Evaluation Prompt

You are a meticulous research auditor specializing in quantitative and qualitative data analysis. Your task is to compare two independent analyses of the same experimental data—one produced by a human expert and one by an AI model—and determine whether they are consistent with each other.
Definition of Consistency: Two analyses are considered consistent if their core findings, conclusions, and key numerical results (e.g., effect sizes, statistical significance, trends, patterns) are in agreement. Minor differences in wording, formatting, or presentation do not affect consistency. However, if the analyses lead to contradictory conclusions, opposite directions of an effect, or substantially different interpretations of the same data, they are inconsistent.
You will receive three pieces of information:
1. Experimental Data – The raw data of the experiment.
2. Human Expert Analysis – The analysis results produced by a human expert.
3. AI Model Analysis – The analysis results produced by an AI model.
Your task include:
1. Review the experimental data to understand the context and expected outcomes.
2. Examine both the human expert analysis and the AI model analysis.
3. Compare the two analyses focusing on: - Main conclusions and key findings - Statistical measures (e.g., p-values, effect sizes, confidence intervals)
```

```
- Trends, patterns, and interpretations - Any critical numerical values or categorical judgments
4. Determine whether the two analyses are consistent or inconsistent according to the definition above.
5. Provide a clear, concise explanation of your judgment, referencing specific points of agreement or disagreement.
You should respond in JSON with the following structure:
```json
{ "consistent": true/false,
"reasoning": "Detailed explanation of the comparison, highlighting where the analyses align or diverge."
}
```

**(1) Analysis results of the BioWater**

I will analyze the data systematically and then validate the hypothesis.

The batch test results reveal distinct kinetic behaviors across the four CODc/CODs ratios, confirming the critical role of mixotrophic balance in MPDA-MBBR. In Batch A (pure autotrophic), rapid sulfide consumption and nitrate reduction were observed, with near-complete nitrite accumulation within 120 min, though later anammox activity led to gradual nitrite depletion. Batch B (CODc/CODs=0.5) showed slower nitrate reduction and significantly reduced nitrite accumulation—only ~0.32 mgN/L—indicating that even a modest organic fraction can shift electron flow, inhibiting autotrophic partial denitrification (APD). In Batch C (CODc/CODs=2.5), nitrate reduction rate increased substantially compared to Batch B, but nitrite accumulated only to ~0.56 mgN/L, and complete denitrification occurred much later (around 540 min). Batch D (pure heterotrophic) exhibited the fastest nitrate reduction rate (up to
```

3.52 mgN/mgCODc·min) and the highest overall nitrate removal (90%), but again, nitrite accumulated only to ~2.74 mgN/L, and residual CODc persisted.

The hypothesis is validated: the optimal mixotrophic ratio is neither pure APD nor pure HPD, but a balanced mix where both autotrophic and heterotrophic denitrification contribute without dominating. Batch C (CODc/CODs=2.5) achieved the best trade-off—nitrite accumulated at a reasonable rate and was sustained for anammox, while CODc was consumed faster than in Batch B, reducing potential inhibition. The data show that increasing CODc/CODs ratio enhances nitrate reduction rate, but at the expense of nitrite accumulation efficiency. Therefore, precise control of the organic-to-sulfur ratio (around 2.5 in this study) is essential to maintain stable nitrite supply for anammox and achieve high total nitrogen removal.

**(2) Analysis results of human experts**

Variation in CODc/CODs strongly regulated nitrogen removal performance and nitrite dynamics in the MPDA system. A pronounced nitrite peak was observed only in the absence of organics (CODc/CODs = 0, batch A), indicating efficient nitrite production via autotrophic denitrification to support anammox. As CODc/CODs increased ($0 \rightarrow 0.5 \rightarrow 2.5 \rightarrow \infty$), ammonium and nitrate removal gradually declined, and overall TN removal decreased (43.3, 38.8, 39.6, and 35.4 mgN/L, respectively), indicating that increased organics adversely affected nitrogen removal in MPDA. Notably, increasing CODc/CODs from 2.5 to $\infty$ led to a marked reduction in TN removal (39.6 to 35.4 mgN/L), underscoring the critical role of CODs in sustaining system performance. Despite variations in carbon structure, the anammox contribution remained consistently high (67–84%) and did not decrease monotonically with increasing CODc/CODs, demonstrating strong competitive capacity of anammox for nitrite and overall process robustness. The absence of nitrite accumulation in batches with organics addition further reflects the competition between heterotrophic denitrification and anammox for nitrite. Collectively, a moderate CODc/CODs range of approximately 0.5–2.5 represents the optimal operational window, preventing nitrite accumulation, maintaining stable TN removal, and sustaining a high anammox contribution.

**(3) Consistency evaluation result of DeepSeek-v4-pro**

{

"consistent": true,

"reasoning": "Both analyses agree on the core findings: (1) the presence of organics strongly influences nitrite dynamics, with a pronounced nitrite peak only in the purely autotrophic Batch A; (2) a mixotrophic condition (moderate CODc/CODs) yields the best overall performance by balancing nitrite supply and consumption, preventing excessive nitrite accumulation while sustaining anammox activity; (3) pure heterotrophic conditions (Batch D) lead to inferior total nitrogen removal despite high nitrate reduction. The human expert identifies an optimal CODc/CODs range of approximately 0.5–2.5, while the AI model pinpoints around 2.5 as the best trade-off. This difference is minor—the AI's specific value falls within the human's broader range, and both highlight that moving to pure heterotrophy (∞) markedly reduces TN removal. The human's TN removal values (43.3, 38.8, 39.6, 35.4 mgN/L) align with the AI's observation that Batch C (2.5) outperforms Batch B (0.5) and Batch D. Both analyses also emphasize the robustness of anammox and the need to control the organic-to-sulfur ratio. No contradictory conclusions, opposite effect directions, or substantially different interpretations are present."

}

**(4) Consistency evaluation result of GLM-5.1**

{

"consistent": true,

"reasoning": "The two analyses are consistent in their core findings, conclusions, and interpretation of the data trends. Both analyses agree on the following key points: 1) **Nitrite Dynamics:** A pronounced nitrite peak (accumulation) only occurs in the absence of organics (Batch A), while the addition of organics (Batches B, C, D) significantly reduces or prevents nitrite accumulation due to competition between heterotrophic denitrification and anammox. 2) **Optimal Operational Window:** Both conclude that a moderate/balanced mixotrophic ratio is optimal. The human specifies an optimal CODc/CODs range of approximately 0.5–2.5, while the AI specifically identifies CODc/CODs = 2.5 (Batch C) as the best trade-off. These conclusions complement rather than contradict each other. 3) **Effect of Organics:** The human notes that increasing organics adversely affects overall TN removal (which decreases), while the AI notes that increasing the ratio enhances the nitrate reduction rate but at the expense of nitrite accumulation efficiency. These are complementary observations (thermodynamic efficiency vs. kinetic rate) rather than contradictory interpretations. Minor differences exist in presentation—such as the human emphasizing anammox contribution percentages and the AI highlighting specific reduction rates—but they do not alter the overarching conclusion that a balanced ratio prevents nitrite accumulation and sustains system performance."

}

**(5) Consistency evaluation result of MiniMax-M3**

{

"consistent": true,

"reasoning": "Both analyses converge on the same core conclusions about the MPDA system: (1) The CODc/CODs ratio critically regulates nitrogen removal and nitrite dynamics; (2) An intermediate mixotrophic balance is optimal, rather than purely autotrophic (Batch A) or purely heterotrophic (Batch D) conditions; (3) Competition between heterotrophic denitrification and anammox for nitrite is the central mechanistic driver; (4) The optimal CODc/CODs range centers on moderate values around 0.5–2.5, with the AI emphasizing 2.5 specifically and the human giving a wider window—both identify this as the practical optimum. Differences are in emphasis rather than in contradiction: the human expert highlights TN removal values (43.3, 38.8, 39.6, 35.4 mgN/L) and anammox contribution stability (67–84%), while the AI focuses on nitrate reduction kinetics, nitrite accumulation levels (~0.32 to ~2.74 mgN/L), and removal rates. The AI's note that Batch D shows the 'fastest nitrate reduction rate' and the human's statement that 'ammonium and nitrate removal gradually declined' with increasing CODc/CODs are not contradictory—they refer to different metrics (kinetic rate vs. cumulative TN removal including anammox-coupled $NH_4$ loss). Both agree that organics addition suppresses nitrite accumulation by redirecting electron flow to heterotrophic denitrification, and both recommend balanced mixotrophic operation. No findings are in direct opposition, and the overall conclusions and recommended operational windows are aligned."

}

**(6) Consistency evaluation result of Kimi-k2.6**

{

"consistent": true,

"reasoning": "Both analyses identify the same overarching patterns and arrive at congruent conclusions. They agree that (1) Batch A (pure autotrophic, CODc/CODs = 0) produces the highest nitrite peak (~8.7 mgN/L) via autotrophic partial denitrification; (2) adding organics suppresses nitrite accumulation in batches B–D because heterotrophic denitrification outcompetes autotrophic pathways for nitrite; (3) pure heterotrophic conditions (Batch D, CODc/CODs = ∞) are suboptimal, leaving residual CODc and yielding the lowest overall nitrogen removal; and (4) the optimal operational window lies at a moderate, mixotrophic CODc/CODs ratio. The human expert specifies an optimal range of ~0.5–2.5 and reports high anammox contribution (67–84%) across this range, while the AI pinpoints the ratio of 2.5 (Batch C) as the 'best trade-off'—a value that falls squarely within the human's recommended window. The AI's emphasis on kinetic rates (e.g., enhanced nitrate reduction with higher CODc/CODs) and the human's emphasis on absolute TN removal amounts (43.3, 38.8, 39.6, 35.4 mgN/L) are complementary rather than contradictory; both recognize that increasing organics alters the nitrogen removal pathway and that a balanced mix is required to sustain anammox and avoid nitrite accumulation. Minor differences in wording, metrics, and specificity (range vs. single ratio) do not constitute opposite effect directions or contradictory core conclusions."

}

**Text S15 Estimation of local grid water intensity and local electrovalence**

Local grid water intensity was estimated using a generation-mix-weighted approach. Annual electricity generation by fuel type was obtained from the Ember Yearly Electricity Data dataset (https://github.com/owid/etl/blob/master/snapshots/ember/2026-04-24/yearly_electricity__global.csv.dvc), which compiles national electricity statistics from official national and international sources. Data for 2021-2025 were used where available; Indonesia was represented by 2021-2024 data and Ukraine by 2021-2022 data because later observations were unavailable. Operational water-consumption factors for individual generation technologies were taken from the literature review[2]. Median factors of 553, 687, 198, 4,491, 672, 1, and 0 US gal/MWh were applied to bioenergy, coal, natural gas, hydropower, nuclear power, solar photovoltaic power, and wind power, respectively. "Other fossil" generation was represented using the coal steam-generation factor, while "other renewables" were assigned a factor of 270 US gal/MWh, corresponding to the median value for the geothermal configurations reported in the same source. For each country $c$, grid water intensity was calculated as:

$$WI_c = \frac{\sum_y \sum_s G_{c,y,s} \cdot W_s}{\sum_y \sum_s G_{c,y,s}} \times \frac{3.7854}{1000} \quad \text{(S3)}$$

where $G_{c,y,s}$ is electricity generation in country $c$, year $y$, and generation source $s$, and $W_s$ is the corresponding operational water-consumption factor in US gal/MWh. The factor 3.7854/1000 converts US gal/MWh to L/kWh. The resulting values therefore represent planning-level estimates of water consumed during electricity generation rather than directly measured national grid values. Electricity imports, differences among individual power plants and cooling systems, and the substantial site-specific variability in reservoir evaporation were not explicitly modeled. The estimates should consequently be interpreted as national generation-mix proxies, particularly for countries with large shares of hydropower[2].

Electricity prices were obtained from the World Bank Doing Business indicator "Price of electricity (US cents per kWh), DB16–20 methodology" (https://databank.worldbank.org/metadataglossary/doing-business/series/IC.ELC.PRI.KH.DB1619). This indicator estimates the electricity tariff paid by a standardized commercial warehouse in the principal business city of each economy, thereby providing a comparable commercial-price proxy rather than a residential tariff or a consumption-weighted national average. To reduce year-specific exchange-rate and tariff fluctuations, the arithmetic mean of all available observations from 2016-2019 was calculated for each country. Values reported in US cents/kWh were divided by 100 to obtain USD/kWh.

**Table S1 Biological wastewater treatment publications from March 1 to June 8, 2026**

| Journal | Number |
|---|---|
| Water Research | 23 |
| Bioresource Technology | 18 |
| Journal of Water Process Engineering | 10 |
| Environmental Science & Technology | 9 |
| Journal of Environmental Chemical Engineering | 8 |
| Journal of Hazardous Materials | 6 |
| Chemical Engineering Journal | 5 |
| Environmental Research | 5 |
| Journal of Environmental Management | 5 |
| Journal of Cleaner Production | 4 |
| Water Science and Technology | 4 |
| Biotechnology and Bioengineering | 2 |
| Nature Water | 1 |
| Total | 100 |

**Table S2 Topic categories of collected biological wastewater treatment publications**

| Topic category | Number |
|---|---|
| Anammox-based technology | 27 |
| Membrane and biofilm technology | 26 |
| General nitrification and denitrification technology | 25 |
| Algal-bacteria technology | 10 |
| Models and intelligent control | 9 |
| Sulfur-based technology | 3 |
| Total | 100 |

**Table S3 Overview of LLMs as targets**

| Model Series | Parameter Size | Pretraining Data Cutoff | Release Date |
| --- | --- | --- | --- |
| LFM2.5-8B-A1B | 8.3B | Unknown | May 2026 |
| Granite4.1:8b | 8B | Unknown | April 2026 |
| Gemma4:e4b | 8B | January 2025 | April 2026 |
| Qwen3.5:9b | 9B | Unknown | February 2026 |
| Ministral-3:8b | 8B | 2023 | December 2025 |
| BioWater | 8B | Finetuned | This study |
| Qwen3.5:397b | 397B | Unknown | February 2026 |

**Table S4 Overview of LLMs as Judges**

| Model Series | Parameter Size | Pretraining Data Cutoff | Release Date |
|---|---|---|---|
| DeepSeek-v4-pro | 1.6T | May 2025 | April 2026 |
| GLM-5.1 | 756B | Unknown | April 2026 |
| MiniMax-M3 | 196B | January 2026 | June 2026 |
| Kimi-k2.6 | 1.04T | April 2025 | April 2026 |
| Gemini3.1 Pro | Unknown (Closed source model) | January 2025 | February 2026 |

**Table S5 List of collected 100 biological wastewater treatment publications**

| No. | DOI | Comment |
|---|---|---|
| 1 | 10.1002/bit.70246 | |
| 2 | 10.1002/bit.70155 | |
| 3 | 10.1016/j.biortech.2026.134898 | |
| 4 | 10.1016/j.biortech.2026.134545 | |
| 5 | 10.1016/j.biortech.2026.134471 | |
| 6 | 10.1016/j.biortech.2026.134678 | |
| 7 | 10.1016/j.biortech.2026.134677 | |
| 8 | 10.1016/j.biortech.2026.134561 | |
| 9 | 10.1016/j.biortech.2026.134895 | |
| 10 | 10.1016/j.biortech.2026.134915 | |
| 11 | 10.1016/j.biortech.2026.134732 | |
| 12 | 10.1016/j.biortech.2026.134499 | Excluded from the comprehension fidelity task |
| 13 | 10.1016/j.biortech.2026.134589 | |
| 14 | 10.1016/j.biortech.2026.134530 | |
| 15 | 10.1016/j.biortech.2026.134416 | |
| 16 | 10.1016/j.biortech.2026.134260 | |
| 17 | 10.1016/j.biortech.2026.134359 | |
| 18 | 10.1016/j.biortech.2026.134799 | |
| 19 | 10.1016/j.biortech.2026.134474 | |
| 20 | 10.1016/j.biortech.2026.134442 | |
| 21 | 10.1016/j.cej.2026.175347 | |
| 22 | 10.1016/j.cej.2026.174684 | |
| 23 | 10.1016/j.cej.2026.175553 | |
| 24 | 10.1016/j.cej.2026.175295 | Excluded from the comprehension fidelity task |
| 25 | 10.1016/j.cej.2026.174471 | Excluded from the comprehension fidelity task |
| 26 | 10.1016/j.envres.2026.124303 | |
| 27 | 10.1016/j.envres.2026.124490 | |
| 28 | 10.1016/j.envres.2026.124456 | |
| 29 | 10.1016/j.envres.2026.124266 | |
| 30 | 10.1016/j.envres.2026.124374 | |
| 31 | 10.1021/acs.est.5c16045 | |
| 32 | 10.1021/acs.est.5c07873 | |
| 33 | 10.1021/acs.est.5c13522 | |
| 34 | 10.1021/acs.est.5c16556 | |
| 35 | 10.1021/acs.est.5c14655 | |
| 36 | 10.1021/acs.est.5c15806 | Excluded from the comprehension fidelity task |
| 37 | 10.1021/acs.est.6c02189 | |
| 38 | 10.1021/acs.est.5c11756 | Excluded from the comprehension fidelity task |
| 39 | 10.1021/acs.est.5c15251 | Excluded from the comprehension fidelity task |
| 40 | 10.1016/j.jclepro.2026.148024 | |
| 41 | 10.1016/j.jclepro.2026.147944 | |
| 42 | 10.1016/j.jclepro.2026.148426 | Trap question |
| 43 | 10.1016/j.jclepro.2026.148067 | Excluded from the comprehension fidelity task |
| 44 | 10.1016/j.jece.2026.122447 | Excluded from the comprehension fidelity task |
| 45 | 10.1016/j.jece.2026.122609 | |
| 46 | 10.1016/j.jece.2026.122710 | |
| 47 | 10.1016/j.jece.2026.122465 | Excluded from the comprehension fidelity task |
| 48 | 10.1016/j.jece.2026.122934 | |
| 49 | 10.1016/j.jece.2026.122660 | Excluded from the comprehension fidelity task |
| 50 | 10.1016/j.jece.2026.122503 | |
| 51 | 10.1016/j.jece.2026.123025 | |
| 52 | 10.1016/j.jenvman.2026.129476 | |
| 53 | 10.1016/j.jenvman.2026.129729 | |
| 54 | 10.1016/j.jenvman.2026.129367 | |
| 55 | 10.1016/j.jenvman.2026.129646 | |
| 56 | 10.1016/j.jenvman.2026.129525 | |
| 57 | 10.1016/j.jhazmat.2026.141472 | |
| 58 | 10.1016/j.jhazmat.2026.142163 | |
| 59 | 10.1016/j.jhazmat.2026.142260 | |
| 60 | 10.1016/j.jhazmat.2026.141801 | |
| 61 | 10.1016/j.jhazmat.2026.142390 | |
| 62 | 10.1016/j.jhazmat.2026.141447 | |
| 63 | 10.1016/j.jwpe.2026.109994 | |
| 64 | 10.1016/j.jwpe.2026.109995 | |
| 65 | 10.1016/j.jwpe.2026.109968 | |
| 66 | 10.1016/j.jwpe.2026.109969 | |
| 67 | 10.1016/j.jwpe.2026.110124 | |
| 68 | 10.1016/j.jwpe.2026.110006 | |
| 69 | 10.1016/j.jwpe.2026.110161 | |
| 70 | 10.1016/j.jwpe.2026.110151 | |
| 71 | 10.1016/j.jwpe.2026.110073 | |
| 72 | 10.1016/j.jwpe.2026.110153 | |
| 73 | 10.1038/s44221-026-00638-5 | |
| 74 | 10.1016/j.watres.2026.125923 | |
| 75 | 10.1016/j.watres.2026.125649 | |
| 76 | 10.1016/j.watres.2026.126003 | |
| 77 | 10.1016/j.watres.2026.125838 | |
| 78 | 10.1016/j.watres.2026.126132 | |
| 79 | 10.1016/j.watres.2026.125892 | |
| 80 | 10.1016/j.watres.2026.125793 | |
| 81 | 10.1016/j.watres.2026.125901 | |
| 82 | 10.1016/j.watres.2026.125775 | |
| 83 | 10.1016/j.watres.2026.125914 | |
| 84 | 10.1016/j.watres.2026.125652 | |
| 85 | 10.1016/j.watres.2026.126021 | Trap question |
| 86 | 10.1016/j.watres.2026.125944 | |
| 87 | 10.1016/j.watres.2026.125922 | |
| 88 | 10.1016/j.watres.2026.125907 | |
| 89 | 10.1016/j.watres.2026.125857 | |
| 90 | 10.1016/j.watres.2026.126015 | |
| 91 | 10.1016/j.watres.2026.125945 | |
| 92 | 10.1016/j.watres.2026.126088 | Excluded from the comprehension fidelity task |
| 93 | 10.1016/j.watres.2026.125926 | |
| 94 | 10.1016/j.watres.2026.125871 | |
| 95 | 10.1016/j.watres.2026.126061 | |
| 96 | 10.1016/j.watres.2026.126087 | |
| 97 | 10.2166/wst.2026.271 | Excluded from the comprehension fidelity task |
| 98 | 10.2166/wst.2026.192 | Excluded from the comprehension fidelity task |
| 99 | 10.2166/wst.2026.200 | |
| 100 | 10.2166/wst.2026.278 | Excluded from the comprehension fidelity task |

**Table S6 The initial concentrations of organic, sulfur, and nitrogen compounds in batch experiments**

| Batch | Sulfide mgS/L | Thiosulfate mgS/L | CODs mg/L | $NH_4^+$ mg N/L | $NO_3^-$ mg N/L | Organic mg COD/L | CODc mg/L |
|---|---|---|---|---|---|---|---|
| A | 30 | 30 | 90 | 30 | 30 | 0 | 0 |
| B | 20 | 20 | 60 | 30 | 30 | 30 | 30 |
| C | 10 | 10 | 30 | 30 | 30 | 60 | 60 |
| D | 0 | 0 | 0 | 30 | 30 | 90 | 90 |

CODs: COD sulfur equivalent
CODc: COD organic equivalent

**Table S7 Distribution of 7,801 wastewater treatment plants in 47 countries (regions)**

| No. | Country (region) | Number of wastewater treatment plants |
|---|---|---|
| 1 | Argentina | 18 |
| 2 | Australia | 83 |
| 3 | Austria | 44 |
| 4 | Belgium | 30 |
| 5 | Brazil | 181 |
| 6 | Bulgaria | 22 |
| 7 | Canada | 178 |
| 8 | Chile | 10 |
| 9 | China | 1842 |
| 10 | Cyprus | 3 |
| 11 | Czech Republic | 30 |
| 12 | Denmark | 43 |
| 13 | Estonia | 5 |
| 14 | Finland | 25 |
| 15 | France | 234 |
| 16 | Germany | 443 |
| 17 | Greece | 38 |
| 18 | Hungary | 19 |
| 19 | India | 330 |
| 20 | Indonesia | 36 |
| 21 | Ireland | 46 |
| 22 | Italy | 269 |
| 23 | Japan | 307 |
| 24 | Latvia | 3 |
| 25 | Lithuania | 7 |

| 26 | Luxembourg | 7 |
|---|---|---|
| 27 | Malaysia | 59 |
| 28 | Malta | 1 |
| 29 | Mexico | 154 |
| 30 | Netherlands | 136 |
| 31 | New Zealand | 22 |
| 32 | Norway | 37 |
| 33 | Poland | 110 |
| 34 | Portugal | 46 |
| 35 | Romania | 46 |
| 36 | Russia | 218 |
| 37 | Slovenia | 11 |
| 38 | South Korea | 79 |
| 39 | Spain | 208 |
| 40 | Sweden | 30 |
| 41 | Switzerland | 126 |
| 42 | Taiwan-China | 25 |
| 43 | Thailand | 35 |
| 44 | Turkey | 82 |
| 45 | Ukraine | 43 |
| 46 | United Kingdom | 296 |
| 47 | United States | 1784 |

**Table S8 Scenario-based LLM application modes for operational decision support in WWTPs**

| LLM application mode | Functional allocation of LLM calls | Rationale, applicability, and operational constraints |
|---|---|---|
| Low-frequency assistance, 80 calls/WWTP/d | Process-state interpretation: 24 calls/d (hourly)<br>Anomaly diagnosis and alarm deduplication: 12 calls/d (every 2 h)<br>Influent, weather, and load forecasting: 8 calls/d (every 3 h)<br>Aeration and return activated sludge recommendations: 6 calls/d (every 4 h)<br>Chemical dosing and sludge-management recommendations: 6 calls/d (every 4 h)<br>Equipment health assessment: 6 calls/d (every 4 h)<br>Energy-planning coordination: 6 calls/d (every 4 h)<br>Compliance checking and report drafting: 4 calls/d<br>Operator interaction and shift handover support: 8 calls/d | Suited to digitally upgraded WWTPs that require continuous decision support while retaining predominantly operator-led control. A complete plant-state interpretation is generated approximately once per hour, while major task-specific assessments are refreshed every 2-4 h. The 80-call scenario is intended to provide coverage across three operating shifts and routine abnormal events without repeatedly invoking the LLM for static reporting or equipment recommendations. |
| Medium-frequency optimization, 180 calls/WWTP/d | Process-state interpretation: 48 calls/d (every 30 min)<br>Anomaly diagnosis and risk classification: 48 calls/d (every 30 min)<br>Aeration and energy-optimization recommendations: 24 calls/d (hourly)<br>Chemical dosing and sludge-line coordination: 12 calls/d (every 2 h) | Suited to WWTPs with relatively mature sensing infrastructure and historical databases. Half-hourly state interpretation and anomaly assessment support trend detection, whereas aeration-optimization |
|  | Critical-asset health assessment: 12 calls/d (every 2 h)<br>Rolling influent and weather forecasting: 12 calls/d (every 2 h)<br>Compliance-boundary and reporting checks: 8 calls/d (every 3 h)<br>Operator interaction and shift handover support: 8 calls/d<br>Cross-process energy coordination: 8 calls/d | recommendations are limited to at most once per hour. Final actuator set-points remain under PLC/MPC execution and should comply with predefined deadbands, ramp-rate limits, and minimum holding times. |
| High-frequency decision support, 360 calls/WWTP/d | Process-state interpretation: 96 calls/d (every 15 min)<br>Anomaly diagnosis and risk classification: 96 calls/d (every 15 min)<br>Aeration and energy-optimization recommendations: 48 calls/d (every 30 min)<br>Chemical dosing and sludge-line coordination: 24 calls/d (hourly)<br>Critical-asset health assessment: 24 calls/d (hourly)<br>Rolling influent and weather forecasting: 24 calls/d (hourly)<br>Cross-process energy coordination: 24 calls/d (hourly)<br>Compliance-boundary and reporting checks: 12 calls/d (every 2 h)<br>Operator interaction and shift handover support: 12 calls/d (every 2 h) | Suited to WWTPs equipped with high-quality online monitoring, a digital twin, redundant computational capacity, and an independent safety-verification layer. Fifteen-minute LLM calls are reserved primarily for plant-state interpretation and anomaly screening, while aeration optimization is limited to one recommendation every 30 min. |

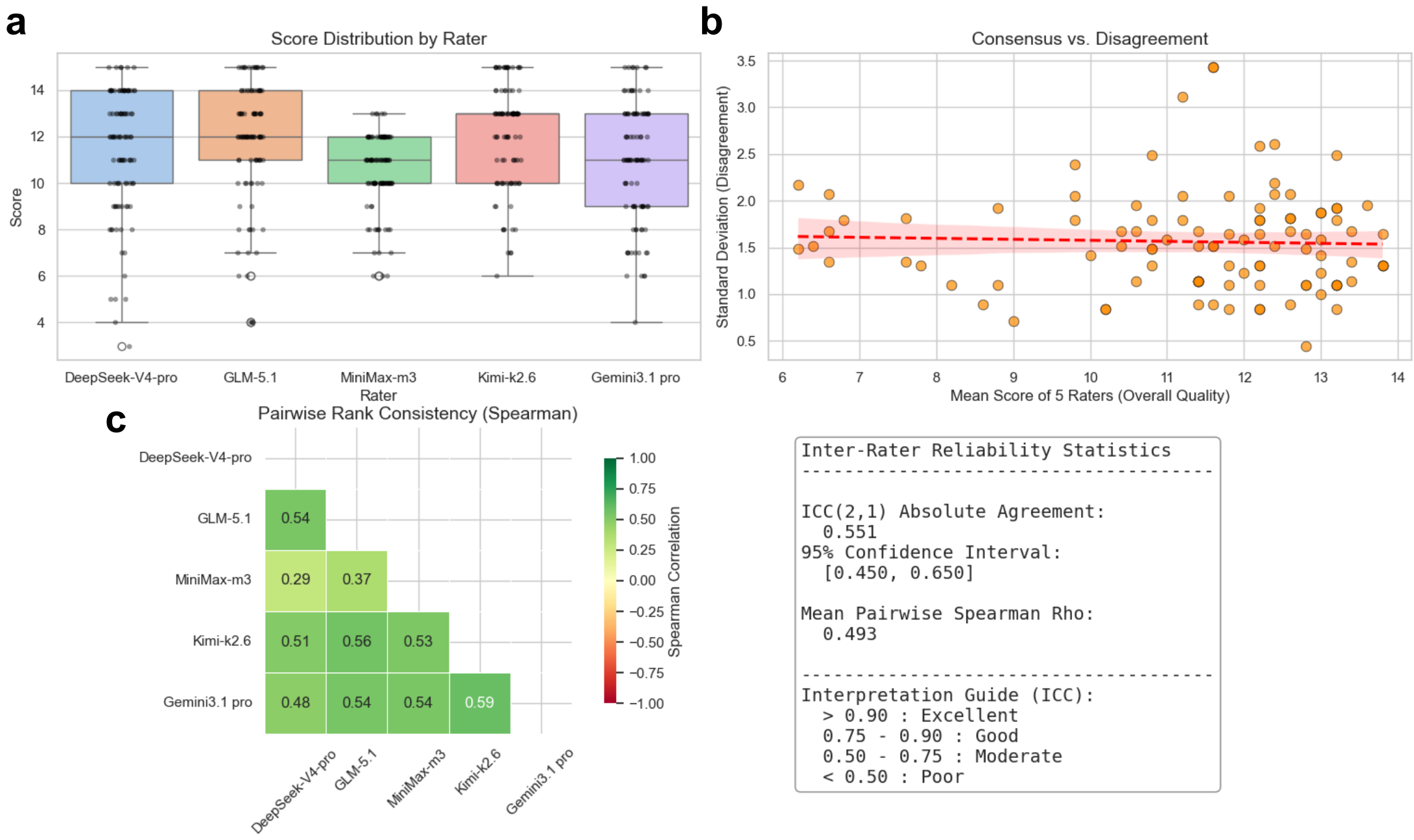

**Fig. S1 Evaluation of inter-rater consistency and scoring behavior across five judge LLMs on retrospective cognition results of LFM2.5-8B-A1B. a**, score distributions across five judge LLMs, reflecting individual strictness or leniency. **b**, relationship between overall sample quality (mean score) and rater disagreement (standard deviation). **c**, pairwise Spearman rank correlation matrix highlighting the ranking consensus among the five judge LLMs.

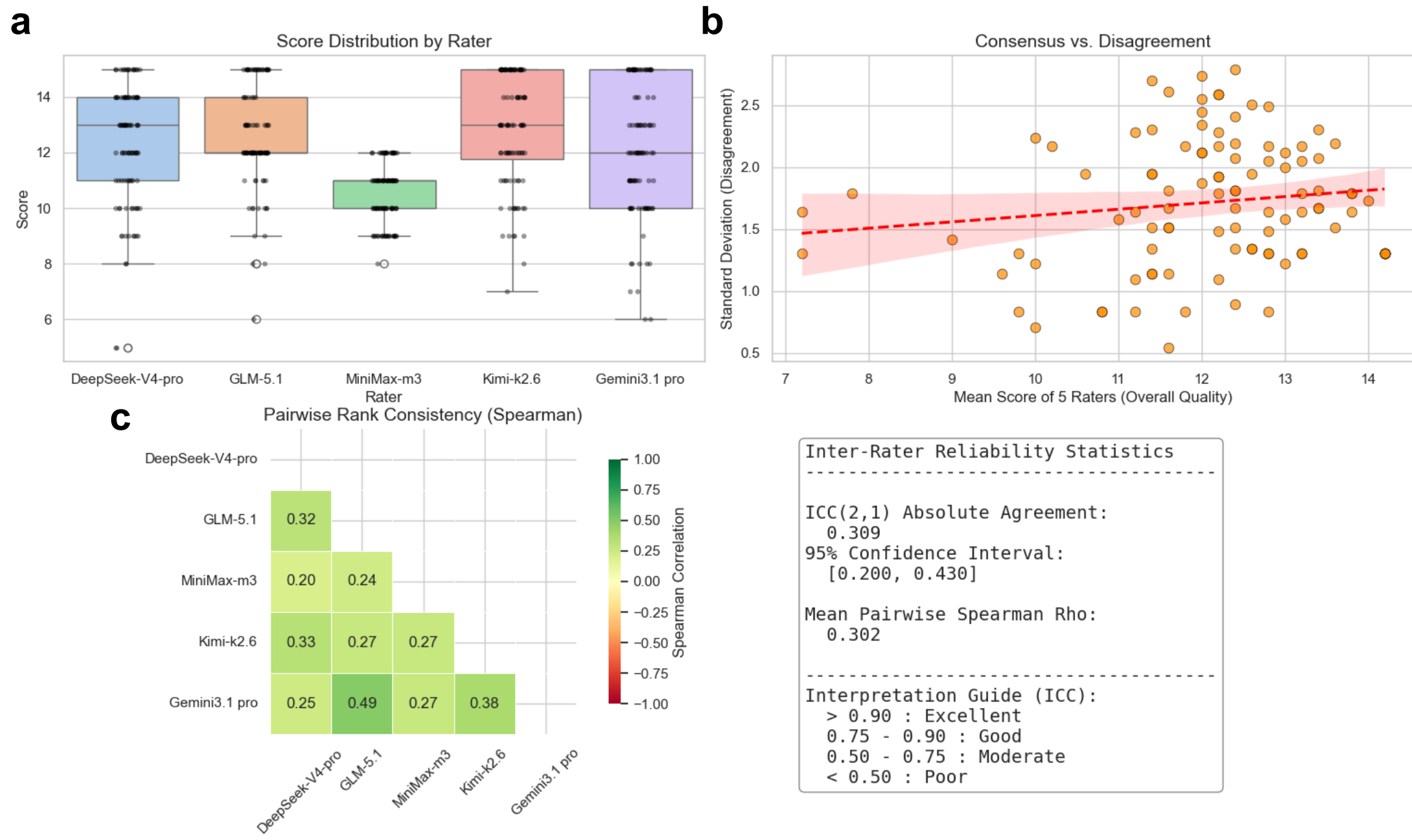

**Fig. S2 Evaluation of inter-rater consistency and scoring behavior across five judge LLMs on retrospective cognition results of Granite4.1:8b. a**, score distributions across five judge LLMs, reflecting individual strictness or leniency. **b**, relationship between overall sample quality (mean score) and rater disagreement (standard deviation). **c**, pairwise Spearman rank correlation matrix highlighting the ranking consensus among the five judge LLMs.

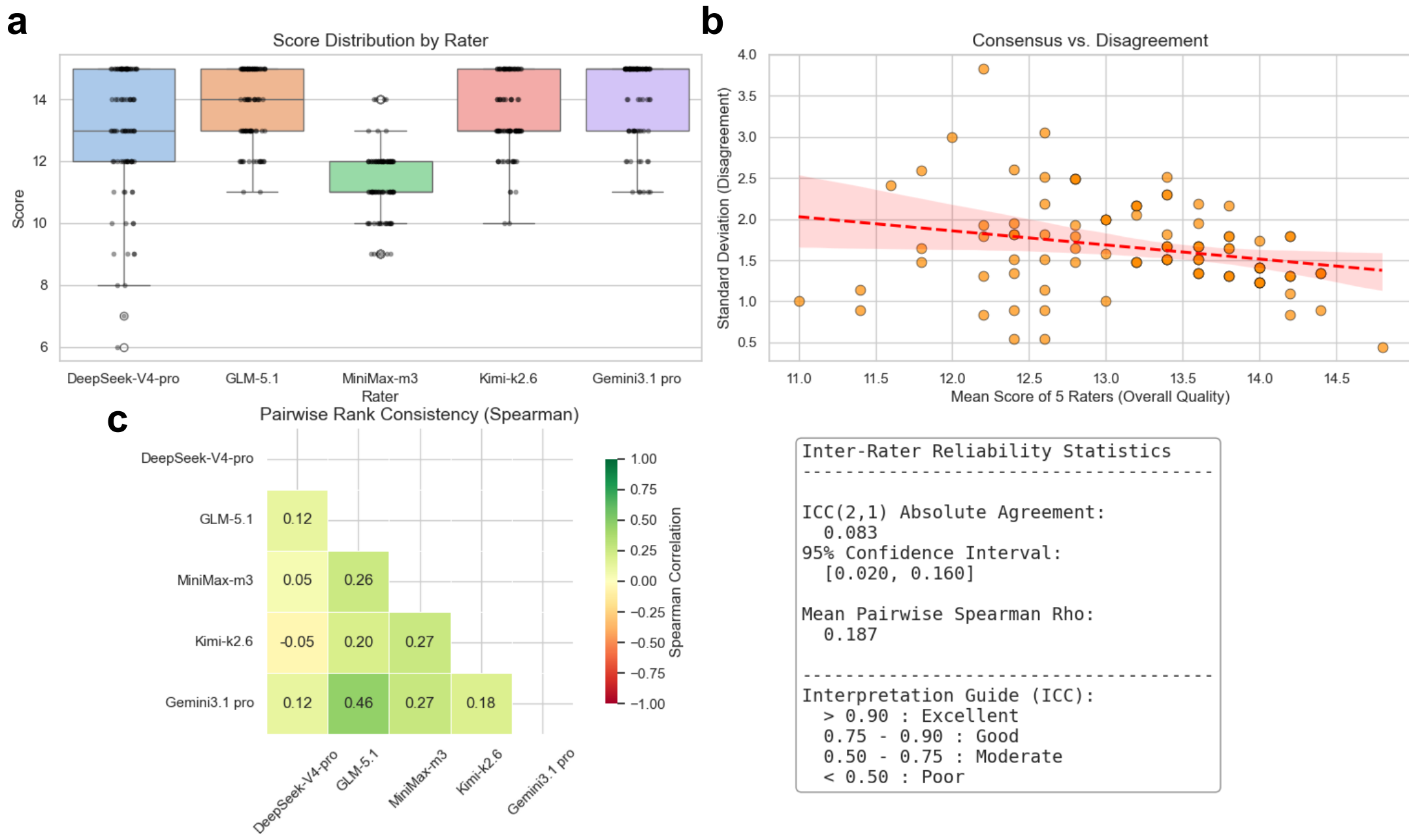

**Fig. S3 Evaluation of inter-rater consistency and scoring behavior across five judge LLMs on retrospective cognition results of Gemma4:e4b. a**, score distributions across five judge LLMs, reflecting individual strictness or leniency. **b**, relationship between overall sample quality (mean score) and rater disagreement (standard deviation). **c**, pairwise Spearman rank correlation matrix highlighting the ranking consensus among the five judge LLMs.

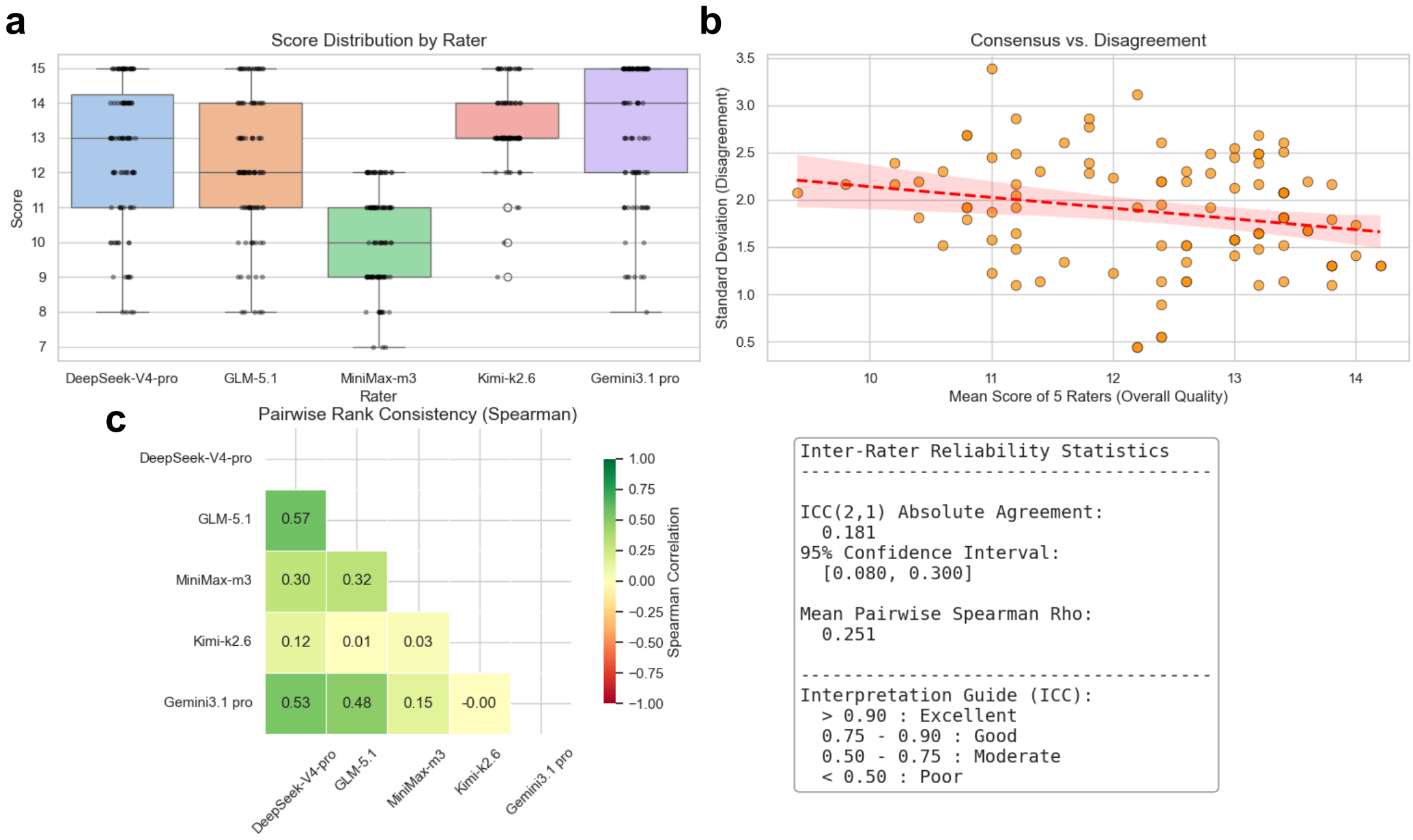

**Fig. S4 Evaluation of inter-rater consistency and scoring behavior across five judge LLMs on retrospective cognition results of Qwen3.5:9b. a**, score distributions across five judge LLMs, reflecting individual strictness or leniency. **b**, relationship between overall sample quality (mean score) and rater disagreement (standard deviation). **c**, pairwise Spearman rank correlation matrix highlighting the ranking consensus among the five judge LLMs.

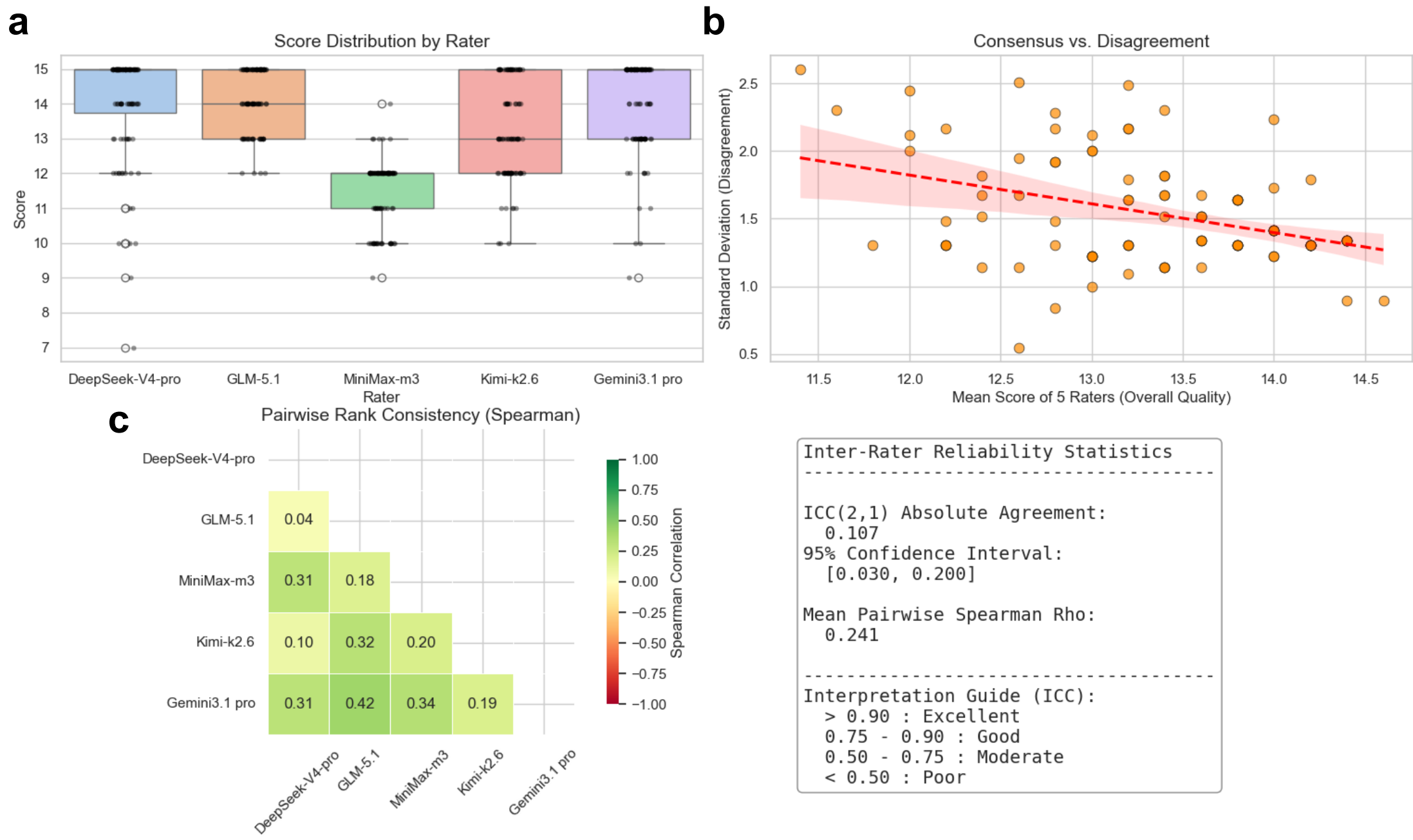

**Fig. S5 Evaluation of inter-rater consistency and scoring behavior across five judge LLMs on retrospective cognition results of Ministral-3:8b. a**, score distributions across five judge LLMs, reflecting individual strictness or leniency. **b**, relationship between overall sample quality (mean score) and rater disagreement (standard deviation). **c**, pairwise Spearman rank correlation matrix highlighting the ranking consensus among the five judge LLMs.

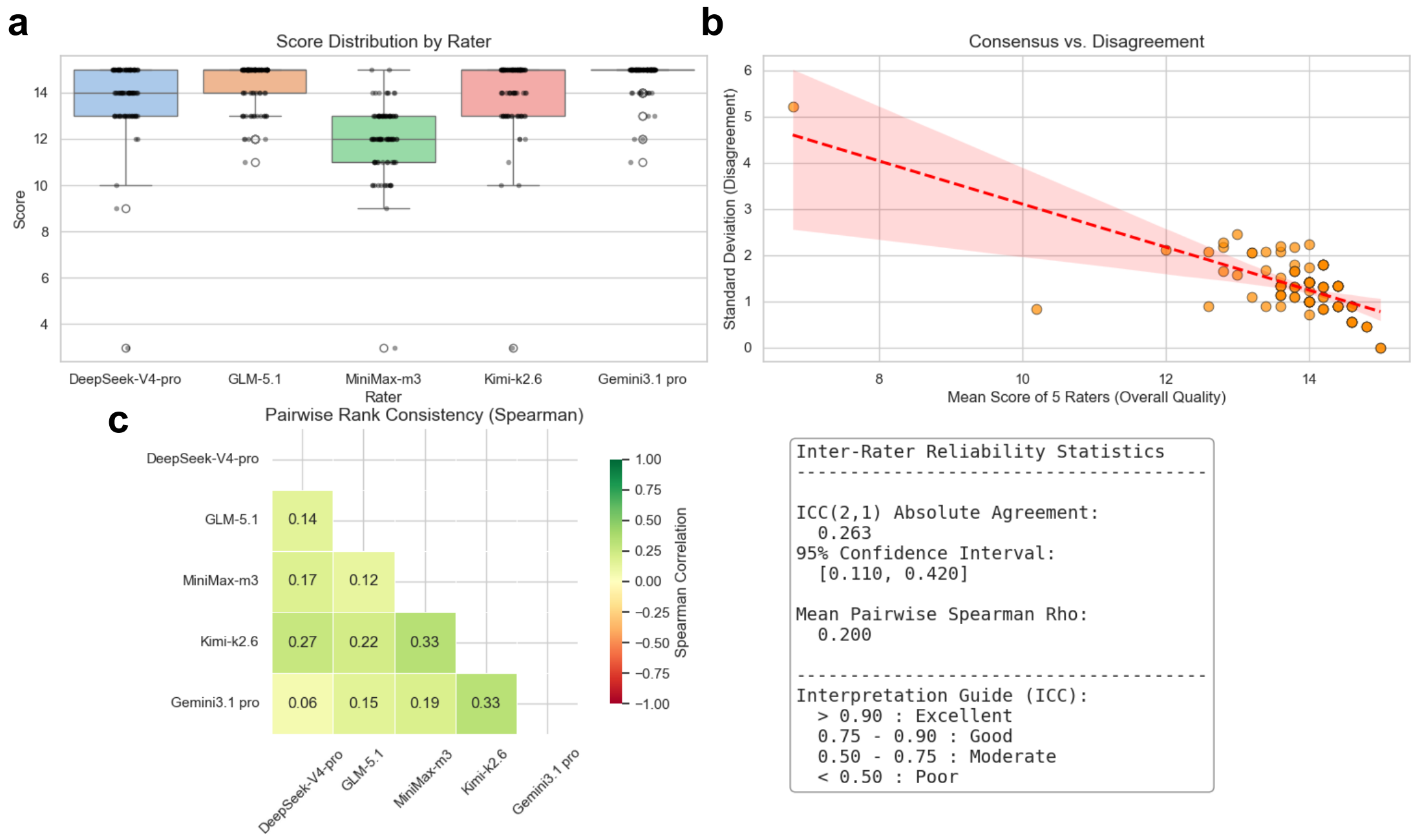

**Fig. S6 Evaluation of inter-rater consistency and scoring behavior across five judge LLMs on retrospective cognition results of Qwen3.5:397b. a**, score distributions across five  judge LLMs, reflecting individual strictness or leniency. **b**, relationship between overall sample quality (mean score) and rater disagreement (standard deviation). **c**, pairwise Spearman rank correlation matrix highlighting the ranking consensus among the five judge LLMs.

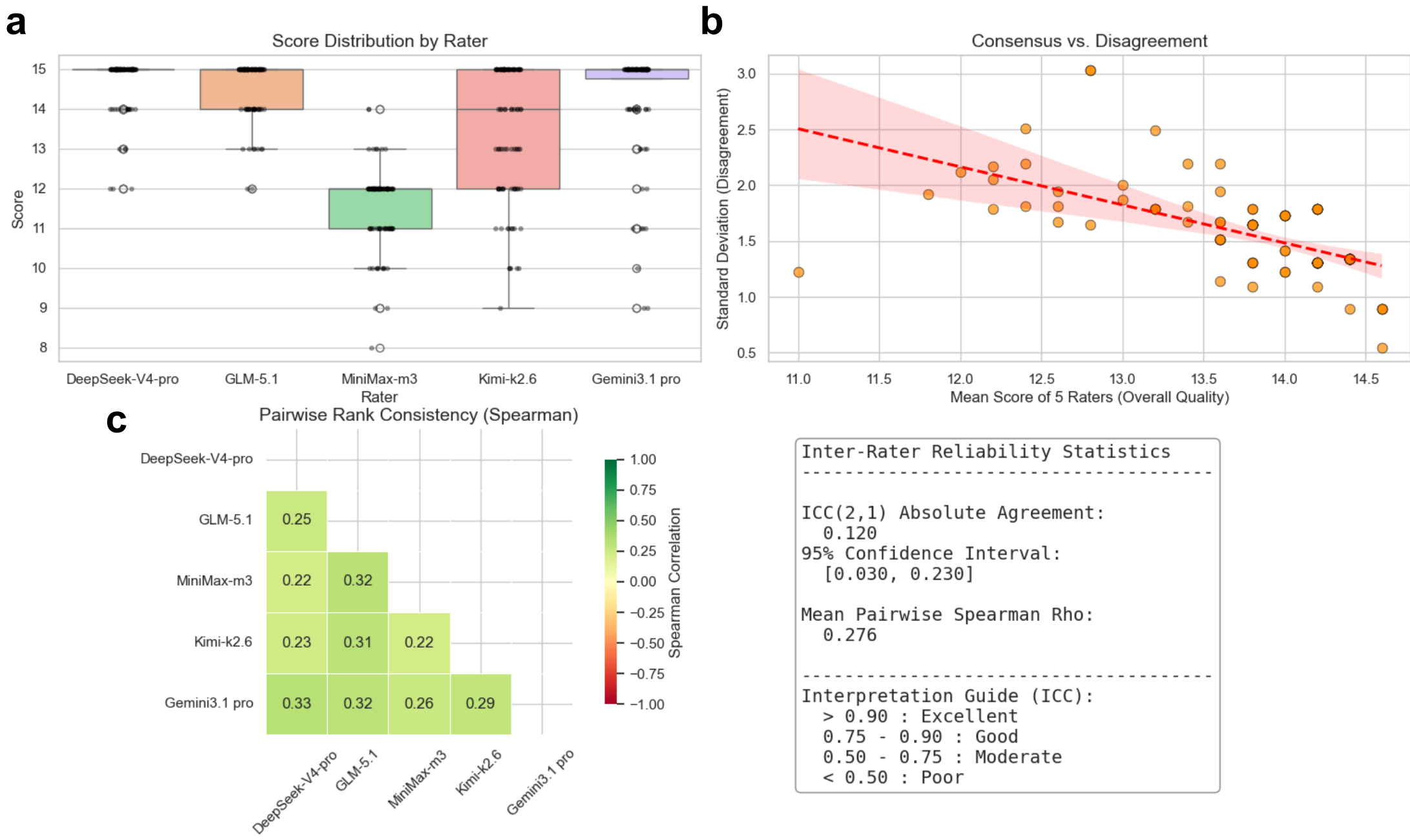

**Fig. S7 Evaluation of inter-rater consistency and scoring behavior across five judge LLMs on retrospective cognition results of BioWater. a**, score distributions across five judge LLMs, reflecting individual strictness or leniency. **b**, relationship between overall sample quality (mean score) and rater disagreement (standard deviation). **c**, pairwise Spearman rank correlation matrix highlighting the ranking consensus among the five judge LLMs.

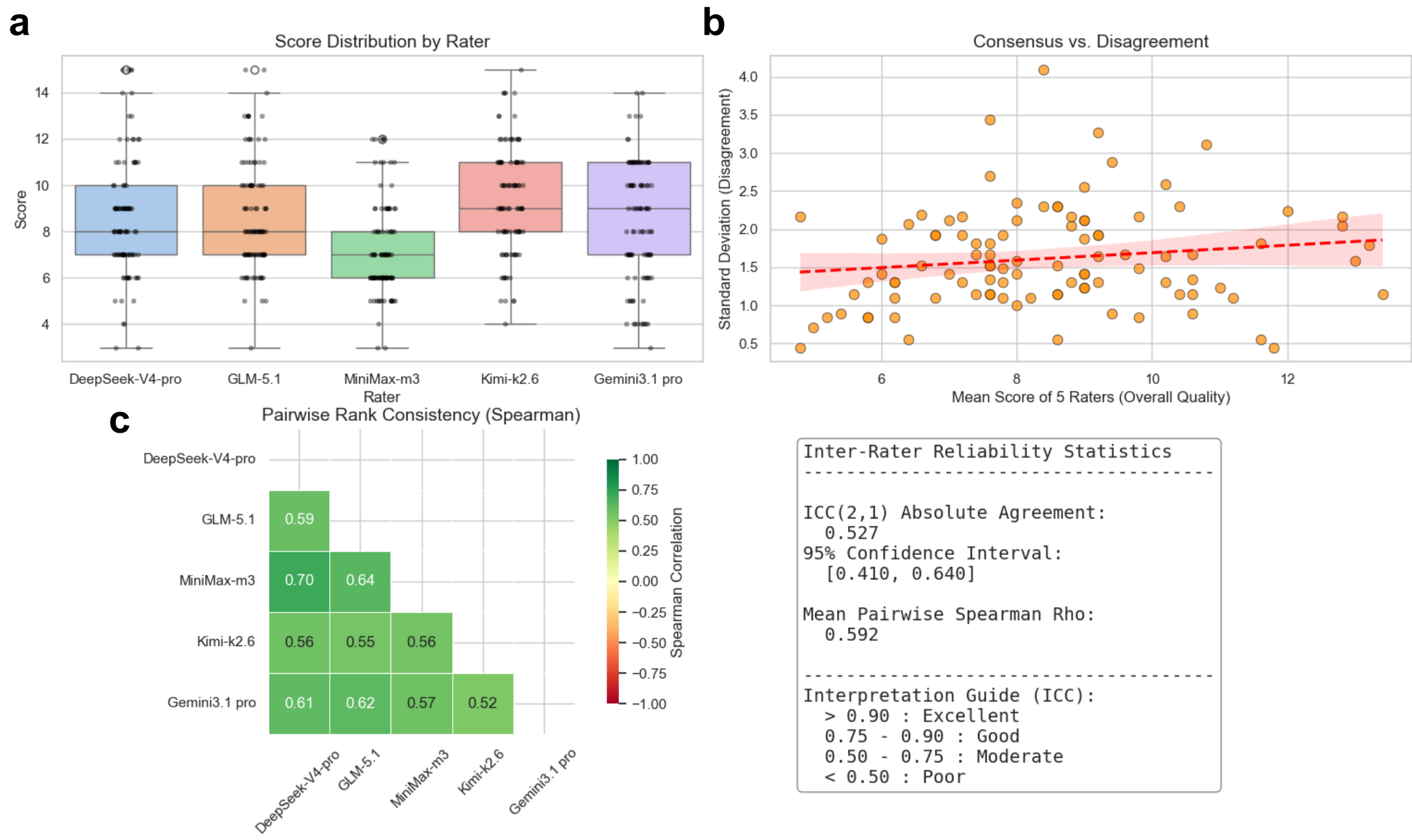

**Fig. S8 Evaluation of inter-rater consistency and scoring behavior across five judge LLMs on prospective extrapolation results of LFM2.5-8B-A1B. a**, score distributions across five judge LLMs, reflecting individual strictness or leniency. **b**, relationship between overall sample quality (mean score) and rater disagreement (standard deviation). **c**, pairwise Spearman rank correlation matrix highlighting the ranking consensus among the five judge LLMs.

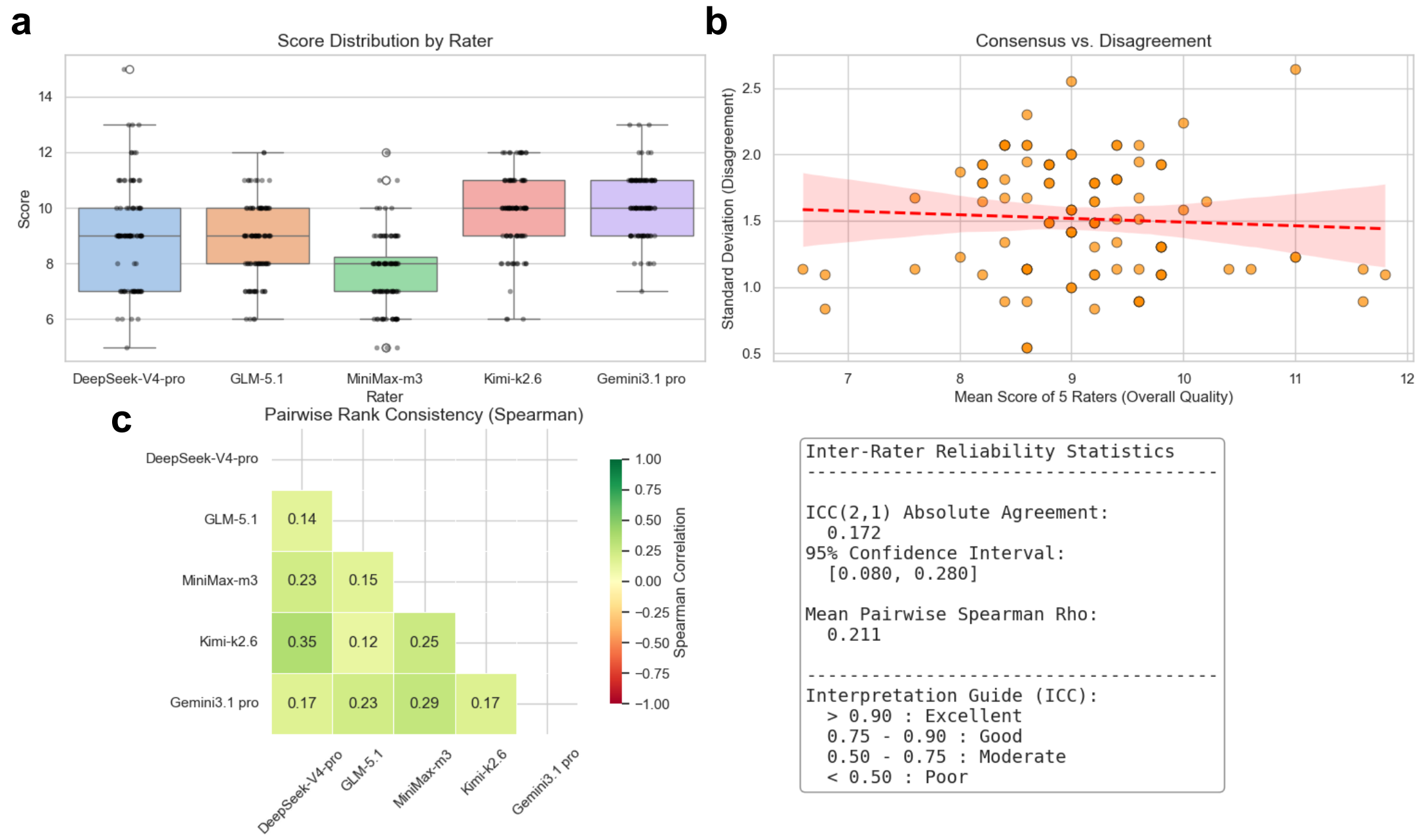

**Fig. S9 Evaluation of inter-rater consistency and scoring behavior across five judge LLMs on prospective extrapolation results of Granite4.1:8b. a**, score distributions across five judge LLMs, reflecting individual strictness or leniency. **b**, relationship between overall sample quality (mean score) and rater disagreement (standard deviation). **c**, pairwise Spearman rank correlation matrix highlighting the ranking consensus among the five judge LLMs.

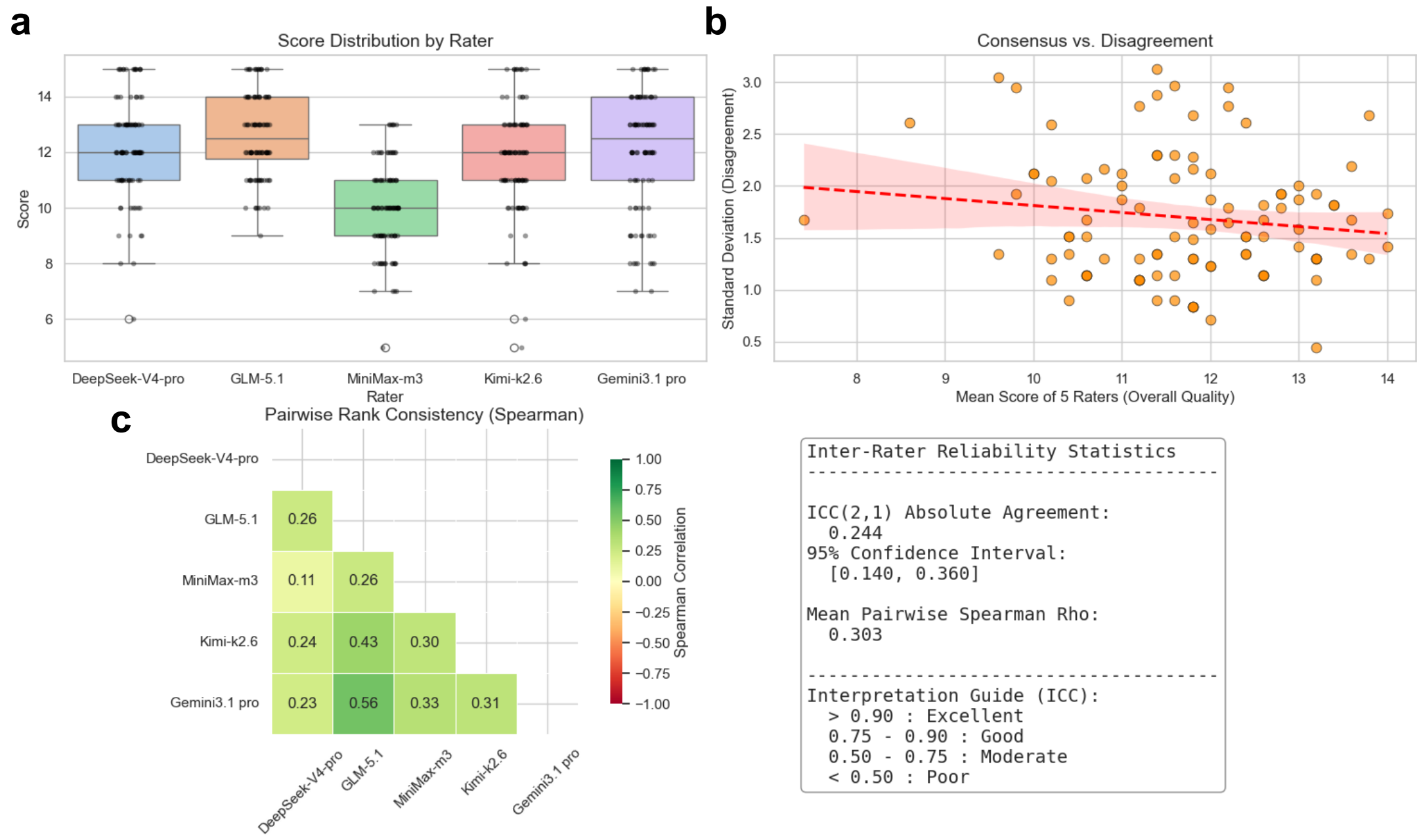

**Fig. S10 Evaluation of inter-rater consistency and scoring behavior across five judge LLMs on prospective extrapolation results of Gemma4:e4b. a**, score distributions across five judge LLMs, reflecting individual strictness or leniency. **b**, relationship between overall sample quality (mean score) and rater disagreement (standard deviation). **c**, pairwise Spearman rank correlation matrix highlighting the ranking consensus among the five judge LLMs.

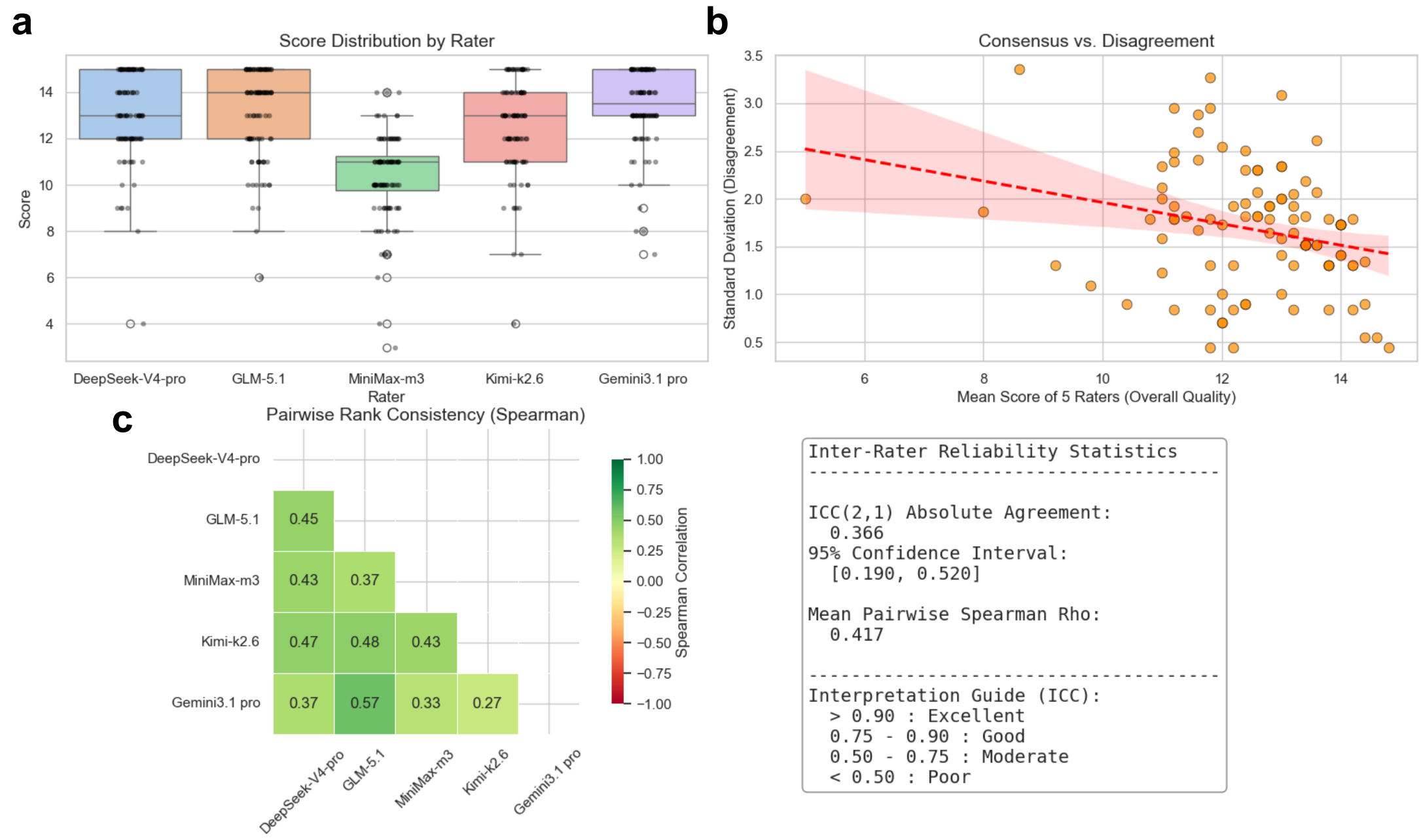

**Fig. S11 Evaluation of inter-rater consistency and scoring behavior across five judge LLMs on prospective extrapolation results of Qwen3.5:9b. a**, score distributions across five judge LLMs, reflecting individual strictness or leniency. **b**, relationship between overall sample quality (mean score) and rater disagreement (standard deviation). **c**, pairwise Spearman rank correlation matrix highlighting the ranking consensus among the five judge LLMs.

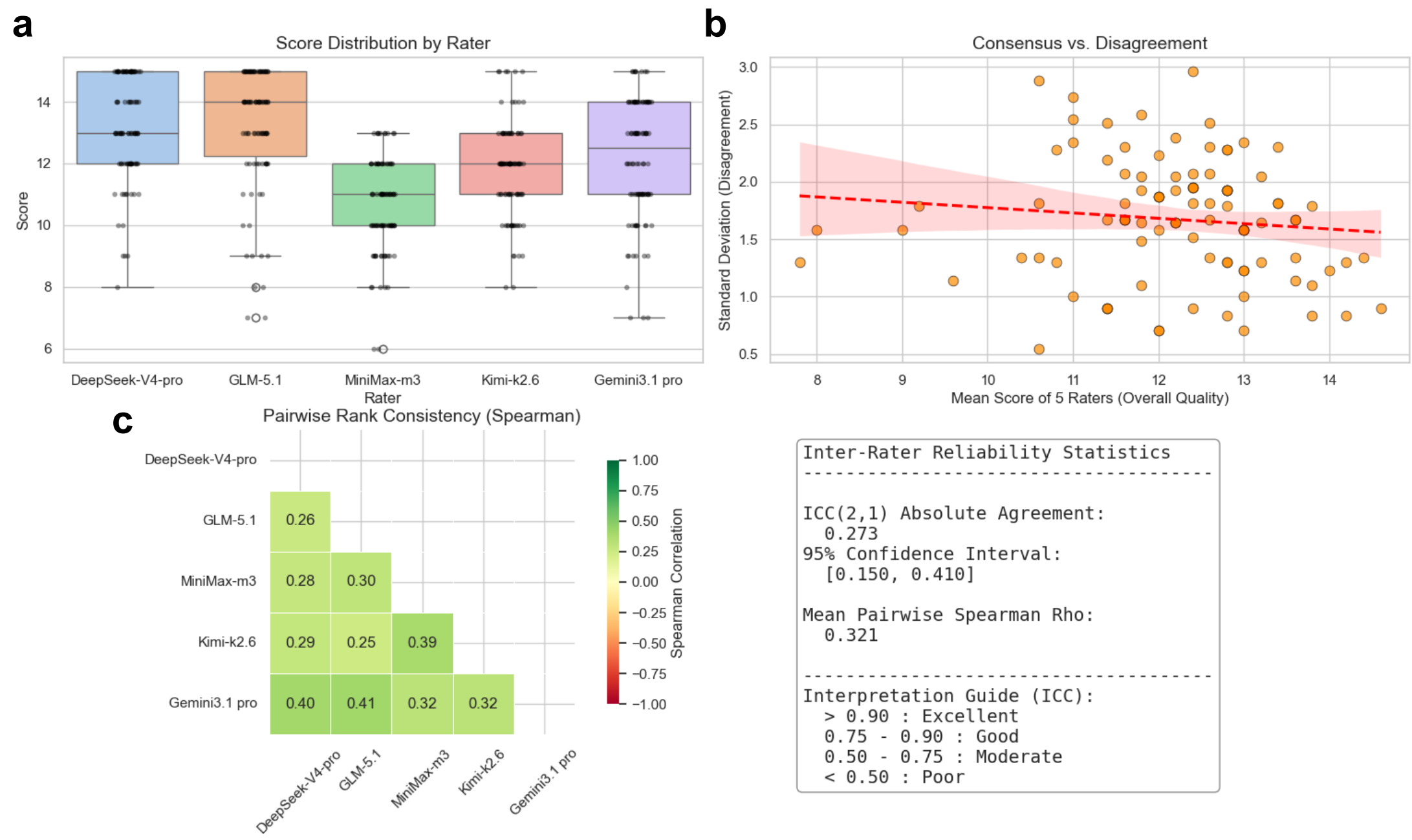

**Fig. S12 Evaluation of inter-rater consistency and scoring behavior across five judge LLMs on prospective extrapolation results of Ministral-3:8b. a**, score distributions across five judge LLMs, reflecting individual strictness or leniency. **b**, relationship between overall sample quality (mean score) and rater disagreement (standard deviation). **c**, pairwise Spearman rank correlation matrix highlighting the ranking consensus among the five judge LLMs.

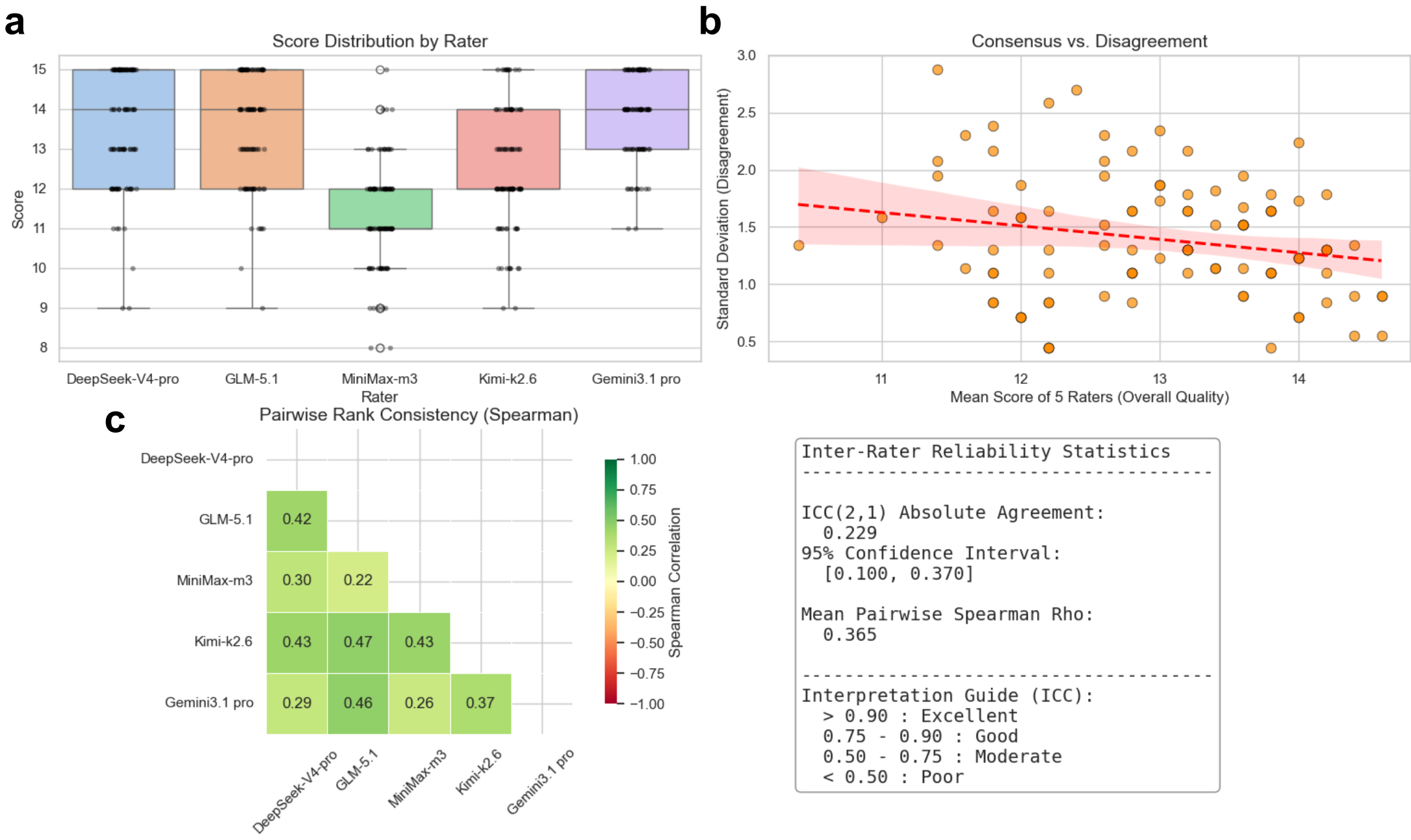

**Fig. S13 Evaluation of inter-rater consistency and scoring behavior across five judge LLMs on prospective extrapolation results of Qwen3.5:397b. a**, score distributions across five judge LLMs, reflecting individual strictness or leniency. **b**, relationship between overall sample quality (mean score) and rater disagreement (standard deviation). **c**, pairwise Spearman rank correlation matrix highlighting the ranking consensus among the five judge LLMs.

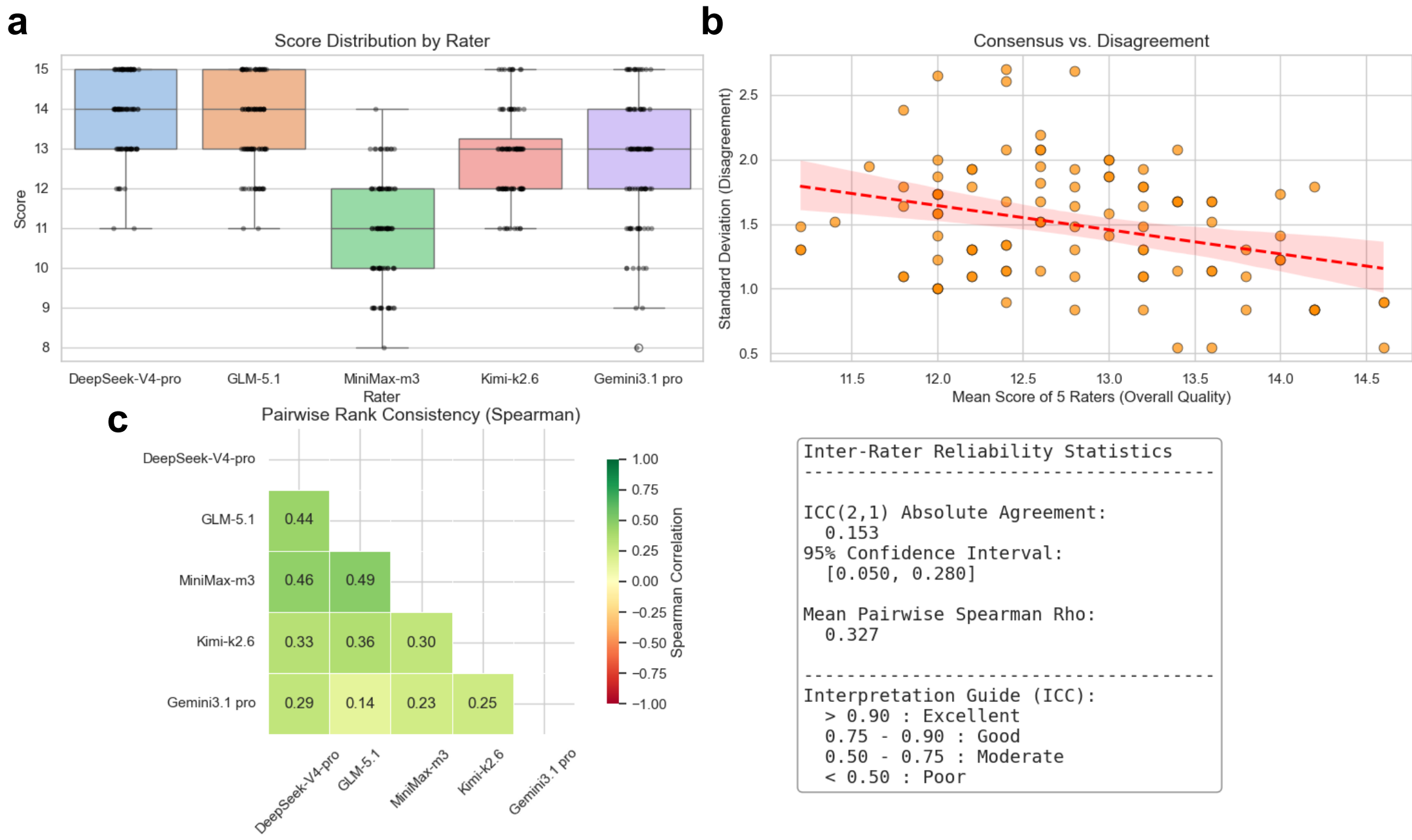

**Fig. S14 Evaluation of inter-rater consistency and scoring behavior across five judge LLMs on prospective extrapolation results of BioWater. a**, score distributions across five judge LLMs, reflecting individual strictness or leniency. **b**, relationship between overall sample quality (mean score) and rater disagreement (standard deviation). **c**, pairwise Spearman rank correlation matrix highlighting the ranking consensus among the five judge LLMs.

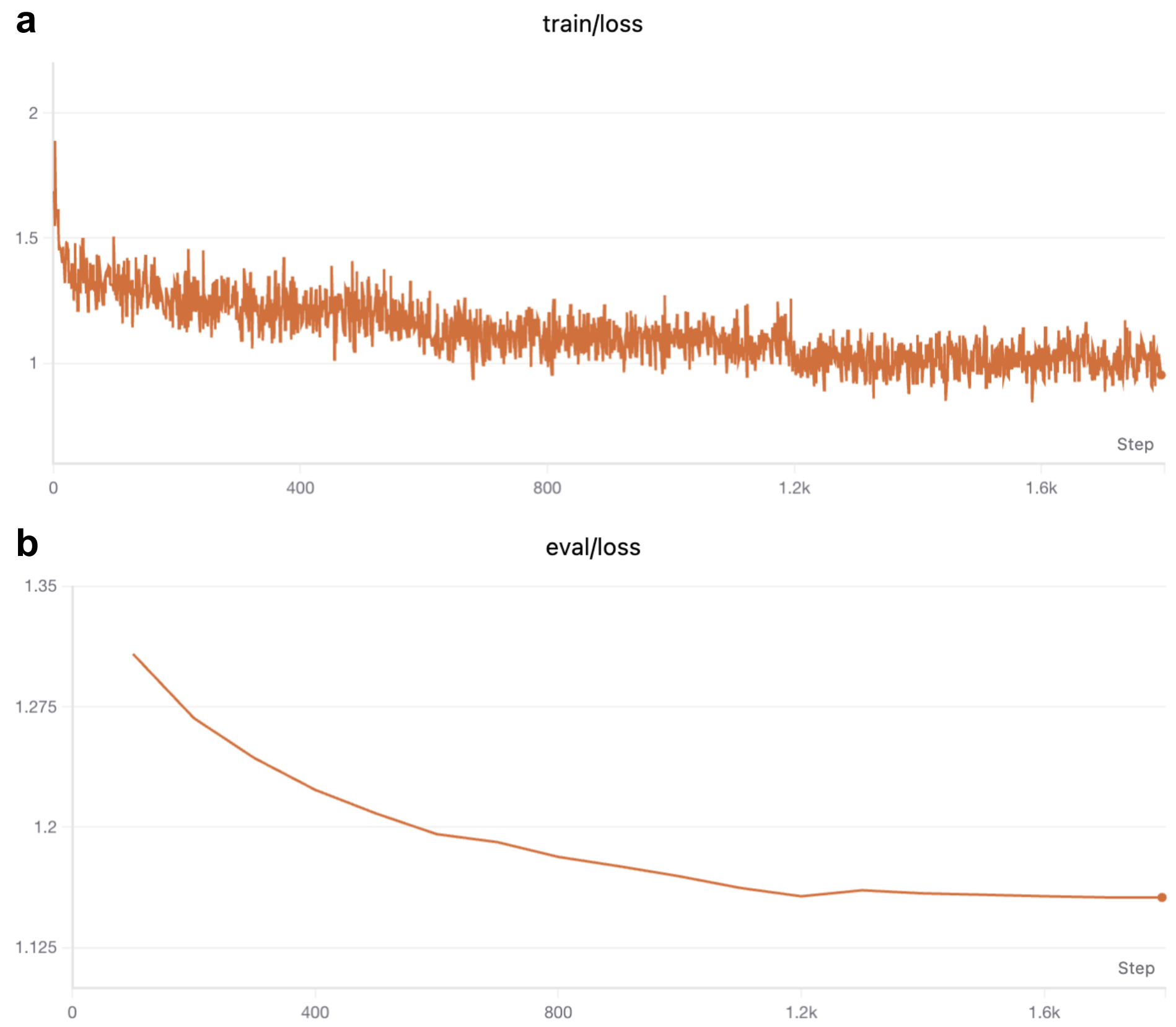

**Fig. S15 The fine-tuning performance of BioWater drawn by the SwanLab automatically. a**, the training loss of BioWater. **b**, the validation loss of BioWater.

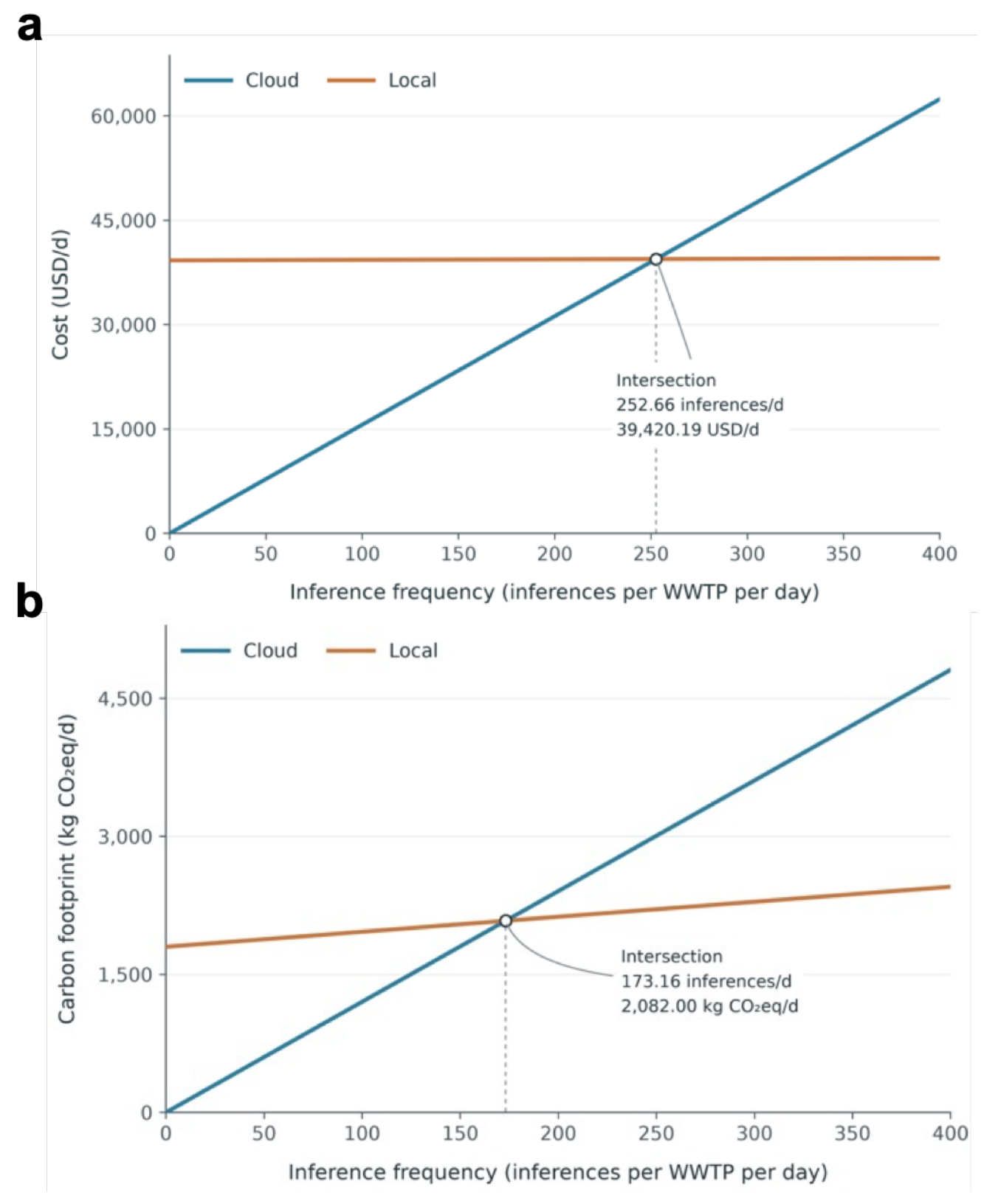

**Fig. S16 Inference-frequency-dependent cost (a) and carbon footprint (b) crossover points between cloud and local LLMs deployment across 7,801 wastewater treatment plants worldwide.** Local deployment costs also include daily equipment depreciation and a fixed annual cost allocated on a daily basis, as specified in the underlying model.

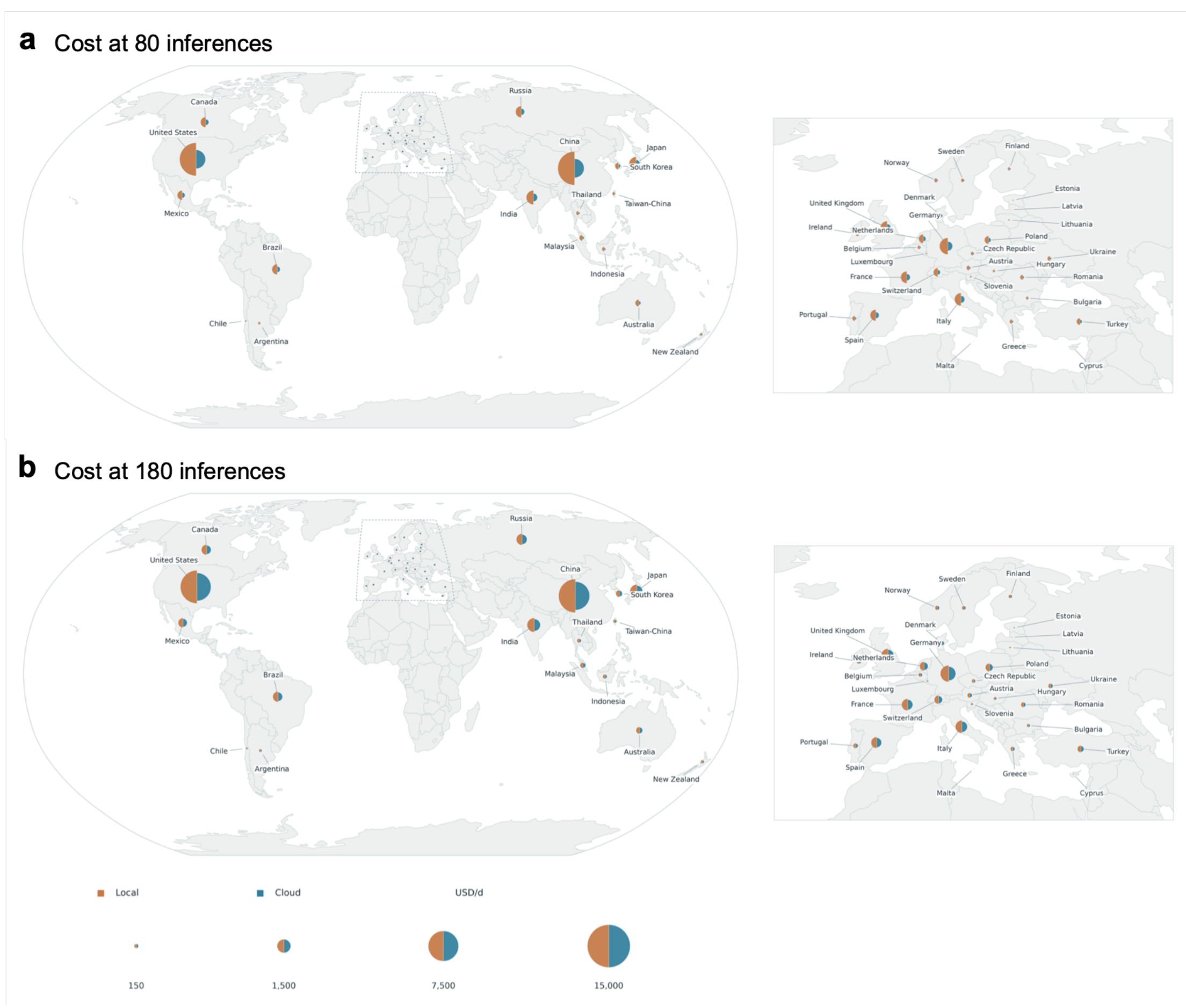

**Fig. S17 Maps compare the daily cost of cloud and local deployment across 47 countries (regions) at 80 and 180 inferences per WWTP per day.**

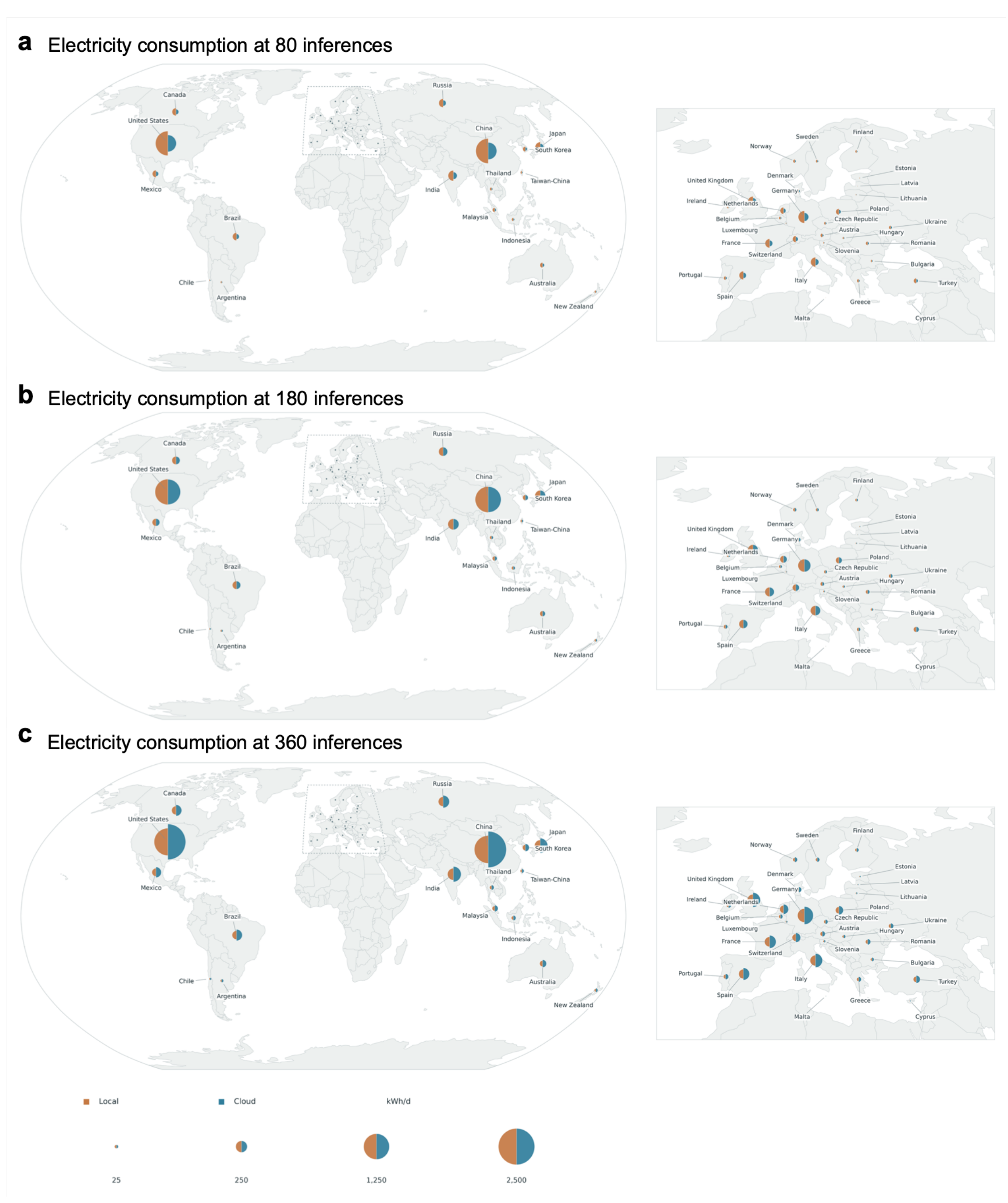

**Fig. S18 Maps compare the daily electricity consumption of cloud and local deployment across 47 countries (regions) at 80, 180 and 360 inferences per WWTP per day.**

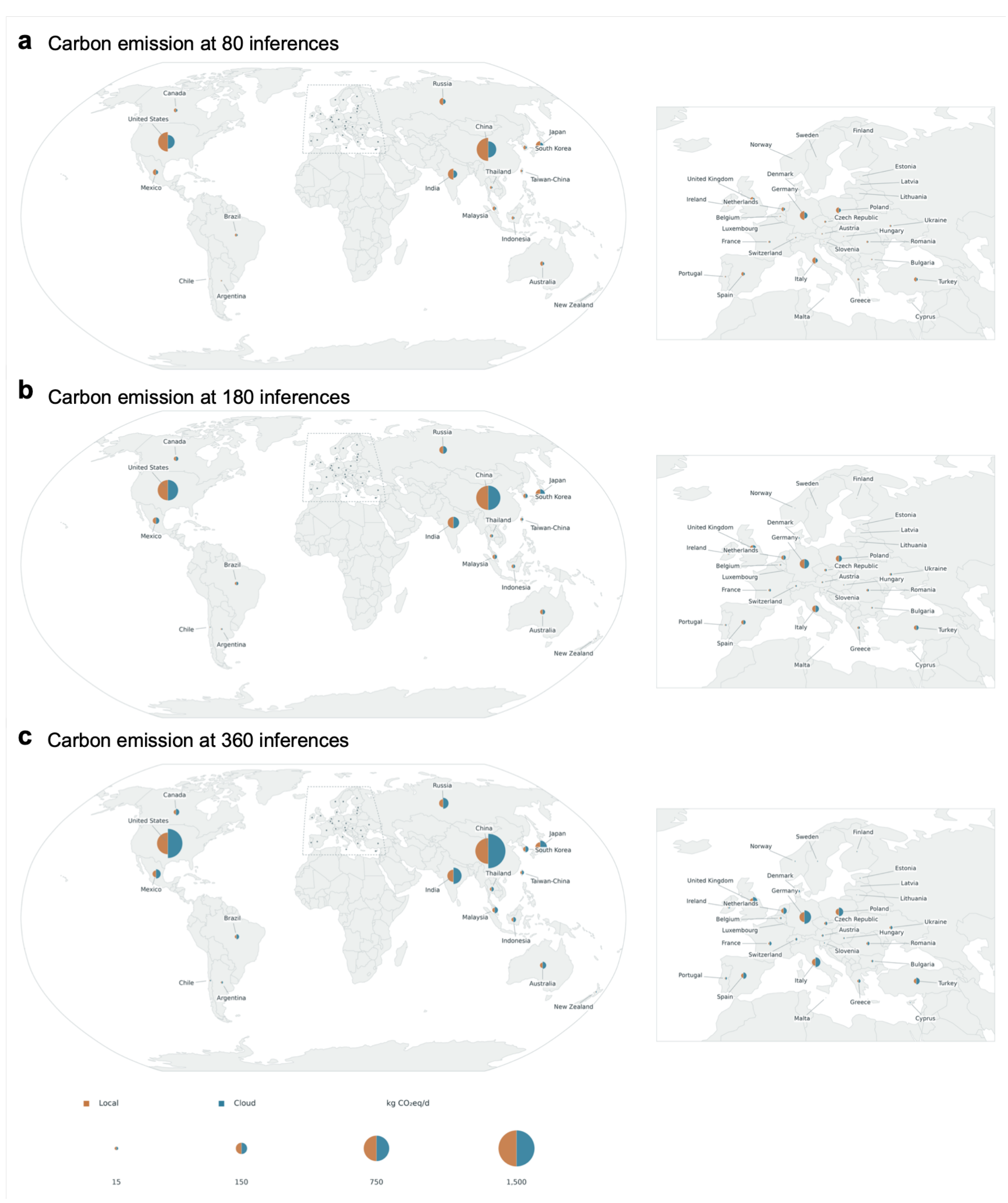

**Fig. S19 Maps compare the daily carbon emission of cloud and local deployment across 47 countries (regions) at 80, 180 and 360 inferences per WWTP per day.**

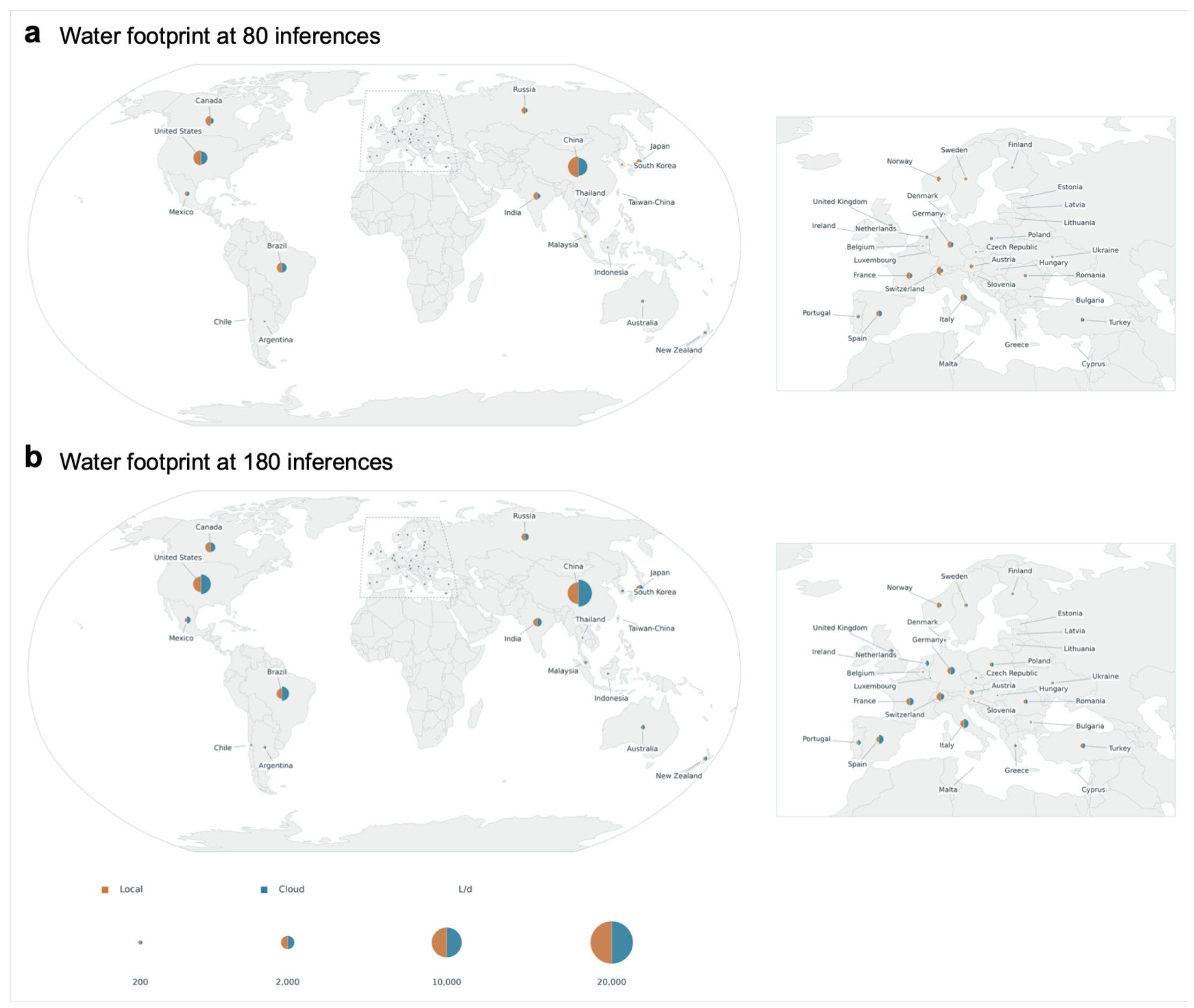

**Fig. S20 Maps compare the daily water footprint of cloud and local deployment across 47 countries (regions) at 80 and 180 inferences per WWTP per day.**